\documentclass[11pt]{article}

\usepackage{adjustbox}
\usepackage{amsfonts,amsmath,amssymb, amsthm}
\allowdisplaybreaks
\usepackage{appendix}
\usepackage{bm}
\usepackage{braket}
\usepackage{xcolor}
    \definecolor{darkgreen}{rgb}{0,0.5,0}
    \definecolor{darkred}{rgb}{0.5,0,0}
    \definecolor{darkblue}{rgb}{0,0,0.6}
    \definecolor{purple}{rgb}{0.4,.2,0.7}
\usepackage{comment}
\usepackage{esint}
\usepackage{euscript}
\usepackage{float}
\usepackage{graphicx}
\usepackage[hyperfootnotes = true, colorlinks = true, linkcolor = darkblue, citecolor = purple]{hyperref}
\usepackage{import}
\usepackage{makecell}
\usepackage{multirow}
\usepackage{subcaption}
\usepackage{tcolorbox}
\usepackage{tikz}
\usepackage[normalem]{ulem}

\makeatletter
\DeclareFontFamily{OMX}{MnSymbolE}{}
\DeclareSymbolFont{MnLargeSymbols}{OMX}{MnSymbolE}{m}{n}
\SetSymbolFont{MnLargeSymbols}{bold}{OMX}{MnSymbolE}{b}{n}
\DeclareFontShape{OMX}{MnSymbolE}{m}{n}{
    <-6>  MnSymbolE5
   <6-7>  MnSymbolE6
   <7-8>  MnSymbolE7
   <8-9>  MnSymbolE8
   <9-10> MnSymbolE9
  <10-12> MnSymbolE10
  <12->   MnSymbolE12
}{}
\DeclareFontShape{OMX}{MnSymbolE}{b}{n}{
    <-6>  MnSymbolE-Bold5
   <6-7>  MnSymbolE-Bold6
   <7-8>  MnSymbolE-Bold7
   <8-9>  MnSymbolE-Bold8
   <9-10> MnSymbolE-Bold9
  <10-12> MnSymbolE-Bold10
  <12->   MnSymbolE-Bold12
}{}

\let\llangle\@undefined
\let\rrangle\@undefined
\DeclareMathDelimiter{\llangle}{\mathopen}    {MnLargeSymbols}{'164}{MnLargeSymbols}{'164}
\DeclareMathDelimiter{\rrangle}{\mathclose}               {MnLargeSymbols}{'171}{MnLargeSymbols}{'171}
\makeatother

\newcommand{\tcr}{\textcolor {red}}
\newcommand{\tcb}{\textcolor {blue}}
\newcommand{\ii}{\mathrm{i}}
\newcommand{\UHP}{\mathrm{UHP}}
\newcommand{\bb}{\mathtt{b}}
\newcommand{\TTT}{\mathcal{T}}
\newcommand{\XXX}{\mathcal{X}}
\newcommand{\NNN}{\mathcal{N}}
\newcommand{\BBB}{\mathcal{B}}
\newcommand{\be}{\begin{equation}}
\newcommand{\ee}{\end{equation}}
\newcommand{\bC}{\mathbb{C}}
\def\p{\partial}
\def\ben{\begin{eqnarray}}
\def\een{\end{eqnarray}}
\def\BB{\mathcal{B}}
\def\MM{\mathcal{M}}
\def\wt{\widetilde}

\newcommand{\ns}{\mathrm{NS}}

\def\tilde{\widetilde}

\numberwithin{equation}{section}

\usepackage[margin = 2.2cm]{geometry}
\usepackage[ragged]{footmisc}
\begin{document}

\thispagestyle{empty}
\begin{center}
    ~\vspace{5mm}
    
    {\Large \bf 

        Instanton-Induced Closed-String Amplitudes in Minimal Superstring Theory at Subleading Order
        
    }
    
    \vspace{0.4in}
    
    {\bf Jyotirmoy Barman,$^1$ Rishabh Kaushik,$^1$ Raghu Mahajan,$^1$\vskip1ex Chitraang Murdia,$^{2}$ and Ashoke Sen.$^{1}$}

    \vspace{0.4in}

    $^1$ International Centre for Theoretical Sciences, Bengaluru, Karnataka 560089, India \vskip1ex 
    $^2$ Department of Physics and Astronomy, University of Pennsylvania, Philadelphia, PA 19104, USA 
    
    \vspace{0.1in}
    
    {\tt jyotirmoy.barman@icts.res.in, rishabh.kaushik@icts.res.in, raghu.mahajan@icts.res.in, murdia@sas.upenn.edu, ashoke.sen@icts.res.in}
\end{center}

\vspace{0.4in}

\begin{abstract}

We compute the disk one-point function, the disk two-point function, and the annulus one-point function of the cosmological constant operator in the type 0A and type 0B minimal superstring theories with $(1,1)$ ZZ instanton boundary conditions.
The moduli-space integrals appearing in the disk two-point function and the annulus one-point function have divergences associated with open-string-channel degenerations, which must be regulated using open-closed string field theory.
The definition of the string field theory interaction vertices requires a choice of locations for the picture-changing operators, which we specify in detail.
After carefully taking into account all contributions, including those from vertical integration, we find that the results precisely match the expectations from DDK-KPZ scaling.
Our technical results on the detailed construction of interaction vertices are a first step toward understanding the analogous quantities in the ten-dimensional type IIB superstring, where one also needs to understand how to treat the bosonic and fermionic collective coordinates at subleading order.

\end{abstract}

\pagebreak

\tableofcontents

\pagebreak

\section{Introduction and summary}
\label{sec:intro}

In string theory, the perturbative S-matrix for closed strings is computed by first evaluating appropriate correlation functions of vertex operators in a two-dimensional conformal field theory (CFT) on punctured Riemann surfaces of various genera, and then integrating these correlation functions over the corresponding moduli spaces \cite{Polchinski:1998rq}.
The choice of the two-dimensional conformal field theory corresponds to a particular choice of background in string theory.
The procedure for computing closed string amplitudes in superstring theory is similar, except that the relevant correlation functions involve insertions of additional operators known as picture-changing operators (PCOs) \cite{Friedan:1985ge, Belopolsky:1997jz, Polchinski:1998rr}.

Worldsheet methods also allow one to compute certain non-perturbative corrections to string amplitudes, known as D-instanton corrections \cite{Green:1997tv, Green:2000ke}.
D-instantons are D-branes with Dirichlet boundary conditions along all non-compact space-time directions and represent finite-action solutions in string theory \cite{Polchinski:1994fq}.
They therefore represent non-trivial saddle points of the string field theory path integral.
The D-instanton contributions to string amplitudes are then calculated as integrals of CFT correlation functions on punctured Riemann surfaces with boundaries, with D-instanton boundary conditions imposed at the worldsheet boundaries.

However, these D-instanton amplitudes often suffer from infrared (IR) divergences that need to be regulated.
String field theory (SFT) provides a natural setting in which these divergences can be treated systematically \cite{deLacroix:2017lif, Sen:2024nfd}.
In particular, SFT techniques have been successfully applied to the computation of the leading and first subleading D-instanton corrections to certain bosonic string theories whose underlying worldsheet CFT consists of either a free boson theory with central charge $c=1$ \cite{Balthazar:2019rnh, Sen:2019qqg, Sen:2020eck,Sen:2021qdk, Alexandrov:2023fvb, Kaushik:2025neu, Alexandrov:2025pzs}, or a minimal model CFT with central charge $c<1$  \cite{Eniceicu:2022nay, Eniceicu:2022dru, Eniceicu:2022xvk}, and a Liouville theory of appropriate central charge. 
SFT techniques have also been used to compute leading-order instanton contributions in type IIB superstring theory in ten space-time dimensions \cite{Agmon:2022vdj, Sen:2021tpp,Sen:2021jbr}, compactifications of type II superstrings \cite{Alexandrov:2021shf,Alexandrov:2021dyl,Alexandrov:2022mmy}, type 0B superstring theory in two dimensions \cite{Balthazar:2022apu,Sen:2022clw,Chakravarty:2022cgj}, type 0B minimal superstring theory \cite{Chakrabhavi:2024szk}, and on D3-branes in type IIB string theory \cite{Scheinpflug:2026vaw}.
Partial results for the subleading corrections in the flat ten-dimensional type IIB background have also been obtained \cite{Agmon:2022vdj, Agmon:2023zhi}.

In this paper we compute subleading terms in certain D-instanton amplitudes in minimal superstring theories \cite{Douglas:2003up, Seiberg:2003nm, Klebanov:2003wg}.
The worldsheet CFT consists of an $N=(1,1)$ superconformal minimal model, an $N=(1,1)$ super-Liouville theory, and the usual superconformal ghosts, which together yield a vanishing total central charge.
For most of the paper, we focus on the theory with the type 0A GSO projection.
The extension to the type 0B case is straightforward, and we carry it out as well.\footnote{
Type 0A minimal superstring theories are dual to double-scaled complex
matrix integrals, whereas type 0B theories are dual to double-scaled
Hermitian or unitary matrix integrals \cite{Morris:1990cq, Dalley:1991qg, Dalley:1991vr, Dalley:1991yi, Dalley:1992br, Gross:1980he, Wadia:1980cp, Periwal:1990gf, Nappi:1990bi, Crnkovic:1990ms, Klebanov:2003wg, Stanford:2019vob, Eniceicu:2023cxn, Johnson:2021owr, Rosso:2021orf}.
While we are studying the subleading corrections to amplitudes, the overall normalization of the instanton amplitudes (given by the exponential of the empty annulus) in the type 0A case has not appeared explicitly in the literature.
For the $(1,1)$ ZZ instanton in type 0A, this normalization can be inferred from the corresponding result in type 0B computed in \cite{Chakrabhavi:2024szk}.
}

This theory has D-instantons, which are (worldsheet) supersymmetric versions of the ZZ brane \cite{Zamolodchikov:2001ah,Ahn:2002ev, Fukuda:2002bv}.
We focus on the (1,1) type ZZ instantons.
Using the general form of the D-instanton contribution to the partition function of the theory, together with DDK-KPZ scaling properties  \cite{David:1988hj,Distler:1988jt,Knizhnik:1988ak}, one can make definite predictions for some of the correlation functions in the theory.
These include the disk one-point function, the disk two-point function and the annulus one-point function.
See equations (\ref{eq: disk_one_KPZ}) and (\ref{eq: f and g_KPZ_expectation}) below.
However, direct worldsheet expressions for the latter two quantities encounter divergences from integration over the moduli space of Riemann surfaces.
We use open-closed superstring field theory to resolve these divergences and find perfect agreement with the predictions of DDK-KPZ scaling.

The eventual goal of these studies is to extend the analysis to critical superstring theories, where the results can be tested against S-duality predictions and then used to make predictions for instanton corrections to amplitudes that are not predicted by supersymmetry and S-duality.
However, attempts to compute these corrections directly have run into difficulties in the past \cite{Agmon:2022vdj,Agmon:2023zhi}.
One difficulty in these computations is that several subtleties enter at once: PCOs and vertical integration, breakdown of the Siegel gauge, integration over the bosonic collective modes, and integration over the fermionic collective modes.
The advantage of working with toy models such as minimal superstring theories is that some of the difficulties are absent.
In particular, the ZZ instantons considered here have neither bosonic nor fermionic collective coordinates, allowing us to isolate the subtleties associated with PCOs and vertical integration.
The next step could be to extend the analysis to two-dimensional superstring theory, where one also needs to deal with bosonic collective modes.
The remaining additional subtlety in critical superstring theories is then the integration over fermionic collective modes.

The rest of the paper is organized as follows.
Section \ref{sbackground} reviews the necessary background, including our conventions for worldsheet correlators and PCOs, super-Liouville theory, D-instantons, and various SFT results.
We first study the type 0A GSO projection.
In section \ref{KPZ scaling}, we review the DDK-KPZ scaling property and use it to derive predictions for the disk one- and two-point functions and the annulus one-point function of the cosmological constant operator with ZZ-instanton boundary conditions.
In section \ref{Vertex_construction}, we review the construction of the interaction vertices of open-closed string field theory used in \cite{Sen:2020eck,Eniceicu:2022xvk}, and supplement this review by making specific choices for the locations of PCOs for each interaction vertex.
In section \ref{sonepoint}, we compute the disk one-point function using worldsheet techniques, and in section \ref{stwopoint}, we repeat this analysis for the disk two-point function, using SFT to regulate open-string-channel divergences.
Finally, in section \ref{sannulus}, we compute the annulus one-point function, again using SFT to tame divergences.
We find perfect agreement with the predictions from DDK-KPZ scaling for all three amplitudes. 
In section \ref{sec:0B}, we show how to extend these results to the type 0B GSO projection.
In appendix \ref{sa}, we find the relation between the normalization constant that appears in disk amplitudes and the tension of the D-instanton.
In appendix \ref{sec: diff_PCO_choice}, we repeat the analysis of sections \ref{sec:disk} and \ref{sannulus} using a different choice of PCO locations and demonstrate that the amplitudes are unaffected by this choice.
This serves as a nontrivial consistency check on our computations.

\section{Background material and conventions} \label{sbackground}

In this section we shall review some background material that will be needed for our analysis and also specify our conventions.
In the five subsections, we discuss the ghost sector, the picture-changing operator, string amplitudes, super-Liouville theory, and ZZ instantons, respectively.

\subsection{The ghost sector} \label{Basic conventions for worldsheet CFT}
 
The worldsheet superconformal algebra of minimal superstring theory is the $N=(1,1)$ super-Virasoro algebra.
Any such worldsheet theory contains a ghost SCFT, namely the $bc\beta\gamma$ system, which has central charge $-15$, and a matter SCFT of central charge $+15$. 
The matter sector is taken to be the $N=(1,1)$ super-Liouville theory \cite{Nakayama:2004vk} coupled to an $N=(1,1)$ minimal model \cite{Friedan:1984rv}. 
In this subsection, we shall describe the ghost sector; the matter sector will be described in section \ref{sec-liouville}.

The ghost sector consists of the Grassmann-odd ghosts $b$, $c$, $\widetilde b$, $\widetilde c$ and the Grassmann-even ghosts $\beta$, $\gamma$, $\widetilde\beta$, $\widetilde\gamma$.
The latter can be bosonized into a set of Grassmann-odd fields $\xi$, $\eta$, $\widetilde\xi$ and $\widetilde\eta$, and a pair of Grassmann-even chiral and anti-chiral scalars $\phi$, $\widetilde\phi$ via the relations \cite{Friedan:1985ge, Polchinski:1998rr, Sen:2024nfd}\footnote{There are some differences in the existing literature about the ordering of various factors in these relations. We follow the conventions of \cite{Sen:2024nfd}.}
\be
\beta = \p \xi \, e^{-\phi}, \qquad 
\gamma=\eta\, e^\phi, \qquad
\widetilde\beta = \bar\p \widetilde\xi \, e^{-\widetilde\phi}, \qquad \widetilde\gamma=\widetilde\eta\, e^{\widetilde\phi}\, .
\label{Bosonisation_conv}
\ee
The leading terms in some important operator product expansions (OPEs) are as follows:
\begin{align}
    b(z)\,c(w) &\sim \frac{1}{z-w} \, , \\
    \xi(z)\, \eta(w)&\sim \frac{1}{z-w} \,,  \label{xi-eta-ope}\\
    \partial \phi(z) \,\partial \phi (w) &\sim -\frac{1}{(z-w)^2} \,, \\
    e^{q_1 \phi}(z)\, e^{q_2 \phi}(w) &\sim (z-w)^{-q_1 q_2} e^{(q_1+q_2)\phi}\,.
\end{align}
The OPEs of the anti-holomorphic fields have similar form with the holomorphic coordinates replaced by anti-holomorphic coordinates.

For most of the paper, we shall work with the type 0A GSO projection in which we keep the $(\text{NS}+,\text{NS}+)$, $(\text{NS}-,\text{NS}-)$, $(\text{R}+, \text{R}-)$ and $(\text{R}-,\text{R}+)$ sectors in the closed-string Hilbert space.
The extension to the type 0B GSO projection is straightforward and is discussed in section \ref{sec:0B}.
The GSO parity of various worldsheet fields is shown in  table \ref{t1}, where we have also described the Grassmann parity, ghost number and picture number carried by various fields. 
We have only displayed the properties of the holomorphic fields; the anti-holomorphic fields satisfy identical properties.

\begin{table}[t]
\begin{center}
\renewcommand{\arraystretch}{1.6}
\begin{tabular}{|c|c|c|c|c|c|}
\hline 
\textbf{Operator} & $\bm{h}$ & 
\makecell{\textbf{Grassmann}\\\textbf{parity}}  & 
\makecell{\textbf{GSO}\\\textbf{parity}} &
\makecell{\textbf{Ghost}\\\textbf{number}} & \makecell{\textbf{Picture}\\\textbf{number}}
\\ \hline
$b$ & $2$ & $-$ & $+$ & $-1$ & $0$ \\ \hline
$c$ & $-1$ & $-$ & $+$ & $+1$ & $0$ \\ \hline
$\beta$ & $\frac{3}{2}$ & $+$ & $-$ & $-1$ & $0$ \\ \hline
$\gamma$ & $-\frac{1}{2}$ & $+$ & $-$ & $+1$ & $0$ \\ \hline
$\xi$ & $0$ & $-$ & $+$ & $-1$ & $+1$ \\ \hline
$\eta$ & $1$ & $-$ & $+$ & $+1$ & $-1$ \\ \hline
$e^{q\phi}$ & $-\frac{1}{2}q(q+2)$ & $(-1)^q$ & $(-1)^q$ & $0$ & $q$ \\ \hline 
PCO $\mathcal{X}$ & $0$ & $+$ & $+$ & $0$ & $+1$ \\
\hline
$e^{\alpha\varphi}$ & $\frac{1}{2}\alpha(Q-\alpha)$ & $+$ & $+$ & $0$ & $0$ \\ \hline
$\psi$ & $\frac{1}{2}$ & $-$ & $-$ & $0$ & $0$ \\ \hline
\end{tabular}
\end{center}
\caption{Some important quantum numbers carried by various worldsheet operators. \label{t1}}
\end{table}

In the upper half plane (UHP), the correlation functions in the ghost sector will be computed using the doubling trick, where we replace the anti-holomorphic fields $\wt b,\wt c,\wt\xi, \wt\eta, \wt\phi$ by their holomorphic counterparts placed at the complex conjugate point, and then evaluate the correlation function of the holomorphic fields in the full complex plane.
Following \cite{Sen:2024nfd},  we choose the normalization for the three-point function of the $c$-ghost,  and the one-point functions of the operators $e^{-2\phi}$ and $\xi$ to be\footnote{
In \eqref{eq:ev_ccc}--\eqref{e210}, the same notation is used for correlators in different sectors; the first correlator is in the $b$-$c$ sector, the second in the $\phi$ sector, and the third in the $\xi$-$\eta$ sector. Ideally, $\langle~\rangle$ should carry a superscript to distinguish between these different correlators, but we shall desist  from doing so for brevity. It should be clear from the context which correlator we are referring to. Most often it will be the full correlator involving all the ghost and matter sectors.}
\begin{align}
\label{eq:ev_ccc}
     \left\langle c(z_1) c(z_2) c(z_3) \right\rangle_{\bC} &=-K\,  (z_1-z_2)(z_2-z_3)(z_1-z_3) \, , \\
     \left\langle e^{-2\phi}(z) \right\rangle_{\bC} &= 1 \,, \\
     \left\langle \xi(z) \right\rangle_{\bC} &= 1\, . \label{e210}
\end{align}
The last correlator is written in the large Hilbert space and the constant $K$ is given below in \eqref{eKT}.
The locations of the operators $e^{-2\phi}$ and $\xi$ are irrelevant since they are dimension zero primaries.
With these conventions, the most general correlation function on the complex plane for the $bc$ system is 
\begin{align}
    \left\langle \prod_{i=1}^nb(x_i)\prod_{a=1}^{n+3}c(y_a) \right\rangle_{\bC}
        &= 
        - K \,
        \frac{\displaystyle\prod_{\substack{i,j=1 \\ i < j}}^n(x_i-x_j)\prod_{\substack{a,b=1\\ a < b}}^{n+3}(y_a-y_b)}{\displaystyle\prod_{i=1}^n\prod_{a=1}^{n+3}(x_i-y_a)}\, , \label{bcgeneralcorrelator}
\end{align}
and that for the $\xi\eta\phi$ system is
{\small
    \begin{align}
        \left\langle \prod_{i=1}^{n+1}\xi(x_i)\prod_{a=1}^n\eta(y_a) \prod_{r=1}^pe^{q_r\phi}(z_r)\right\rangle_{\bC}^{\text{large}}
        &=
        \frac{\displaystyle\prod_{\substack{i,j=1\\ i < j}}^{n+1}(x_i-x_j)\prod_{\substack{a,b=1\\ a < b}}^n(y_a-y_b)\prod_{\substack{r,s=1 \\ r < s}}^{p}(z_r-z_s)^{-q_rq_s}}{\displaystyle\prod_{i=1}^{n+1}\prod_{a=1}^n(x_i-y_a)}
        \, \delta_{\sum_r q_r,-2}\,.
        \label{exietaphi}
    \end{align}
}
\par\noindent
Although we have given the correlator \eqref{exietaphi} in the large Hilbert space, we shall write most of our subsequent formulas in the small Hilbert space \cite{Friedan:1985ge}. 
This means that in any correlator, all $\xi$ insertions either have derivatives acting on them or enter through differences of the $\xi$ field between two points.
To compute those from the large Hilbert space correlation function given above, we simply need to insert an extra $\xi$ at the \emph{leftmost} position of the correlator and compute this new correlator using \eqref{exietaphi}.
The result is independent of the location of this extra $\xi$ \cite{Friedan:1985ge,Verlinde:1987sd}.

The D-instanton contribution to any amplitude is accompanied by a
multiplicative factor of $e^{-\TTT}$ where $\TTT$ is the D-instanton tension.
The constant $K$ appearing in \eqref{eq:ev_ccc} 
is related to $\TTT$ via the relation
\be \label{eKT}
K= \frac{1}{2}\, \eta_c^{-1/2}\,  g_s \TTT, \qquad \text{where } \quad  \eta_c := \frac{\ii}{2\pi}\, .
\ee
The string coupling $g_s$ is defined implicitly via the precise definition of the string amplitudes, see (\ref{eapp1}) and section \ref{sec-sft-amplitudes} for more details.
The $\ii$ in the expression for $\eta_c$ should be interpreted as $e^{\ii\pi/2}$, and $\ii^\alpha$ for any number $\alpha$ should be interpreted as $e^{\ii\pi\alpha/2}$.
The relation \eqref{eKT} differs from that given in \cite{Sen:2024npu} by a minus sign. 
The reason for this has been described in appendix \ref{sa} where it is shown that the sign and the normalization in the relation between $K$ and $\TTT$ depends on the overall normalization of the picture-changing operator---we are using the normalization given in Eq.(\ref{pco}) below.

\subsection{The picture-changing operator (PCO)}

In this subsection we shall describe the precise definition of the picture-changing operator (PCO) that we shall use and also lay out some relevant facts related to picture-changing.

The holomorphic BRST current in superstring theories is given by
\begin{align}
j_\text{B} = c (T_m + T_{\xi\eta} + T_\phi) + b\, c\, \partial c + \gamma T_F - \frac{1}{4}\gamma^2 b.
\label{eq: J_b_expression}
\end{align}
Using \eqref{Bosonisation_conv}, in terms of the $\eta,\xi, \phi$ fields, we have $\gamma = \eta \, e^{\phi}$ and $\gamma^2 = \eta \, \partial \eta \, e^{2\phi}$.
Here, $T_m$ is the matter stress tensor and $T_F$ is the matter supercurrent, which together generate a super-Virasoro algebra with central charge 15. 
The explicit expressions for $T_{\xi \eta}$ and $T_\phi$ are
\begin{equation}
    T_{\xi \eta}= -\eta \, \partial \xi \,, \quad 
    T_\phi=-\frac{1}{2}\, (\partial \phi)^2-\partial^2\phi.
    \label{eq: stress_tensor_etaxiphi}
\end{equation}
We denote the corresponding BRST charge by $Q_\text{B}$.

Given these definitions, we take the PCO to be
\begin{align}
\mathcal{X}(z) := 2 \{Q_\text{B}, \xi(z)\}  = 2c \, \partial \xi + 2e^{\phi}\, T_F 
- \frac{1}{2} \left( 
\partial \eta\, e^{2\phi}\, b + \partial (\eta\, e^{2\phi}\,  b)
\right).
\label{pco}
\end{align}
Our definition \eqref{pco} differs from the usual one by a factor of two, but since the overall normalization of the PCO can be absorbed into a redefinition of the string coupling (and normalizations of the Ramond sector vertex operators), no physical results are affected by this choice. 
The anti-holomorphic counterpart $\widetilde{\mathcal{X}}(\bar{z})$ is defined similarly.
Note that the PCO carries picture number $1$.

For a surface with genus $g$ and $b$ boundaries, the total picture number of the insertions must be $2(2g + b-2 )$ to yield a non-vanishing correlator.
To achieve this, we need to insert PCOs on the Riemann surface, since the picture numbers of the external operators involved in the correlator might not add up to this required value. 
The final amplitude remains unaffected by the precise locations of the PCOs, provided they are placed in a manner that avoids spurious poles and satisfies factorization constraints near the moduli space boundaries \cite{Verlinde:1987sd, deLacroix:2017lif}. 
The latter translates into the condition that, in any NS-sector degeneration, an open-string state propagating through the degeneration must have picture number $-1$, whereas a closed-string state propagating through the degeneration must have picture number $(-1,-1)$.
Similar rules exist for R-sector degeneration.
These constraints dictate the distribution of PCOs across different Riemann surface components and the neck in the degeneration limit \cite{deLacroix:2017lif}. 

\subsection{String field theory amplitudes} 
\label{sec-sft-amplitudes}

In this section, we shall write down the expression for a general interaction term in open-closed SFT following \cite{Sen:2024nfd}.

We choose the sign of the SFT action $S$ such that the path integral is weighted by $e^{+S}$. 
Thus there is no relative sign between an amplitude and the corresponding interaction vertex of SFT.

A general Riemann surface can be described as a collection of basic components like spheres with three holes, disks around closed-string punctures, semi-disks around open-string punctures, etc. 
These components are glued  along closed curves $C_s$ and open curves $L_m$ (with the ends of $L_m$ lying on a worldsheet boundary). 
Let $\sigma_s$ be the coordinate system to the left of $C_s$ and $\tau_s$ the coordinate system to the right of $C_s$.
Similarly, let $\sigma_m$  be the coordinate system to the left of $L_m$ and $\tau_m$ the coordinate system to the right of $L_m$.
The worldsheet moduli $\{u^i\}$ are contained in the transition functions that relate $\sigma_s$ to $\tau_s$ and $\sigma_m$ to $\tau_m$.
Concretely, we have functional relations of the form
\be
\sigma_s = F_s(\tau_s, \vec u), \qquad \sigma_m = G_m(\tau_m, \vec u)\,.
\ee
Since we are considering superstring theory, we also need to specify the locations of the PCOs $\XXX$ and $\widetilde{\XXX}$. 
Let us suppose that the PCO insertions take the form $\prod_\alpha \XXX_\alpha(y_\alpha) \prod_\beta \widetilde{\XXX}_\beta(\widetilde y_\beta)$, where $y_\alpha$ and $\widetilde y_\beta$ are typically functions of the moduli $u^i$. 
We define
\begin{align}
\BB_i := &\phantom{+}
\sum_s \ \biggl[ \ 
\ointop_{C_s} d\sigma_s \frac{\p F_s}{\p u^i} b(\sigma_s) + 
\ointop_{C_s} d\bar\sigma_s \frac{\p \bar F_s}{\p u^i} \wt{b}(\bar \sigma_s) \ \biggr] \nonumber \\
&+
\sum_m \ 
\biggl[ \ 
\intop_{L_m}  d\sigma_m \frac{\p G_m}{\p u^i}  b(\sigma_m) + 
\intop_{L_m} d\bar\sigma_m  \frac{\p \bar G_m}{\p u^i} \wt{b}(\bar \sigma_m) \ 
\biggr]\nonumber \\
&- 2 \,\sum_\alpha \frac{1}{\XXX(y_\alpha)}  \p\xi(y_\alpha) {\p y_\alpha\over \p u^i}
-2\, \sum_\beta \frac{1}{\widetilde{\XXX}(y_\beta)} \bar\p\widetilde\xi(\widetilde y_\beta) {\p 
\widetilde y_\beta\over \p u^i}\, .
\label{eapp0}
\end{align}
The extra factors of two in the last line, compared to what was given in \cite{Sen:2024nfd}, can be traced to the extra factors of two in the definitions of $\XXX$ and $\wt\XXX$ that we are using; see (\ref{pco}).
Each of the integrals $\ointop_{C_s} d\sigma_s$ and $\int_{L_m} d\sigma_m$ is understood to include an intrinsic factor of $1/(2\pi \ii)$, while each anti-holomorphic integral includes an intrinsic factor of $-1/(2\pi \ii)$.
The $1/\XXX$ and $1/\wt{\XXX}$ factors in the last line of (\ref{eapp0}) simply mean that the corresponding PCOs should be dropped from (\ref{eapp1}) below.

For $n_c$ external closed-string states $A^c_1,\ldots,A^c_{n_c}$ and $n_o$ external open-string states $A^o_1,\ldots,A^o_{n_o}$, define a $p$-form on the moduli space $\MM_{g,b,n_c,n_o}$ of genus-$g$ Riemann surfaces with $b$ boundaries and the specified punctures by
\begin{align}
\Omega^{(g,b,n_c,n_o)}_p (A_1^c,\cdots, A_{n_c}^c; & A_1^o, 
\cdots , A_{n_o}^o )
:= g_s^{2g-2+b+n_c + \frac{1}{2} n_o}
\eta_c^{3g-3+n_c+{3\over 2} b
+{3\over 4} n_o}\, 
\frac{1}{p!} 
du^{i_1}\wedge \cdots \wedge du^{i_p}  \times \nonumber \\
& \times \Bigg\langle \BB_{i_1} \cdots
\BB_{i_p} \, \prod_\alpha \XXX_\alpha(y_\alpha)
\prod_\beta \widetilde{\XXX}_\beta(\widetilde y_\beta)\, 
A_1^c \cdots A_{n_c}^c
A_1^o \cdots A_{n_o}^o  
\Bigg \rangle_{\Sigma_{g,b,n_c,n_o }}
\, ,
\label{eapp1}     
\end{align}
where 
$\langle \cdots \rangle$ denotes the correlation function on the punctured Riemann surface $\Sigma_{g,b,n_c,n_o }$. 
Recall that $\eta_c = \frac{\ii}{2\pi}$ is simply a constant.
The vertex operators $A^c_i$ and $A^o_i$ are inserted
using the local coordinate system on the disks or the semi-disks on which the
corresponding puncture lies. 
An exception to the above formula is the disk one-point function of closed strings for which we use \cite{Sen:2024npu, Sen:2024nfd}
\be \label{ea4new}
\Omega^{(0,1,1,0)}_0(A^c) = \eta_c^{1/2} \langle c_0^- A^c\rangle_{\Sigma_{0,1,1,0}}\, .
\ee
Here $c_0^- = \frac{1}{2} (c_0 - \wt{c}_0)$.
Up to overall signs (discussed below), the string amplitudes are obtained by integrating  $\Omega^{(g,b,n_c,n_o)}_p$  over the moduli space $\MM_{g,b,n_c,n_o}$, with $p$ given by the dimension of 
the moduli space. 
However, this definition is formal since, typically, there are divergences in the relevant integrals coming from the boundaries of moduli space; in such cases the integral must be interpreted using the SFT prescription that we shall review below.

The interaction terms of SFT are obtained by integrating the same forms over an appropriate \emph{subspace} of the moduli space $\MM_{g,b,n_c,n_o}$, setting all the $A_i^o$'s to the open string field $\Psi_o$, all the $A_i^c$'s to the closed string field $\Psi_c$, and dividing the result by the combinatorial factor  $n_c!n_o!$.

To get the overall sign of the amplitudes we need to specify the precise definition of a positive integration measure on moduli space \cite{Sen:2024nfd}.
For amplitudes involving purely closed strings and without boundaries, the moduli space has a complex structure, and for a complex modulus $u=u_1+ \ii u_2$, we take $du_1\wedge du_2$ to have positive measure, and there is no additional sign in the amplitude.
In the presence of open strings or boundaries there is no such natural choice of the sign.  
The prescription arrived at in \cite{Sen:2024npu, Sen:2024nfd} is as follows.
For the disk amplitude with one open and one closed string, there is an additional minus sign besides the normalization constants given above in (\ref{eapp1}). 
Every additional open-string vertex operator on the boundary, with location parametrized by a modulus $u$, is accompanied by a factor of $-\BB_u$ inserted to the immediate left of the corresponding vertex operator. 
The integration over $u$ is taken to have positive measure if increasing $u$ moves the vertex operator in a direction that keeps the worldsheet to the left. 
Every additional closed-string insertion, say at $u=u_1+\ii u_2$, is accompanied by $du_1\wedge du_2\, \BBB_{u_1}\BBB_{u_2}$ with $du_1\wedge du_2$ having positive integration measure.

In our analysis, we shall also encounter vertical-integration contributions, which arise when the location of a PCO jumps as we move through moduli space.
These contributions can be regarded as limits of the moduli-space integral
defined above, where $y_\alpha$ or $\widetilde y_\beta$ varies sharply
from $y_1$ to $y_2$ as one of the moduli $u$ changes by a small amount.
Using \eqref{eapp0} and \eqref{eapp1}, we see that this corresponds to the insertion of
\be
-2\, (\xi(y_2) - \xi(y_1))\, ,
\ee
in place of $\BBB_u$ and dropping the corresponding PCO $\XXX(y_\alpha)$. 
A similar result holds when the location of the anti-holomorphic PCO jumps discontinuously.

Another pair of useful results that we shall borrow from \cite{Sen:2025xaj} are  the overall normalizations of the one-point function of closed strings and the one-point function of open strings on the annulus.\footnote{
See (5.3) and (5.47) of \cite{Sen:2025xaj}; the only change from the bosonic-string case treated there is the presence of PCOs.} 
We label the annulus by a complex coordinate $w$, with
\begin{equation}
    w = 2\pi (x + \ii y), \qquad 
  0 \leq x \leq \frac{1}{2}, \qquad 
  y \equiv y + t\, .
  \label{annulus_coord}
\end{equation}
Then the annulus one-point amplitude of an NS-sector closed-string vertex operator $V_C$ is given by
\ben 
\label{e2.47}
  && -2\pi \mathcal{N}_{\text{C}}\int_0^\infty d t \, \int_0^{1/4} d x \, \Bigg\langle
    \left(\int_0^{\pi}b(w)d w+\int_0^{\pi} \tilde{b}(\bar{w})d\bar{w}\right)
     \left(\oint_{2\pi x}b(w')d w'+\oint_{2\pi x}\tilde{b}(\bar{w}')d \bar{w}'\right)\nonumber \\
     && \hskip 2in \mathcal{X}(w_1)
     \mathcal{X}(w_2) \, V_C(w, \bar{w}) \Bigg\rangle_{\rm A} \, ,
\een
where $\ointop_{2\pi x}$ denotes integration around an anti-clockwise contour around $2\pi x$, $\langle \cdots \rangle_{\rm A}$ denotes the correlation function on the annulus, and
\be \label{eq: NC} 
\mathcal{N}_{\text{C}} = g_s \eta_c\, .
\ee
For an external open-string vertex operator $V_O$, the one-point function on the annulus is
\be \label{e253xx}
- \mathcal{N}_{\text{O}}\int_0^\infty d t \,  \Bigg\langle
    \left(\int_0^{\pi}b(w)d w+\int_0^{\pi} \tilde{b}(\bar{w})d\bar{w}\right)
      \mathcal{X}(w_1)
      \, V_O(w, \bar{w}) \Bigg\rangle \, ,
\ee
where
\be\label{eq: NO} 
\mathcal{N}_{\text{O}}=g_s^{1/2}\eta_c^{3/4}\, .
\ee

\subsection{Super-Liouville theory}
\label{sec-liouville}

Minimal superstring theory is based on an $N=(1,1)$ superconformal algebra on the worldsheet \cite{Seiberg:2003nm}. 
The matter sector of the worldsheet theory contains a super-Liouville theory \cite{Nakayama:2004vk} and a minimal SCFT \cite{Friedan:1984rv}, with both SCFTs together comprising a central charge of $+15$.
In our analysis we shall consider only those vertex operators for which the minimal SCFT component is the identity operator, and hence we do not need any information about it. 
In this subsection we shall review the Liouville SCFT part of the matter sector.

The Euclidean action of super-Liouville theory is given by\footnote{We remind the reader that, in our convention, the Euclidean path integral is weighted by $\exp(+S)$.}$^{,}$\footnote{We are writing the curvature term explicitly since it will be needed below. We have specialized to a flat metric in all the other terms.}
\begin{equation}
\label{eq:SL_action_2}
    S = -\frac{1}{2\pi} \int d^2 z \left[ \partial \varphi \, \bar{\partial} \varphi + \frac{1}{4}\, QR\varphi +\psi \, \bar{\partial} \psi +\widetilde{\psi} \, \partial \widetilde{\psi} - F^2 \right] - 2\mu \int d^2 z \left[ - \bb F \, e^{\bb \varphi} + \bb^2 \, \ii \psi \widetilde{\psi} \, e^{\bb \varphi} \right],
\end{equation}
where $\varphi$ is the Liouville field, $\psi$ and $\widetilde{\psi}$ are its superpartner Majorana-Weyl fermions and $F$ is the auxiliary field of this supermultiplet.
We are using the convention that $d^2 z = d x d y$ without any factors of two.
The quantity $\mu$ is the Liouville cosmological constant, and $Q$ and $\bb$ are parameters of the Liouville theory related to each other via
\be\label{eQb}
Q = \bb^{-1} + \bb\, .
\ee
Note that the last term in (\ref{eq:SL_action_2}) can be written as
\begin{align}
    (-2\pi \ii \mu) \cdot 
    \int \frac{-d^2 z}{\pi} \, (-\bb^2 \psi \widetilde{\psi} \, e^{\bb \varphi}) \,.
    \label{nice-form-mu}
\end{align}
The significance of this form lies in the fact that the precise combination $-c \widetilde{c}\, \bb^2 \psi \widetilde{\psi}\, e^{\bb \varphi}$ appears as one of the terms in $\mathcal{X}\widetilde{\mathcal{X}} \, c \widetilde{c} \, e^{-\phi} e^{-\widetilde{\phi}} \, e^{\bb \varphi}$ (see (\ref{PCO_action}) below), and our goal is to compute amplitudes involving the closed-string operator $c \widetilde{c} \, e^{-\phi} e^{-\widetilde{\phi}} \, e^{\bb \varphi}$.
The choice of measure $-d^2z/\pi$ has the following significance \cite{Sen:2024nfd}: turning on a background string field $\epsilon c\wt c V$ in the $(0,0)$ picture corresponds to deforming the worldsheet action by $-\epsilon\int \frac{d^2 z}{\pi} V$; see also \eqref{unintegrated-operator} and \eqref{integrated-action} below.
Motivated by the form of (\ref{nice-form-mu}), we introduce the quantity $\wt\mu$ related to $\mu$ via
\be\label{econvention}
\wt\mu := -2\pi \ii \mu\, .
\ee
One should think of $\mu$ as real and $\wt\mu$ as purely imaginary.

The equations of motion for $\varphi$ and $F$ on a flat background are
\begin{align}
\label{eq:EOM_1}
    \partial \bar{\partial} \varphi &= -2\pi \mu \bb^2 F \, e^{\bb\varphi}+2\pi\mu \bb^3 \, \ii \psi \widetilde{\psi}\, e^{\bb\varphi}, \\
\label{eq:EOM_4}    
    F &= -2 \pi \mu \bb \, e^{\bb\varphi}.
\end{align}
Note that since the right-hand side of (\ref{eq:EOM_1}) is non-zero, the operator $\partial \varphi$ is not holomorphic (and, similarly, $\bar{\partial} \varphi$ is not anti-holomorphic).

The OPEs of the $\varphi$ and $\psi$ are
\be
\partial \varphi(z) \,\partial \varphi (w) \sim -\frac{1}{(z-w)^2}, \qquad
\psi(z) \psi(w) \sim \frac{1}{z-w}\, .
\label{liouville-ope}
\ee
The stress tensor $T(z)$ and the supercurrent $T_{F}(z)$ for the super-Liouville SCFT are\footnote{These are free-field linear dilaton expressions and, as in standard in Liouville theory, they are supposed to be valid only in the asymptotic $\varphi \to - \infty$ limit.}
\begin{align}
    T(z) &= - \frac{1}{2} \, (\partial \varphi)^2 + \frac{Q}{2} \, \partial^2 \varphi - \frac{1}{2}\, \psi \, \partial \psi \, , \label{tliouville} \\
    T_F(z) &= \frac{\ii}{2} \,( \psi \, \partial \varphi  -  Q\, \partial \psi) \, .
\end{align}
Using the OPEs \eqref{liouville-ope}, we can calculate the OPEs between $T$ and $T_F$, which form the standard super-Virasoro algebra:
\begin{align}
    T(z)T(0) &\sim \frac{3(1 +2 Q^2)}{4z^4}+\frac{2T(0)}{z^2}+\frac{\partial T(0)}{z} \,,
    \label{TT OPE}  \\
     T(z) T_{F}(0) &\sim \frac{3T_{F}(0)}{2 z^2}+\frac{\partial T_{F}(0)}{z}\,,
    \label{TT_F OPE}  \\
    T_{F}(z)T_{F}(0) &\sim \frac{(1+2 Q^2)}{4z^3}+\frac{\,T(0)}{2z}.
    \label{T_FT_F OPE}
\end{align}
The first term in (\ref{TT OPE}) determines the central charge for the super-Liouville CFT to be $c = \frac{3}{2}+3Q^2$.
In the literature, conventions differ about the overall normalization of $T_F(z)$.
We have picked a choice convenient for us, which is reflected in the coefficient of the $z^{-3}$ term in (\ref{T_FT_F OPE}).

We will also need the transformation law of $\partial \varphi$ under the coordinate change $z \to f(z)$, which follows from its OPE with the stress tensor (\ref{tliouville})
\begin{equation}
    \partial \varphi(z, \bar{z}) \rightarrow \frac{Q}{2} \, \frac{f''(z)}{f'(z)} + f'(z) \,\partial \varphi\left(f(z), \overline{f(z)}\right)\,.
    \label{Coord_trs}
\end{equation}
There is a similar transformation for $\bar{\partial} \varphi$. 
Observe that $\partial \varphi$ and $\bar{\partial} \varphi$ are not conformal primaries due to the extra term proportional to $Q$.\footnote{This arises because there is a third-order pole in the OPE between the stress tensor and $\partial \varphi$. The coefficient of this third-order pole can be computed using free-field formulas.}

In the NS-NS sector, super-Liouville theory contains scalar primary operators of the form $V_\alpha=e^{\alpha \varphi}$ with scaling weights\footnote{The R-R sector vertex operators also contain spin fields, but they will not be relevant for our discussion.} 
\be\label{ehalpha}
h_\alpha= \widetilde{h}_\alpha =\frac{1}{2}\alpha (Q-\alpha)\, .
\ee
Using this expression for the scaling weights together with \eqref{eq:EOM_4}, we see that the interaction terms in the action \eqref{eq:SL_action_2} are marginal, since for $\alpha=\bb$, the operator $e^{\bb \varphi}$ has weights $\left(\frac{1}{2},\frac{1}{2}\right)$.

We want to compute the correlation functions of the cosmological constant vertex operator, which lies in the NS-NS sector, and is hence naturally expressed in the $(-1,-1)$ picture. 
Including the ghost factors, the cosmological constant vertex operator is
    \begin{align}
        V := c \widetilde{c}\, e^{-\phi}e^{-\widetilde{\phi}}\, e^{\bb \varphi}.
        \label{Cosmo_op}
    \end{align}  
The matter part of this vertex operator, which lies in the Liouville SCFT sector, will be denoted by $V_\bb = e^{\bb \varphi}$.
We need the following OPEs of $V_\bb$:
\begin{align}
    \partial{\varphi}(z)V_\bb(w, \bar{w}) \, &\sim \, \frac{-\bb}{z-w}
    V_\bb(w, \bar{w}) \,,
    \label{V OPE} \\
    T_F(z)\,  V_\bb(w, \bar{w}) &\sim \frac{-\ii \bb/2}{(z-w)}\, \psi \, 
    V_\bb(w, \bar{w}) \,, \\
    \widetilde T_F(\bar z)\,  V_\bb(w, \bar{w}) &\sim \frac{-\ii \bb/2}{(\bar z-\bar w)}\, \widetilde\psi  \,  
    V_\bb(w, \bar{w})\, .
\end{align}
Having discussed the bulk properties of the worldsheet SCFT, we now turn to the properties of the boundary state of interest.

\subsection{ ZZ instantons}
\label{sec-zz-inst}

We shall be computing D-instanton corrections to the amplitudes in minimal superstring theory.
The analog of D-instanton boundary conditions in Liouville theory are the ZZ boundary conditions \cite{Zamolodchikov:2001ah}.
Basic observables for the ZZ branes in super-Liouville theory were computed using the bootstrap approach in \cite{Ahn:2002ev, Fukuda:2002bv}, closely following the original analysis in the bosonic case.
See \cite{Nakayama:2004vk} for a review.
For simplicity, we consider the basic $(1,1)$ ZZ brane, which has the property that the Hilbert space of an open string with both ends on this brane contains only the identity operator \cite{Zamolodchikov:2001ah}.
In this subsection, we shall review a few more important features that will be relevant for our analysis.

With the type 0A GSO projection, the super-Liouville sector of the ZZ instanton boundary state is given by \cite{Ahn:2002ev, Fukuda:2002bv, Seiberg:2004ei, Irie:2007mp}
\begin{align}
     \ket{(1,1)} &= \sqrt{2} \int_{0}^{\infty}  d P \, \psi^{\ns}_{1,1}(P) \,|\ns, P , -\rrangle  \, , \label{11-state-0a} \\
     \Psi^\ns_{1,1}(P) &= \left( \pi \mu \gamma\left( \frac{\bb Q}{2} \right) \right)^{-\ii P \bb^{-1}}  \frac{\sqrt{2}\,\pi}{P\,\Gamma \left( -\ii P \bb^{-1} \right)\,\Gamma \left( -\ii P \bb \right)} \, ,
\end{align}
where the state on the right side of the top line is the NS sector Ishibashi state.
Also, $\Gamma(x)$ is the gamma function and we use the standard definition $\gamma(x) = \tfrac{\Gamma(x)}{\Gamma(1-x)}$.
The minimal SCFT wavefunction corresponds to identity state.
The type 0B GSO projection is discussed in section \ref{sec:0B}.

It turns out that the partition functions of super-Liouville theory, of the $(2,4k)$ minimal SCFT,\footnote{Here $k$ is an integer greater than or equal to one.} and of the $bc\beta\gamma$ system are \cite{Ahn:2002ev, Goddard:1986ee,  Chakrabhavi:2024szk}
\begin{align}
    Z_{\text{L}}(v) &= \sqrt{\frac{\vartheta_3(v)}{\eta(v)^3}} \left( v^{-\frac{(2k+1)^2}{16k}} - v^{-\frac{(2k-1)^2}{16k}} \right) \, , \\
    Z_{\text{M}}(v) &= \sqrt{\frac{\vartheta_3(v)}{\eta(v)^3}} \sum_{j\in\mathbb{Z}}\left( v^{\frac{(8kj+2k-1)^2}{16k}} - v^{\frac{(8kj+2k+1)^2}{16k}} \right) \, , \\
    Z_{\text{G}}(v) &= \frac{\eta(v)^3}{\vartheta_3(v)} \, , 
\end{align}
where we define the ``nome" $v$ as
\begin{equation}
\label{eq:nome_def}
    v := e^{-2\pi t} \, .
\end{equation}
Recall that $2\pi t$ is the open-string time (\ref{annulus_coord}).
The full annulus partition function in string theory is a product of these three contributions:
\begin{equation}
\label{eq:Zv_def}
    Z(v) = Z_{\text{L}}(v) Z_{\text{M}}(v) Z_{\text{G}}(v) = \sum_{j\in\mathbb{Z}}\left( v^{\frac{(4jk-1)(2j+1)}{2}} - v^{(4jk+2k+1)j}  - v^{(4jk+2k-1)j} + v^{\frac{(4jk+1)(2j+1)}{2}}  \right) \, .
\end{equation}
In the limit $v \to 0$, the leading terms in $Z(v)$ are
\begin{align}
    Z(v) = v^{-1/2} - 2 + O(v^{1/2})\, ,
    \label{zv-leading}
\end{align}
which come from the $j=0$ term in the sum in (\ref{eq:Zv_def}).

The leading term $v^{-1/2}$ in (\ref{zv-leading}) shows that the open-string spectrum on the $(1,1)$ ZZ instanton includes a tachyon with $L_0 = -1/2$.
The tachyon is described by the vertex operator $\zeta\, c\, e^{-\phi}$, where $\zeta$ can be regarded either as a  Chan-Paton factor \cite{Sen:1999mg} or a boundary fermion mode \cite{Witten:2023snr} that squares to 1 and behaves as a Grassmann odd object.
The presence of this tachyon causes the worldsheet integrals to diverge at the boundaries of moduli space. 
This happens since moduli space integrals are obtained from SFT amplitudes by replacing the $1/L_0$ factor in the propagator by its Schwinger parametrization
\be\label{elzeroq}
L_0^{-1} = \int_0^1 dq \, q^{L_0-1}\, .
\ee
For tachyonic states with negative $L_0$ the right-hand side diverges. 
The remedy is to replace it by the left-hand side. 
This leads to the following replacement rule:
\be \label{ereprule}
\int_0^1 dq\, q^{\alpha-1} \, \longrightarrow \, \alpha^{-1}\, ,
\qquad \text{when } \alpha < 0\, .
\ee
The quantity $q$ is also known as the sewing parameter, since from the worldsheet perspective, the propagator arises by identifying the local coordinates $w_1$, $w_2$ around two open-string punctures (at the two ends of the propagator, respectively) via the plumbing-fixture relation $w_1w_2=-q$, thereby sewing the semi-disks to which these punctures belong.

In the Siegel gauge of SFT, which the worldsheet theory uses, there are also a pair of $L_0=0$ states associated with the vertex operators $\eta \, c$ and $\p\xi \, c\, e^{-2\phi}$.
These give rise to the $-2$ contribution in the annulus partition function (\ref{zv-leading}).
For these states, both the left and right hand sides of (\ref{ereprule}) diverge, and so we need to examine the physical origin of these $L_0=0$ states more closely. 

It turns out \cite{Sen:2020cef, Sen:2021qdk, Sen:2021tpp} that, on a D$p$ brane with $p \geq 0$, these modes, now also carrying momentum parallel to the D$p$-brane worldvolume, are the vertex operators for the Faddeev-Popov ghosts associated with gauge fixing the U(1) gauge symmetry on the brane. 
However, since the U(1) is not local symmetry for the D-instanton (the worldvolume is a single point), fixing the U(1) gauge is meaningless. 
Hence, the remedy is to remove these Faddeev-Popov ghosts from the path integral and instead integrate over the field associated with the vertex operator $ \p\xi \, c\, \partial c  \, e^{-2\phi}$ that was erroneously set to zero using the U(1) gauge 
transformation \cite{Sen:2020cef, Sen:2021qdk, Sen:2021tpp}. 
We will refer to this field as the out-of-Siegel gauge (OSG) field.
So we supplement the rule \eqref{ereprule} by
\be \label{ereprule1}
\int_0^1 dq\, q^{-1} \, \longrightarrow \, 0\, ,
\ee
and allow the OSG field to propagate along the internal open-string propagators.
When we do this, we also need to divide the path integral by the volume of the gauge group \cite{Sen:2021qdk}.
For this, we need to find the appropriate Jacobian factor that arises from the relation between the SFT gauge transformation parameter and the rigid U(1) transformation parameter.
At subleading orders in the perturbation expansion, this induces additional terms in the effective action and hence additional contributions to string amplitudes \cite{Sen:2020eck, Eniceicu:2022xvk}.

For reference, table~\ref{t2} summarizes the properties of the vertex
operators that are important in our analysis.

\begin{table}[t]
\begin{center}
\renewcommand{\arraystretch}{1.6}
\begin{tabular}{|c|c|c|c|c|c|c|}
\hline 
\makecell{\textbf{Vertex}\\\textbf{operator}}  & 
$\bm{h}$ & 
\makecell{\textbf{Grassmann}\\\textbf{parity}}  & 
\makecell{\textbf{GSO}\\\textbf{parity}} &
\makecell{\textbf{Ghost}\\\textbf{number}} & \makecell{\textbf{Picture}\\\textbf{number}} & 
\textbf{Remarks}
\\ \hline
$c \, e^{-\phi}$ & $-\frac{1}{2}$ & $+$ & $-$ & $1$ & $-1$ & Tachyon \\ \hline
$\partial \xi \, c \, e^{-2\phi}$ & $0$ & $+$ & $+$ & $0$ & $-1$ & Faddeev-Popov `c'-ghost\\ 
\hline
$\eta \, c$ & $0$ & $+$ & $+$ & $2$ & $-1$ & Faddeev-Popov `b'-ghost \\ 
\hline
$\partial \xi \, c \, \partial c\, e^{-2\phi}$ & $0$ & $-$ & $+$ & $1$ & $-1$ & Out-of-Siegel-gauge mode\\ 
\hline
$\eta \, c\, \partial c$ & $0$ & $-$ & $+$ & $3$ & $-1$ & Antifield of $\partial \xi\, c\, e^{-2\phi}$ \\ 
\hline
\end{tabular}
\end{center}
\caption{Properties of some vertex operators that play an important role in our analysis.} \label{t2}
\end{table}

We now review the disk one-point functions in super-Liouville theory with $(1,1)$ ZZ boundary conditions. 
For this boundary condition, the one-point functions of $V_\alpha$ and $\ii\psi\widetilde{\psi}V_\alpha$ on the upper half-plane are \cite{Ahn:2002ev, Fukuda:2002bv}
\begin{align} 
    \langle V_{\alpha}(z)\rangle_{\UHP} &=\frac{U(\alpha)}{|z-\bar{z}|^{2h_{\alpha}}} \,, \label{e219x}\\
    \langle\ii \psi \widetilde{\psi} V_{\alpha}(z)\rangle_{\UHP} &= (Q-\alpha) \, \alpha^{-1}\frac{U(\alpha)}{|z-\bar{z}|^{2h_{\alpha}+1}}\, ,\quad \text{where}
    \label{eq: Liouville_one_pt_fermion} \\
    U(\alpha) &= 
    \left( \pi \mu \gamma \left(\frac{\bb Q}{2}\right) \right)^{-\frac{\alpha}{\bb}} 
    \frac{\Gamma \left( \frac{\bb Q}{2} \right) \Gamma \left( \frac{Q}{2\bb} \right) \frac{Q}{2}}
    {\Gamma \left( -\alpha \bb + \frac{\bb Q}{2} \right) 
    \Gamma \left( -\frac{\alpha}{\bb} + \frac{Q}{2\bb} \right) \left( \frac{Q}{2} - \alpha \right)}.
    \label{eq: U_alpha_expression}
\end{align}
Specializing to $\alpha = \bb$, the one-point functions of $V_{\bb}$ and $\ii \psi \widetilde{\psi} V_{\bb}$ on the upper half plane are
\begin{align}
    \langle V_{\bb}(\ii)\rangle_{\UHP} &= \frac{Q}{2\bb\, \ii\wt\mu}\, ,   \label{one-pt-vb}\\
    \langle \ii \psi \widetilde{\psi} V_{\bb}(\ii)\rangle_{\UHP} &=
    \frac{Q}{4 \bb^3 \, \ii \wt\mu}\, ,
    \label{eq: Liouville one-point function}
\end{align}
where recall that $\widetilde{\mu}$ is a rescaling of $\mu$, defined in (\ref{econvention}).

Using the equation of motion \eqref{eq:EOM_1} on a flat surface after neglecting the $F$ term,
\begin{equation} \label{epartial}
    \partial \bar{\partial} \varphi = 2\pi\mu \bb^3 \, \ii \psi \widetilde{\psi}\, e^{\bb\varphi}\, .
\end{equation}
Using \eqref{eq: Liouville_one_pt_fermion} for $\alpha = \bb$, we now get
\begin{equation} \label{epartial1}
        \langle \partial \bar{\partial} \varphi (z,\bar{z}) \rangle_{\UHP}=\frac{Q}{|z-\bar{z}|^2}\, .
\end{equation}
Dropping the term proportional to $F$ in (\ref{epartial}) can be justified as follows. 
Since the equation of motion of $F$ does not contain derivatives acting on $F$, the correlation functions of $F$ are contact terms, i.e. are non-vanishing only when the argument of $F$ approaches another vertex operator or the boundary. 
In the approach based on string field theory, we never let the argument of a vertex operator approach another vertex operator or a boundary; these are treated using Feynman diagrams involving lower order vertices. 
Hence, in the region where we will apply \eqref{epartial}, the terms proportional to $F$ can be dropped.

Upon integrating \eqref{epartial1} and noting that the two lowest-dimension operators that can appear in the bulk-boundary OPE are the identity operator and the stress tensor, we find
\begin{equation} \label{e2.27}
    \partial \varphi (z,\bar{z})=-\frac{Q}{z-\bar{z}}+\mathcal{O}(z-\bar{z}),
\end{equation}
\begin{equation} \label{e2.28}
    \bar{\partial} \varphi (z,\bar{z})=+\frac{Q}{z-\bar{z}}+\mathcal{O}(z-\bar{z}).
\end{equation}
The absence of terms of order $(z-\bar{z})^0$ in the OPE can be traced to the absence of dimension-one boundary operators on the $(1,1)$ ZZ brane in Liouville theory. 
Using \eqref{e2.27} and \eqref{e2.28}, we obtain the one-point functions
\be \label{ephiexpect}
\langle \partial \varphi (z,\bar{z}) \rangle_{\UHP} = -\frac{Q}{z-\bar{z}},
\qquad \langle \bar{\partial} \varphi (z,\bar{z})\rangle_{\UHP} = \frac{Q}{z-\bar{z}}\, .
\ee
Similarly, from \eqref{e219x} we can get the leading term in the bulk-boundary OPE of $V_\alpha$:
\be \label{e2.19OPE}
V_\alpha(z, \bar{z}) \sim {U(\alpha)\over |z-\bar z|^{2h_\alpha}} \, \left(1 + \mathcal{O}|z-\bar z|^2 \right)\,.
\ee

For the sake of completeness, we note that, in the matrix-integral dual, the $(1,1)$ ZZ brane contributes in the ungapped phase \cite{Seiberg:2003nm}.
On the string theory side side, the sign of $\mu$ determines whether the theory is in the gapped or the ungapped phase.

\section{Predictions from DDK-KPZ scaling}
\label{KPZ scaling}

In this section, we will derive the value of the disk one-point, the disk two-point, and the annulus one-point function of the Liouville cosmological constant operator (all suitably normalized) using the DDK-KPZ scaling argument \cite{David:1988hj,Distler:1988jt,Knizhnik:1988ak}.
We closely follow the discussion in the bosonic case \cite{Eniceicu:2022xvk}.

The partition function of the minimal superstring, including both the perturbative and the one-instanton contributions, is given by
\begin{align}
    Z(\tilde{\mu}, g_s) &= Z^{(0)}(\tilde{\mu}, g_s) + Z^{(1)}(\tilde{\mu}, g_s) + \dots,
    \label{eq: partition_function} \\
    Z^{(1)}(\tilde{\mu}, g_s) &= Z^{(0)}(\tilde{\mu}, g_s) \exp \left( g_s^{-1} A(\tilde{\mu}) + \frac{1}{2} \log g_s + B(\tilde{\mu}) + g_s C + \dots \right),
    \label{eq: one_instanton_partition_fn}
\end{align}
where $\tilde{\mu}$ is a simple rescaling of $\mu$ defined in (\ref{econvention}).
Here $Z^{(0)}(\tilde{\mu}, g_s)$ is the perturbative contribution to the partition function and $Z^{(1)}(\tilde{\mu}, g_s)$ is the one-instanton contribution.  
The $Z^{(0)}$ factor on the right-hand side of (\ref{eq: one_instanton_partition_fn}) can be traced to the fact that we need to include contributions from disconnected worldsheets, some of which may have boundaries ending on the instanton and others may have no boundaries.
The coefficient $\frac{1}{2}$ multiplying the $\log g_s$ term is special to the $c<1$ minimal string/superstring theory, for generic instantons.\footnote{The instanton lying at the origin of eigenvalue plane in the gapped phase of the complex matrix integral is an exceptional case. In the string theory language, this corresponds to the $(1,2k)$ ZZ instanton in the $(2,4k)$ minimal string.} 
It can be traced to the breakdown of the Siegel gauge for open strings living on D-instantons \cite{Sen:2021qdk, Eniceicu:2022nay}.

Taking derivatives of $Z^{(1)}(\tilde{\mu}, g_s)$ with respect to $\tilde{\mu}$, we can get the one-instanton contribution to the $n$-point function of the Liouville cosmological constant operator. 
In taking the derivatives, there will be terms in which one or more derivatives hit the $Z^{(0)}(\tilde{\mu}, g_s)$ factor in \eqref{eq: one_instanton_partition_fn}. 
Such terms will produce closed-string worldsheet components without boundaries and will not be the subject of interest in our work, so we shall not write them. 
So the one instanton contribution to the $n$-point function of the cosmological constant operator will be given by,
\begin{align}
   g_s^n\, \frac{1}{Z^{(0)}} 
   \frac{\partial^n Z^{(1)}}{{\partial \,\tilde{\mu}}^n} 
    &\supset 
    e^{g_s^{-1} A(\wt\mu) + \frac{1}{2} \log g_s + B(\wt\mu)} 
    \left( \frac{\partial A}{\partial \tilde{\mu}} \right)^n \nonumber \\
    &\times \left(
    1 + {n (n-1)\over 2}\, g_s  \frac{\partial^2 A}{{\partial \tilde{\mu}}^2} 
    \bigg/ \left( \frac{\partial A}{\partial \tilde{\mu}} \right)^2
    + n\,  g_s  \frac{\partial B}{\partial \tilde{\mu}} \bigg/ \frac{\partial A}{\partial \tilde{\mu}} 
    + g_s C + O(g_s^2)
    \right).
    \label{eq: partition_fn_derivative}
\end{align}
In our convention, each closed-string vertex operator carries a factor of $g_s$, as in \eqref{eapp1}; this accounts for the factor of $g_s^n$ on the left-hand side of \eqref{eq: partition_fn_derivative}.

We now compare various terms in \eqref{eq: partition_fn_derivative} with the expected string amplitudes. 
The quantity $g_s^{-1} A(\tilde{\mu})$ is the instanton action and $\exp\left(\frac{1}{2} \log g_s + B(\tilde{\mu})\right)$ is the exponential of the annulus partition function (also known as the ``normalization" of the instanton-contribution).
The D-instanton tension $\TTT$ is given by
\be \label{etension}
\TTT = K_0\, g_s^{-1}, \qquad \text{with } \quad K_0 := - A(\tilde\mu)  \, .
\ee
The value of $K_0$ can be computed in principle, but we shall not need it.

In the worldsheet calculation, the leading non-perturbative contribution to $n$-point function of closed string operators is given by $e^{-\mathcal T}$ times the annulus partition function times the product of $n$ disk one-point functions. 
Since we do not consider terms where an external closed string is inserted on a Riemann surface without boundary, the $Z^{(0)}$ factor drops out.
Comparing this with \eqref{eq: partition_fn_derivative}, we get the one-point function of the cosmological constant operator on the disk as:
\begin{equation}
    A_{\text{disk}}(V) = \frac{\partial A}{\partial \tilde{\mu}}\, .
    \label{eq: disk one-point definition}
\end{equation}

The first subleading correction to the non-perturbative contribution to the disk $n$-point function has several components. 
First, in the product of $n$ disk one-point functions, any pair may be replaced by a disk two-point function.
This comes with a combinatorial factor of $\binom{n}{2}=\frac{n(n-1)}{2}$. 
Second, one of the disk one-point functions can be replaced by an annulus one-point function. This carries a combinatorial factor of $n$.
Finally, the product of $n$ disk one-point functions can be multiplied by a 0-point function on a three-holed sphere or a handle-disk. 
This will not carry any $n$-dependent factor.
If we denote by $g_s f$  the ratio of the disk two-point function to the square of the disk one-point function and by $g_s g$ the ratio of the annulus one-point function to the disk one-point function, 
then comparison with \eqref{eq: partition_fn_derivative} yields
\begin{equation}
    f = \frac{\partial^2 A}{\partial \tilde{\mu}^2} \bigg/ \left( \frac{\partial A}{\partial \tilde{\mu}} \right)^2 \,, \qquad
    g = \frac{\partial B}{\partial \tilde{\mu}} \bigg/ \frac{\partial A}{\partial \tilde{\mu}} \,.  
    \label{eq: f and g_definition}
\end{equation}
The contribution  $C$ in \eqref{eq: partition_fn_derivative} comes from a three-holed sphere and a handle-disk, but we will not consider these topologies in this work.

We shall now determine the form of $A_{\text{disk}}(V)$, $f$ and $g$ using the DDK-KPZ  scaling argument \cite{David:1988hj,Distler:1988jt,Knizhnik:1988ak}.
We begin by noting that the super-Liouville CFT partition function on a given Riemann surface with Euler number $\chi$  takes the form
\begin{align}
Z_\chi(\mu) &= \int [\mathcal{D}\varphi] \, 
\exp (S)\, , 
\label{Liouville}
\end{align}
where the action of Liouville theory is given in (\ref{eq:SL_action_2}).
We now shift the zero mode of the Liouville field using a field
redefinition $ \varphi \to \varphi - \frac{1}{\bb} \log \mu$, which allows us to determine the relation between $Z_\chi(\mu)$ and $Z_\chi(1)$.
The linear shift leads to a term in the exponent which is proportional to $ R \log \mu$ and the surface integral of the Ricci scalar gives us the Euler characteristic of the surface.\footnote{As presented, this argument is only for closed surfaces. Nevertheless, it continues to hold for surfaces with boundary as well.}
This leads to
\begin{align}
    Z_\chi(\mu) &= 
    \left( \mu^{{\frac{Q}{2\bb}}} \right)^{\chi} Z_\chi(1) \, .
\end{align}
Thus the partition function has a simple power-law dependence on $\mu$.
On the other hand, the full string partition function depends on the string coupling $g_s$ via an overall multiplicative factor of $g_s^{-\chi}$. 
It follows that the dependence of the partition function on $g_s$ and $\mu$ occurs through the single combination $g_s^{-1}\mu^{Q/2\bb}$.
Therefore in \eqref{eq: one_instanton_partition_fn}, we must have
\begin{align}
    g_s^{-1} A(\tilde{\mu}) &= - A_*\, g_s^{-1} \tilde{\mu}^{Q/2\bb}, 
    \label{eq: g_s definition} \\
    \frac{1}{2} \log g_s + B(\tilde{\mu}) &= \frac{1}{2} \log g_s - \frac{Q}{4\bb} \log \tilde{\mu} + B_*, \label{eq: annulus_KPZ}
\end{align}
where the constants $A_*$ and $B_*$ are independent of $\tilde{\mu}$ and $g_s$. 
Crucially, these equations give us the full $\tilde{\mu}$-dependence of $A(\tilde{\mu})$ and $B(\tilde{\mu})$.

We can now determine the disk one-point function by plugging (\ref{eq: g_s definition}) into the definition (\ref{eq: disk one-point definition}):
\begin{equation}
    \boxed{
    A_{\text{disk}}(V) = 
    -\frac{Q}{2 \bb\wt\mu} \, K_0\, .}
    \label{eq: disk_one_KPZ}
\end{equation}
Here $K_0 = g_s \mathcal{T}$, as defined in (\ref{etension}).
Plugging (\ref{eq: g_s definition}) and (\ref{eq: annulus_KPZ}) into \eqref{eq: f and g_definition}, we obtain the DDK-KPZ prediction for $f$ and $g$
\begin{equation}
    \boxed{
    f = \frac{1}{K_0}\, \left(\frac{2\bb}{Q} -1\right) , 
    \qquad
    g = \frac{1}{2\, K_0}.}
    \label{eq: f and g_KPZ_expectation}
\end{equation}
The main goal of this paper is to verify (\ref{eq: disk_one_KPZ}) and (\ref{eq: f and g_KPZ_expectation}) by explicit worldsheet computations while using SFT to regularize divergences from the boundaries of the moduli space.

\section{String vertices and PCO locations}
\label{Vertex_construction}
In this section, we specify the SFT interaction vertices needed for our computations.
Firstly, we need to specify the local coordinates at the open-string punctures and the range of moduli parameters associated with the interaction vertex.
We take this data to be the same as in \cite{Sen:2020eck, Eniceicu:2022xvk, Sen:2025xaj}.\footnote{The detailed form of the OOOO vertex on the upper half-plane was not needed in \cite{Sen:2020eck, Eniceicu:2022xvk}. Here, it is needed only for the alternate choice of PCO locations discussed in appendix \ref{sec: diff_PCO_choice}.
}
Since external closed strings are always on-shell, we do not need to specify local coordinates at the closed-string punctures.

Secondly, since we are working with superstrings as opposed to the bosonic string analysis in \cite{Sen:2020eck, Eniceicu:2022xvk, Sen:2025xaj}, we further need to specify the location of PCOs required in each interaction vertex.
The number of PCOs required for an amplitude with only NS-sector vertex operators is
\begin{align}\label{eq: PCO_number}
 n_{\text{pco}}=2n_{c}+n_{o}-2(2-2g-b)\,,
\end{align}
where $n_{c}$ is the number of closed string insertions,  $n_o$ is the number of open string insertions,  $g$ is the genus, and $b$ is the number of boundaries. 
A concise summary of our choice of PCO locations is given in table \ref{ta2}. 
In the rest of this section, we shall concisely review the interaction vertex conventions of \cite{Sen:2020eck, Eniceicu:2022xvk, Sen:2025xaj} and elaborate on table \ref{ta2}.

\begin{table}
\begin{center}
\renewcommand{\arraystretch}{1.6}
\begin{tabular}{|c|c|c|}
\hline 
Vertices & PCOs required & PCO location \\ \hline
CO UHP& $1$ & \shortstack{$\frac{\mathcal{X}+\tilde{\mathcal{X}}}{2}$ on top of the C-insertion at $\ii$}\\ \hline
CC UHP & $2$ & $\mathcal{X}\tilde{\mathcal{X}}$ on top of the C-insertion at $\ii y$ \\ \hline
OOO UHP& $1$ & $\mathcal{X}$ at the cyclic-permutation-invariant point $z_p=e^{\ii\pi/3}$ \\ \hline
COO UHP & $2$ & $\mathcal{X}\tilde{\mathcal{X}}$ on top of the C-insertion at $\ii $ \\ \hline
O annulus& $1$ & $\mathcal{X}$ at $\hat{z}_p$ in the $\hat{z}$-coordinate on O annulus \\ \hline
C annulus& $2$ & $\mathcal{X}\tilde{\mathcal{X}}$ on top of the C-insertion  in C annulus \\ \hline
OOOO UHP & $2$ & Points $\hat y_1$ and $\hat y_2$ on the OOOO UHP, see Eq.~(\ref{eperminv}) \\ \hline
COOO UHP & $3$ & $\frac{\mathcal{X}+\tilde{\mathcal{X}}}{2}$ on the C-insertion and $\mathcal{X}$ on  two of the O-insertions \\ \hline
\end{tabular}
\end{center}
\caption{The choice of PCO locations for different interaction vertices used in the main body of the paper.
We redo the calculations using a different choice of PCO locations in appendix \ref{sec: diff_PCO_choice}.
}
\label{ta2}
\end{table}

\subsection{CO vertex on UHP}
\label{sec-co-uhp}

We describe the CO vertex on the UHP using the complex coordinate $z$.
Since there are no moduli in this geometry, we place the closed-string puncture at $ z = \ii $ and the open-string puncture at $ z = 0 $. 
The local coordinate $ w $ in a neighbourhood of the open-string puncture is related to $z$ as
\begin{equation}\label{eq: CO_trans_func}
    w = \lambda z \, ,
\end{equation}
where $ \lambda $ is a large positive real parameter \cite{Sen:2020eck}.

According to \eqref{eq: PCO_number}, this vertex requires one PCO. 
We place the symmetric combination $(\mathcal{X}+\tilde{\mathcal{X}})/2$ on the closed-string puncture at $z=\ii$. 
One could instead choose either $\mathcal{X}$ or $\tilde{\mathcal{X}}$ alone, but the symmetric choice leads to substantial simplifications in the subsequent computations.
This simplification arises because, for this choice of PCO, the CO vertex with either a tachyon or an OSG insertion vanishes.
In appendix \ref{sec: diff_PCO_choice}, we verify that repeating the analysis with only the holomorphic PCO $\mathcal{X}$ at $z=\ii$ leaves the amplitudes unchanged.

\subsection{OOO vertex on UHP}
\label{sec: OOO_Vertex}
We describe the OOO vertex on the UHP using the complex coordinate $\check{z}$.
The three open-string punctures are located at $ \check{z} = 0 $, $ \check{z} = 1 $, and $ \check{z} = \infty $. 
A large positive real parameter $ \alpha $ is introduced, and the following local coordinates are chosen at these punctures \cite{Sen:2020eck}:
\begin{equation}
    w_1 = \alpha \, \frac{2\check{z}}{2 - \check{z}}, \quad w_2 = \alpha \, \frac{2(\check{z} - 1)}{\check{z} + 1}, \quad w_3 = \alpha \, \frac{2}{1 - 2\check{z}}. 
    \label{O-O-O coord}
\end{equation}
These local coordinates are chosen to maintain symmetry under cyclic permutations of $(0,1,\infty)$ under the map $\check{z}\rightarrow \frac{1}{1-\check{z}}$. 

According to \eqref{eq: PCO_number}, we need one PCO, and we choose to place $\mathcal{X}$ at 
\begin{align}
    z_p := e^{\frac{\ii \pi}{3}}\, .
    \label{def-zp}
\end{align}
This location satisfies $z_p=\frac{1}{1-z_p}$, and is hence invariant under the cyclic permutations discussed above. 
This choice is convenient since it implies that we don't have to explicitly average over different cyclic permutations. 
We could have also chosen to put $\widetilde{\mathcal{X}}$ at $z_p$, but this choice does not affect the final amplitudes.

\subsection{COO vertex on UHP}
The COO vertex on the UHP has a one-dimensional moduli space. 
Denoting the UHP coordinate by $z$, we put the closed-string insertion at $z=\ii$, the two open-string insertions at $z=\pm \beta$,  and restrict $\beta$ to lie in the range
\begin{align}
(2\widetilde\lambda)^{-1} &\le \beta\le 1\, ,
\quad\quad  \text{where} \\
\wt\lambda &:= \lambda\, \alpha\, .
\end{align}
As discussed in \cite{Sen:2020eck}, the complementary region $\beta\in [0,(2\tilde{\lambda})^{-1}]$ is described by a CO vertex connected to an OOO vertex by an open-string propagator. 
The region $\beta>1$ can be mapped to the $\beta<1$ region by the $\text{SL}(2,\mathbb{R})$ map $z\to -1/z$, which moves the open-string punctures to the $\beta>1$ configuration while leaving the closed-string puncture fixed at $\ii$. 
The local coordinates at the open-string punctures are taken to be \cite{Sen:2020eck}
\begin{align}
w_a &=
\alpha \tilde{\lambda}
\frac{4\tilde{\lambda}^2+1}{4\tilde{\lambda}^2}
\frac{z-z_a}
{(1+z_a z)+\tilde{\lambda} f(z_a)(z-z_a)},
\qquad
a=1,2,   \label{eq: local_coord_COO} \\
\text{with }\quad  z_1 &=-\beta \,,\quad z_2=\beta. 
\end{align}
The real function $f(\beta)$ satisfies
\begin{equation}
f\!\left(\frac{1}{2\tilde{\lambda}}\right)
= \frac{4\tilde{\lambda}^2-3}{8\tilde{\lambda}^2},
\qquad
f(1)=0,
\qquad
f(-\beta)=f\left(\frac{1}{\beta}\right)=-f(\beta).
\label{f-values}
\end{equation}
The value of $f(\beta)$ at $\beta=\frac{1}{2\tilde{\lambda}}$ is fixed by the requirement that the local coordinates match the ones induced from the gluing of CO and OOO interaction vertices.
The other restrictions on $f(\beta)$ simplify many calculations, see \cite{Sen:2020eck}.

According to \eqref{eq: PCO_number}, we need two PCOs in this case. We choose to insert $\mathcal{X}\tilde{\mathcal{X}}$ on the closed-string vertex operator at $z=\ii$. 
With this choice, there is a mismatch of PCO prescription in going from the CO-OOO degeneration region to the COO vertex region, and we need to do vertical integration at $\beta=\frac{1}{2\tilde{\lambda}}$ to account for the movement of $\mathcal{X}$ from $z_p$ in OOO UHP to a $\tilde{\mathcal{X}}$ or $\mathcal{X}$ at the point $\ii$ in the COO interaction vertex. 
As a matter of convention, we include the contribution from vertical integration in the definition of the COO interaction vertex. 
This will also be implicitly understood for the other interaction vertices.

\subsection{CC vertex on UHP} \label{sCCvertex}

\begin{figure}[t]
\centering
\setlength{\unitlength}{0.8mm}

\begin{picture}(140,30)(0,0)

% Diagram (a)
\linethickness{0.5mm}
\put(10,20){\tcb{\line(1,0){20}}}
\linethickness{0.3mm}
\put(30,20){\tcr{\line(1,0){20}}}
\linethickness{0.5mm}
\put(50,20){\tcb{\line(1,0){20}}}

\put(30,20){\makebox(0,0){$\times$}}
\put(50,20){\makebox(0,0){$\times$}}
\put(40,8){\makebox(0,0){(a)}}

% Diagram (b)
\linethickness{0.5mm}
\put(90,20){\tcb{\line(1,0){20}}}
\put(110,20){\tcb{\line(1,0){20}}}

\put(110,20){\makebox(0,0){$\times$}}
\put(110,8){\makebox(0,0){(b)}}

\end{picture}

\caption{The two Feynman diagrams contributing to the disk amplitude with two external closed strings. The thick blue lines denote closed strings, while the thin red line denotes an open string.}
\label{figthree}
\end{figure}
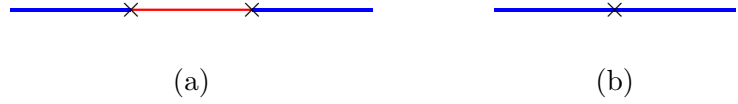

The CC amplitude on the UHP has a one-dimensional moduli space.
We place one closed-string insertion at $z=\ii$ and the other at $z=\ii y$, with $y\in[0,1]$. 
The region $y\in[0,\lambda^{-2}]$ is covered by the degeneration in Fig.~\ref{figthree}(a), in which two CO vertices are joined by an open-string propagator \cite{Sen:2020eck}. 
The remaining region, $y\in[\lambda^{-2},1]$, defines the fundamental CC vertex shown in Fig.~\ref{figthree}(b).

According to \eqref{eq: PCO_number}, this geometry requires two PCOs, and we choose to place $\mathcal{X}\tilde{\mathcal{X}}$ on the closed-string insertion at $z=\ii y$. 
At $y=\lambda^{-2}$, one must include a vertical integration contribution to match the PCO locations inherited from the degeneration region. 
We include this contribution as part of the CC interaction vertex.

\subsection{O vertex on annulus}
\label{sec-o-annulus}

We now consider the annulus amplitude with a single open-string insertion.
Let $w$ be the complex coordinate on the annulus, with the identification \eqref{annulus_coord}.
The Killing vector on the annulus allows us to fix the open-string puncture at any chosen point on the boundary, so its position is not a modulus.
Thus the moduli space is one-dimensional, parametrized by $v=e^{-2\pi t}\in(0,1)$.

One contribution to this amplitude comes from sewing two open-string punctures of an OOO vertex with an open-string propagator. 
This gives the degeneration region.
We glue the O insertions at $\check{z} = 1$ and $\check{z} = \infty$, and define new coordinates $\hat{z}$ and $w$ as follows:
\begin{equation}
\label{eq: zt_zh_SL2R}
    \check{z} =: \frac{(4 + 3u ) \hat{z} - 4 + 3u }{(4 + u ) \hat{z} - 2u }, \quad 
    u  :=\frac{q}{\alpha^2}, \quad 
    0 \leq u\leq \alpha^{-2}, \quad 
    w := \frac{1}{\ii} \log \hat{z} -{3\over 2}\, \ii\, u, \quad 
    0 \leq \text{Re}\, w \leq \pi\, .
\end{equation}
Here $q$ is the Schwinger parameter associated with the open string propagator.
These coordinates exhibit the following identifications as a consequence of the gluing of the two open-string punctures \cite{Sen:2020eck}
\begin{equation}
    \hat{z} \equiv u^{-1} \left( 1 - \tfrac{1}{2} u  \right) \hat{z}, \qquad w \equiv w - \ii \log \left( u \left( 1 - \tfrac{1}{2} u  \right)^{-1} \right)\, . 
\end{equation}
Comparison with \eqref{annulus_coord} gives $v = u ( 1 - \tfrac{1}{2} u )^{-1}$. 
Hence, the range $v \in \left(0, (\alpha^2 - \tfrac{1}{2})^{-1}\right]$ is covered by this Feynman diagram. 
The remaining region $v\in \left[(\alpha^2 - \tfrac{1}{2})^{-1}, 1\right)$ is thus the range covered by the fundamental O-vertex on the annulus. 

In the degeneration region, the open-string puncture located at 
$ \check{z} = 0 $ is mapped to
\begin{equation}\label{eq: O_ann_O_location}
    \hat{z} \approx 1 - \tfrac{3}{2} u , \qquad 
    w \approx 0\,  .
\end{equation}
In the vertex region, we continue to choose the open-string puncture to be located $w = 0$ and choose the local coordinate $w_o$ around it to be the same as the one used for the open-string puncture at $\check{z} = 0$ with $ u = \alpha^{-2} $, i.e.
\begin{equation}
    w_o = 2\alpha {\check{z}\over 2-\check{z}}= 2\alpha \frac{(4 + 3\alpha^{-2}) \hat{z} - 4 + 3\alpha^{-2}}{(4 - \alpha^{-2}) \hat{z} + 4 - 7\alpha^{-2}}. 
    \label{Local_coord_O_vertex}
\end{equation}

From \eqref{eq: PCO_number}, the O-vertex on the annulus requires one PCO. 
In parallel with the choice of local coordinate around the O insertion, we choose the PCO location to be inherited from the degeneration region.
Namely, we set $\check{z}=z_p=e^{\ii\pi/3}$, as in section \ref{sec: OOO_Vertex}, and map it to the $\hat z$-coordinate using \eqref{eq: zt_zh_SL2R} with $u=\alpha^{-2}$. 
We denote the resulting point by $\hat z_p$ in the $\hat z$-coordinate and by $w_p$ in the $w$-coordinate. 
Consequently, no vertical integration is required at the interface $v=\left(\alpha^2-\frac{1}{2}\right)^{-1}$, since the PCO location has no discontinuity.

\subsection{C vertex on annulus} 

\label{sCannulus}

\begin{figure}[!t]
\centering
\begin{adjustbox}{width=0.95\textwidth}
\begin{tikzpicture}[thick, scale=1.0, every node/.style={scale=0.8}]
% Style definitions
\tikzset{
    arrow/.style={->, >=latex, thick},
    disk/.style={circle, draw, minimum size=1.2cm},
    closed/.style={thick},
    open/.style={thick}
}

% Bounding box for clean centering
\path[use as bounding box] (-1.8,-1.3) rectangle (12.8,1.0);

% (a)
\begin{scope}[xshift=0cm]
    \draw[blue, very thick] (-1.5,0) -- (-0.5,0);
    \draw[red] (-0.5,0) -- (0.5,0);
    \draw[red] (1,0) circle (0.5);
    \node at (0,0.25) {$q_1$};
    \node at (1,0.75) {$q_2$};
    \node at (-0.5,0) {$\times$}; 
    \node at (0.5,0) {$\times$};
    \node at (0,-1) {(a)};
\end{scope}

% (b)
\begin{scope}[xshift=4cm]
    \draw[blue, very thick] (-1,0) -- (0,0);
    \draw[red] (0,0) -- (1,0);
    \node at (0,0) {$\times$};
    \node at (0.5,0.25) {$q_1$};
    \node at (1,0) {$\otimes$};
    \node at (0,-1) {(b)};
\end{scope}

% (c)
\begin{scope}[xshift=8cm]
    \draw[blue, very thick] (-1,0) -- (0,0);
    \draw[red] (0.5,0) circle (0.5);
    \node at (0.5,0.75) {$q_2$};
    \node at (0,0) {$\times$};
    \node at (0,-1) {(c)};
\end{scope}

% (d)
\begin{scope}[xshift=12cm]
    \draw[blue] (-0.5,0) -- (0.5,0);
    \node at (0.5,0) {$\otimes$};
    \node at (0,-1) {(d)};
\end{scope}

\end{tikzpicture}
\end{adjustbox}
\caption{Feynman diagrams contributing to the annulus amplitude with one external closed string. Vertices marked by $\times$ and $\otimes$ represent UHP and annulus amplitudes, respectively.}
\label{fig:feynman-diagrams_2}
\end{figure}
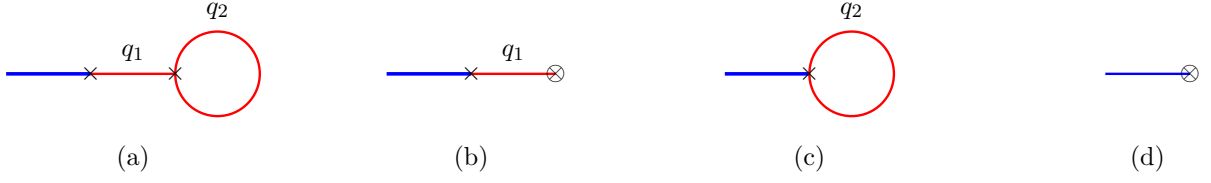

We now consider the annulus amplitude with a single closed-string insertion. 
The moduli space is two-dimensional. 
We continue to use the annulus coordinate $w$ introduced in \eqref{annulus_coord}. 
The closed-string puncture is located at $w_c=2\pi x$. 
The moduli $(v,x)$ range over $0<v<1$ and $0\leq x\leq \frac14$.

\begin{figure}[t]
    \centering
    \tikzset{every picture/.style={line width=0.75pt}} % Set default line width

    \tikzset{every picture/.style={line width=0.75pt}} %set default line width to 0.75pt

    \scalebox{0.75}{%
        \begin{tikzpicture}[x=0.75pt,y=0.75pt,yscale=-1,xscale=1]
            %uncomment if require: \path (0,321); %set diagram left start at 0, and has height of 321

            %Shape: Rectangle [id:dp9309442929369403]
            \draw (210,14.03) -- (631.9,14.03) -- (631.9,286.42) -- (210,286.42) -- cycle;

            %Straight Lines [id:da2438747903896621]
            \draw (311.5,242.42) -- (631.53,242.02);

            %Straight Lines [id:da3261827700481601]
            \draw (311.5,242.42) -- (312.25,286.42);

            %Straight Lines [id:da9759165970362289]
            \draw (208.7,238.02) -- (311.5,242.42);

            %Curve Lines [id:da11415599157340717]
            \draw (215.46,14.43) .. controls (212.83,131.22) and (235.72,229.62) .. (311.5,242.42);

            % Text Node
            \draw (227.4,206.12) node [anchor=north west][inner sep=0.75pt] [] {(c)};

            % Text Node
            \draw (410.81,105.35) node [anchor=north west][inner sep=0.75pt] [] {(d)};

            % Text Node
            \draw (189.46,153.93) node [anchor=north west][inner sep=0.75pt] [font=\small] {$x$};

            % Text Node
            \draw (424.28,292.84) node [anchor=north west][inner sep=0.75pt] [font=\small] {$v$};

            % Text Node
            \draw (196.49,288.78) node [anchor=north west][inner sep=0.75pt] [font=\small] {$0$};

            % Text Node
            \draw (642.58,291.24) node [anchor=north west][inner sep=0.75pt] [font=\small] {$1$};

            % Text Node
            \draw (185.04,1.26) node [anchor=north west][inner sep=0.75pt] [font=\small] {$\frac{1}{4}$};

            % Text Node
            \draw (254.71,256.67) node [anchor=north west][inner sep=0.75pt] [] {(a)};

            % Text Node
            \draw (446.85,256.67) node [anchor=north west][inner sep=0.75pt] [] {(b)};

            % Text Node
            \draw (298.30,292.84) node [anchor=north west][inner sep=0.75pt] [font=\small] {$v\approx \alpha^{-2}$};

            % Text Node
            \draw (640.46,228.93) node [anchor=north west][inner sep=0.75pt] [font=\small] {$x\approx \left(2\pi \tilde{\lambda}\right)^{-1}$};
        \end{tikzpicture}
    }

    \caption{Moduli space of the annulus with a single closed-string insertion, showing its decomposition into the bulk region and boundary degenerations. Region (d) represents the bulk of the moduli space, while regions (a), (b), and (c) correspond to distinct degeneration limits. Each region is associated with the respective Feynman diagram (a)--(d) in Fig.~\ref{fig:feynman-diagrams_2}.}
    \label{fig:moduli-feynman}
\end{figure}
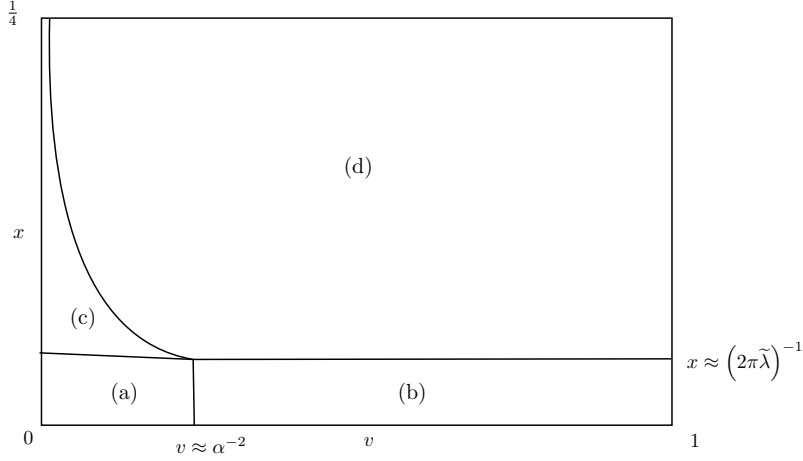

The four Feynman diagrams contributing to this amplitude are shown in Fig.~\ref{fig:feynman-diagrams_2} and the corresponding regions in the moduli space covered by these diagrams are shown in Fig.~\ref{fig:moduli-feynman}. 

The Feynman diagram in Fig.~\ref{fig:feynman-diagrams_2}(a) is obtained by sewing the O insertion of the CO-vertex to an O insertion of the OOO-vertex with sewing parameter $q_1$, and then sewing the remaining two O insertions of the OOO-vertex with sewing parameter $q_2$.
We get the following  relation between $(q_1,q_2)$ and the moduli $(v,x)$ \cite{Sen:2020eck}:
\begin{align}\label{eq: reg(a)_xv}
    v= \frac{q_2}{\alpha^2}\left(1-\frac{q_2}{2\alpha^2}\right)^{-1}, \qquad 2\pi x=\frac{q_1}{\tilde{\lambda}}\left(1-\frac{q_2}{\alpha^2}\right) \, .
\end{align}
Since $q_1,q_2\in[0,1]$, these relations tell us that region (a) is defined by the inequalities
\begin{align}
     0\leq v\leq \left(\alpha^2-\frac{1}{2}\right)^{-1}, \qquad 0\leq 2\pi x \leq \tilde{\lambda}^{-1}\frac{2-v}{2+v} \ .
 \end{align}

The Feynman diagram in Fig.~\ref{fig:feynman-diagrams_2}(b) is obtained by sewing the O insertion of the CO-vertex to the O insertion of the annulus O-vertex.
This leads to the following relation between the coordinates $z$ and $\hat{z}$ on the two vertices (see sections \ref{sec-co-uhp} and \ref{sec-o-annulus})
\begin{equation} \label{eq:sewing_annulus_b}
    2\alpha \frac{(4 + 3 \alpha^{-2}) \hat{z} - 4 + 3 \alpha^{-2}}{(4 - \alpha^{-2}) \hat{z} + 4 - 7 \alpha^{-2}} \cdot \lambda z = -q_1, \qquad 0 \leq q_1 \leq 1.
\end{equation}
Recall also the relation between $\hat{z}$ and $w$ given in (\ref{eq: zt_zh_SL2R}).
It follows that the closed-string puncture at $ z = \ii $ is located in the $w$-coordinate at
\begin{equation} \label{eq:x_annulus_b_approx}
    w_c 
    \approx u \left( 1 - \alpha^{-2}  \right), \qquad u = \frac{q_1}{\tilde{\lambda}}, \quad 0 \leq u \leq \tilde{\lambda}^{-1}.
\end{equation}
Hence, region (b) is characterized by 
\begin{equation} \label{eq:region_b}
    \left( \alpha^2 - \tfrac{1}{2} \right)^{-1} < v < 1, \qquad 0 < 2\pi x < \tilde{\lambda}^{-1} \left( 1 - \alpha^{-2} \right).
\end{equation}

The Feynman diagram in Fig.~\ref{fig:feynman-diagrams_2}(c) is obtained by sewing together the two O insertions in the COO-vertex. 
This region is parametrized by \cite{Sen:2020eck}
\begin{align} 
    \frac{1}{2\tilde{\lambda}} \leq \beta \leq 1, \quad 
    0 \leq u \leq \alpha^{-2} 
    \left( 1 + \frac{1}{4 \tilde{\lambda}^2} \right)^{-2}, \quad 
    \text{where} \quad 
    u=\frac{q_2}{\alpha^2}\left(1+\frac{1}{4\tilde{\lambda}^2}\right)^{-2}\, .
    \label{euqrel}
\end{align}
The relation between $(v,x)$ and $(\beta, u)$ is \cite{Sen:2020eck}
\begin{equation}
\begin{aligned}
    2\pi x(\beta,u)&=2\tan^{-1}\beta-\frac{u}{\beta\tilde{\lambda}^2}(1-\beta^2-2\beta\tilde{\lambda} f(\beta))\\
    v(\beta,u)&=\frac{u(1+\beta^2)^2}{4\beta^2\tilde{\lambda}^2}\left(1+\frac{u}{2\beta^2\tilde{\lambda}^2}\left(1-\beta^2-2\beta f(\beta)\tilde{\lambda}\right)^2\right) \ .
    \label{eq:xv_simplified}
\end{aligned}
\end{equation}

We define the rest of the moduli space i.e. region (d) as the fundamental C-vertex on the annulus. 
It covers the following region in  $(x,v)$ plane
\begin{equation} \label{eq:region_d_approx}
\begin{aligned}
    x&\in\left[\frac{1}{2\pi\tilde{\lambda}}(1 - \alpha^{-2}) ,  \frac{1}{4}\right], \\
    v&\in\left[\frac{\alpha^{-2} \tilde{\lambda}^{-2}\left( 1 + \frac{1}{4 \tilde{\lambda}^2} \right)^{-2}}{\sin^2{2\pi x}} \left( 1 - \frac{2 \left( \cot^2(2\pi x) - \tilde{\lambda}^2 f(\tan{\pi x})^2 \right)}{\alpha^{2} \tilde{\lambda}^{2} \left( 1 + \frac{1}{4 \tilde{\lambda}^2} \right)^{2}} \right), 1\right]\, .
\end{aligned}
\end{equation}

Equation \eqref{eq: PCO_number} shows that the annulus C-vertex requires two PCOs; we place them as $\mathcal{X}\tilde{\mathcal{X}}$ on the C insertion.
As a consequence, no vertical integration is needed at the interface of regions (c) and (d) as the PCOs $\mathcal{X}\tilde{\mathcal{X}}$ are inserted on C in both regions. 
However, vertical integration is required at the interface between regions (b) and (d). In region (b), the PCO combination $(\mathcal{X}+\widetilde{\mathcal{X}})/2$ is placed at the C insertion of the CO-vertex, and the PCO $\mathcal{X}$ is placed at a particular point on the annulus, as described in section \ref{sec-o-annulus}.

\subsection{OOOO vertex on UHP} 
\label{soooo}

We now consider the OOOO amplitude on the UHP. 
Its moduli space is one-dimensional.
We fix three open-string insertions at $0$, $1$, and $\infty$; the position $x\in(0,1)$ of the fourth insertion parametrizes the moduli space.

There are two degeneration regions, $x\in[0,\epsilon]$ and $x\in[1-\epsilon,1]$, where $\epsilon$ will be specified below. 
The first degeneration is obtained by taking two OOO interaction vertices on the UHP, as described in section \ref{sec: OOO_Vertex}, and gluing them at the punctures located at their respective origins.
Let $z'$ and $z$ be the coordinates on the two OOO interaction vertices. 
The local coordinates around the punctures are exactly as in (\ref{O-O-O coord})
\begin{align}
     w'&= \alpha\frac{2z'}{2-z'}, \quad w_1 = -2\alpha\frac{1-z'}{1+z'},\quad w_2=\alpha\frac{2}{1-2z'}\\
    w&= \alpha\frac{2z}{2-z}, \quad w_3 = -2\alpha\frac{1-z}{1+z},\quad w_4=\alpha\frac{2}{1-2z}\, .
\end{align}
We glue using the plumbing fixture relation $ww'=-q$ and consider the resulting UHP in the coordinate $z$. 
We further introduce the coordinate $\check{z} = (az+b)/(cz+d)$ to bring the points $w_1=0$, $w_3=0$ and $w_4=0$ to $0,1$ and $\infty$ in the
$\check{z}$ plane, respectively.
The point $w_2=0$ is located at $x$ in the $\check z$-plane.
Then we have the following relations between $\check z$ and $w_1,w_2,w_3,w_4,x$:
\begin{align}\label{eq: gluing_x_eps}
    \check{z}= \widetilde F_1(w_1, x):=\frac{2 x  w_1}{2 \alpha\sqrt{1-x} +\left(2-\sqrt{1-x}\right) w_1}, \quad
    &\check{z}= \widetilde F_2(w_2, x):=\frac{x (2 \alpha +w_2)}{2 \alpha +\left(1-2 \sqrt{1-x}\right) w_2} , \nonumber\\
    \check{z}= \widetilde F_3(w_3, x):= 1-\frac{2 \sqrt{1-x} w_3}{w_3-2 \alpha } , \quad
    &\check{z}= \widetilde 
    F_4(w_4, x):= -\frac{\alpha  \sqrt{1-x}}{w_4}-\frac{\sqrt{1-x}}{2}+1 , \nonumber\\
    x =\frac{16q\alpha^2}{(q+4\alpha^2)^2}, \qquad
    & 0\le x\le  \epsilon:=
     \alpha^{-2} \left(1+\frac{1}{4\alpha^2}\right)^{-2}\, .
\end{align}

To get the local coordinates for the second degeneration where $x \in [1-\epsilon,1]$, we use cyclic symmetry of the OOO interaction vertex.
The $\text{SL}(2,\mathbb{R})$ transformation
\be \label{efourcyclic}
\check{z}\to {1-x\over 1-\check{z}}
\ee
takes the points $(0,x,1,\infty)$ to $(1-x,1,\infty,0)$. 
This produces new local coordinates
$\widehat F_i(w_i,x)$ 
\be
\widehat F_{i+1}(w,1-x) = {1 - x\over 1-\widetilde F_i(w,x)}, \qquad i=1,2,3,4, \qquad \widehat
F_5\equiv \widehat F_1\, .
\ee
Note that the arguments of $\widehat F$ and $\widetilde F$ differ: $\widehat F$ is evaluated at $1-x$, whereas $\widetilde F$ is evaluated at $x$.
Thus, for $x\in [1-\epsilon,1]$ we get the following local coordinates:
 \begin{align}\label{eq: gluing_x_1-eps}
    \check{z}= \widehat F_1(w_1, x)\equiv\frac{2 \sqrt{x} w_1}{2 \alpha +w_1}, \quad 
    &\check{z}= \widehat F_2(w_2, x)\equiv\frac{x \left(2 \alpha +\left(\frac{2}{\sqrt{x}}-1\right) w_2\right)}{2 \alpha +\left(2 \sqrt{x}-1\right) w_2}\, ,\nonumber\\
    \check{z} = \widehat F_3(w_3, x)\equiv1-\frac{ 2(1-x) w_3}{ \left(\left(2-\sqrt{x}\right) w_3-2 \alpha\sqrt{x} \right)}, \quad
    &\check{z}= \widehat F_4(w_4, x)\equiv\sqrt{x}\left(\frac{1}{2}-\frac{\alpha}{w_4}\right)\, .
 \end{align}

The fundamental OOOO interaction vertex will cover the remaning region 
$x\in [\epsilon, 1-\epsilon]$. 
In this region, we can make any choice of local coordinates interpolating between the above two, i.e.\
\be
F_i(w,\epsilon) = \widetilde F_i(w,\epsilon), \qquad F_i(w,1-\epsilon) = \widehat F_i(w,1-\epsilon)\, .
\ee
However, a generic interpolation is not invariant under cyclic permutations. 
In that case, constructing the full amplitude would require an explicit average over all $4!$ permutations of the external states. 
We can avoid this by choosing the interpolating local coordinates to be cyclically invariant, in the sense that the four functions are mapped into one another under the transformation \eqref{efourcyclic}, as follows: 
\be\label{eq: OOOO_vertex_trans_func_1}
F_{i+1}(w,1-x) = {1 - x\over 1-F_i(w,x)}
\qquad \text{for} \quad x\in [\epsilon,1-\epsilon].
\ee
With this choice, the cyclic average is built into the vertex, and we need only sum over the inequivalent cyclic orderings rather than all permutations.
With (\ref{eq: OOOO_vertex_trans_func_1}), we only need to specify $F_1(w,x)$. 
A possible choice is
\be\label{eq: OOOO_vertex_trans_func_2}
F_1(w_1, x):=\frac{2x}{\sqrt{1-\epsilon}}\frac{w_1}{2 \alpha+s(x)w_1} \, ,
\ee
where $s(x)$ is an arbitrary function with the particular values
\be\label{eq: s_limits}
s(\epsilon)=\frac{2}{\sqrt{1-\epsilon}}-1\, ,\qquad 
s(1-\epsilon)=1\, .
\ee
Computing $F_2, F_3$ and $F_4$ explicitly gives
\begin{align}\label{eq: OOOO_trans_func_explicit}
    \check z=F_1(w_1,x)&=\frac{2 x w_1}{\sqrt{1-\epsilon } (2 \alpha +w_1 s(x))},\nonumber\\
    \check z=F_2(w_2,x)&=\frac{x \sqrt{1-\epsilon } (2 \alpha +w_2 s(1-x))}{2 \alpha  \sqrt{1-\epsilon }+w_2 \sqrt{1-\epsilon } s(1-x)+2 (x-1) w_2},\nonumber\\
    \check z=F_3(w_3,x)&=\frac{2 \alpha  \sqrt{1-\epsilon }+w_3 \sqrt{1-\epsilon } s(x)-2 x w_3}{2 \alpha  \sqrt{1-\epsilon }+w_3 \sqrt{1-\epsilon } s(x)-2 w_3},\nonumber\\
    \check z=F_4(w_4,x)&=-\frac{1}{2} \sqrt{1-\epsilon } s(1-x)-\frac{\alpha  \sqrt{1-\epsilon }}{w_4}+1\, .
\end{align}

According to  \eqref{eq: PCO_number}, we need two PCOs in this case. 
We choose to put two $\mathcal{X}$'s at locations $\hat y_1(x)$ and $\hat y_2(x)$ given by
\be \label{eperminv}
\hat y_1(x) = x+ \ii \sqrt{x-x^2}, \qquad 
\hat y_2(x) = x-\ii\sqrt{x-x^2}\, ,
\ee
which satisfy
\be
\hat y_i(1-x) = {1-x\over 1 - \hat y_i(x)}\, ,\qquad i = 1,2\, .
\ee
Vertical integrations are required at the interfaces $x=\epsilon$ and $x=1-\epsilon$ to move the PCOs from the locations inherited from the point $z_p$ on the OOO vertices (see Eq.~(\ref{def-zp}) and section \ref{sec: OOO_Vertex}) to the bulk OOOO locations $\hat y_{1}, \hat y_2$ on the UHP.

\def\figcooo{

\def\JPicScale{0.8}
\ifx\JPicScale\undefined\def\JPicScale{1}\fi
\unitlength \JPicScale mm
\begin{picture}(200,70)(0,0)

%---------------- Figure (a) ----------------%
\linethickness{0.5mm}
\put(10,50){\tcb{\line(1,0){15}}}
\linethickness{0.3mm}
\put(25,50){\tcr{\line(1,0){15}}}
\linethickness{0.3mm}
\multiput(40,50)(0.12,-0.12){83}{\tcr{\line(1,0){0.12}}}
\linethickness{0.3mm}
\multiput(40,50)(0.12,0.12){83}{\tcr{\line(1,0){0.12}}}
\linethickness{0.3mm}
\multiput(50,60)(0.12,0.24){42}{\tcr{\line(0,1){0.24}}}
\linethickness{0.3mm}
\put(50,60){\tcr{\line(1,0){10}}}

\put(25,30){\makebox(0,0)[cc]{(a)}}
\put(32,53){\makebox(0,0)[cc]{$q_2$}}
\put(43,58){\makebox(0,0)[cc]{$q_1$}}

%---------------- Figure (b) ----------------%
\linethickness{0.5mm}
\put(70,50){\tcb{\line(1,0){15}}}
\linethickness{0.3mm}
\multiput(85,50)(0.12,0.12){83}{\tcr{\line(1,0){0.12}}}
\linethickness{0.3mm}
\multiput(85,50)(0.12,-0.12){83}{\tcr{\line(1,0){0.12}}}
\linethickness{0.3mm}
\multiput(95,60)(0.12,0.24){42}{\tcr{\line(0,1){0.24}}}
\linethickness{0.3mm}
\put(95,60){\tcr{\line(1,0){10}}}

\put(85,30){\makebox(0,0)[cc]{(b)}}
\put(88,58){\makebox(0,0)[cc]{$q_1$}}

%---------------- Figure (c) : OLD (d) ----------------%
\linethickness{0.5mm}
\put(120,50){\tcb{\line(1,0){10}}}
\linethickness{0.3mm}
\put(130,50){\tcr{\line(1,0){10}}}
\linethickness{0.3mm}
\multiput(140,50)(0.12,0.12){83}{\tcr{\line(1,0){0.12}}}
\linethickness{0.3mm}
\put(140,50){\tcr{\line(1,0){10}}}
\linethickness{0.3mm}
\multiput(140,50)(0.12,-0.12){83}{\tcr{\line(1,0){0.12}}}

\put(140,30){\makebox(0,0)[cc]{(c)}}
\put(137,53){\makebox(0,0)[cc]{$q_2$}}

%---------------- Figure (d) : OLD (c) ----------------%
\linethickness{0.5mm}
\put(160,50){\tcb{\line(1,0){15}}}
\linethickness{0.3mm}
\multiput(175,50)(0.12,0.12){83}{\tcr{\line(1,0){0.12}}}
\linethickness{0.3mm}
\put(175,50){\tcr{\line(1,0){15}}}
\linethickness{0.3mm}
\multiput(175,50)(0.12,-0.12){83}{\tcr{\line(1,0){0.12}}}

\put(175,30){\makebox(0,0)[cc]{(d)}}

\end{picture}
}

\subsection{COOO vertex on UHP}
Lastly, we consider the COOO amplitude on the UHP, whose moduli space is two-dimensional.
We can parametrize the moduli space by the positions of the first two O insertions, with the C insertion and the third O insertion fixed at $\ii$ and $0$, respectively.
However, we shall use a more general parametrization in which the closed-string vertex operator is fixed at $\ii$ and we take the locations of the three open-string vertex operators as functions of two independent variables $\beta_1,\beta_2$, as in \cite{Sen:2020eck}.
The local coordinates around the punctures are taken to be $w_1, w_2$ and $w_3$ with transition functions
\begin{equation} \label{elocalCOOO}
    z=F_a(w_a, \vec{\beta})=f_a(\vec{\beta})+g_a(\vec{\beta}) \, w_a+ \frac{1}{2}h_a(\vec{\beta})\, w_a^2 +\mathcal{O}(w_a^3), \qquad a=1,2,3\, .
\end{equation}
The locations of the three open-string punctures are 
$\{f_1(\vec\beta), f_2(\vec\beta), f_3(\vec\beta)\}$.

\begin{figure}
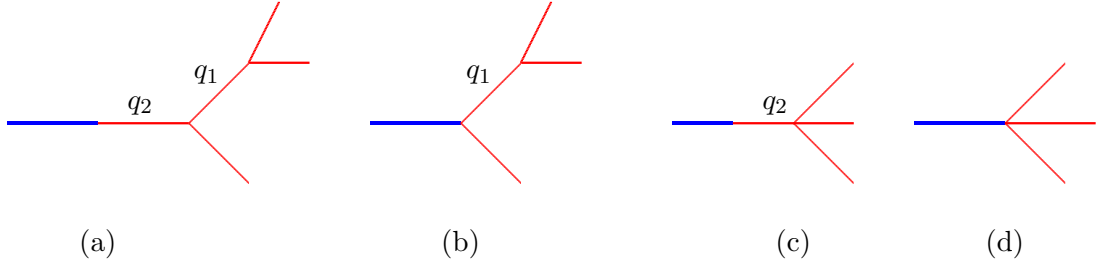

\begin{center}

\hbox{\figcooo}

\vskip -.8in

\caption{This figure shows four Feynman diagrams contributing to the disk amplitude with 
one external
closed string and three external open strings. $q_1$ and $q_2$ represent
the sewing parameters of the corresponding open string propagators.
\label{figcooo}
}
\end{center}
\end{figure}

The range of $\beta_1,\beta_2$ for the fundamental COOO vertex can be determined as usual by first determining the region of the moduli space covered by the lower order Feynman diagrams shown in Fig.~\ref{figcooo}(a), (b) and (c), and then requiring that the remaining region is covered by the COOO interaction vertex.
We shall determine this range when we need it.

For our use in appendix \ref{sec: diff_PCO_choice}, we shall need the expressions for the $F_a$'s induced from Fig.~\ref{figcooo}(c). 
By cyclic symmetry of the OOOO-vertex, the O insertion of the CO-vertex may be sewn to any of its four O insertions with the same result.\footnote{To deal with permutations that change the cyclic ordering, we shall sum over different diagrams where the external O's are permuted.} 
We choose to sew it to the O insertion at $x$.
Using \eqref{eq: CO_trans_func} we see that this leads to the identification $\lambda z \, w_2=-q_2$.
We think of the coordinate $z$  on the CO-vertex to also be the global coordinate on the COOO vertex.
This leads to the following relation between the local coordinates $w_i$ at the open-string punctures and $z$:
\begin{align}\label{eq: SymmOOOO_vertex_gluing}
    z &=F^x_0(w_1, u,x)=\frac{u}{2 \alpha}\left(\frac{4 (x-1) w_1}{2 \alpha  (\epsilon -1)+w_1 (\epsilon -1) s(x)+2 w_1 \sqrt{1-\epsilon }}+s(1-x)\right) \, ,\nonumber\\
    z &=F^x_1(w_3, u,x)=\frac{u}{2 \alpha}\left(-\frac{4 x w_3}{(\epsilon -1) (2 \alpha +w_3 s(x))}+s(1-x)-\frac{2}{\sqrt{1-\epsilon }}\right) \, , \nonumber\\
    z &=F^x_{\infty}(w_4, u,x)=\frac{u}{2\alpha}\frac{2 s(1-x) \left(\alpha  +\frac{2 (x-1) w_4}{\sqrt{1-\epsilon }}\right)+w_4 s(1-x)^2+\frac{4 (1-x)}{\sqrt{1-\epsilon }} \left(\frac{w_4}{\sqrt{1-\epsilon }}-\alpha \right)}{2 \alpha +w_4  s(1-x)+\frac{2 (x-1) w_4}{\sqrt{1-\epsilon }}}\, ,
\end{align}
where $u=q_2/\lambda$. 
We can relabel these transition functions as $F_1,F_2,F_3$ to match the notation of \eqref{elocalCOOO}.

According to  \eqref{eq: PCO_number}, we need three PCOs for the COOO vertex. 
We shall describe the choice in section \ref{esbulkcooo}.
As usual, we must also include vertical-integration contributions from any mismatch in the PCO locations across the boundaries between different Feynman-diagram regions of moduli space.

\section{Disk one- and two-point functions}
\label{sec:disk}

\subsection{Disk one-point function} \label{sonepoint}
The cosmological constant vertex operator is $V=c\tilde{c} e^{-\phi}e^{-\tilde{\phi}} e^{\bb \varphi}$.
To calculate the disk one-point function of this operator, we need the one-point function of the operator $e^{\bb \varphi}$ on the disk with ZZ boundary conditions, which is given in \eqref{one-pt-vb} \cite{Ahn:2002ev, Fukuda:2002bv}.
The corresponding UHP one-point function, with the insertion at $z=\ii$, is
\begin{equation}
    A_{\text{disk}}(V)    =\frac{1}{2}\eta_c^{1/2} \,\left\langle \left( \partial c-\bar{\partial}\widetilde{c} \right) \, c \widetilde{c} \, e^{-\phi}e^{-\widetilde{\phi}} \, e^{\bb\varphi}(\ii) \right\rangle _{\UHP} 
    = \frac{1}{2} \eta_c^{1/2} \cdot 8 K \cdot (2\ii)^{-1} \cdot \frac{Q}{2\bb \, \ii \tilde{\mu}}
        = -  K_0\, \frac{Q}{2\bb\wt\mu} \, .
    \label{3.11}
\end{equation}
In the first step, we made use of (\ref{ea4new}). 
In the second step, the factor $8K$ is the contribution from the $c$-ghost correlator (\ref{eq:ev_ccc}), the factor $(2\ii)^{-1}$ is the contribution from the $\phi$ correlator (\ref{exietaphi}), and the factor $Q/(2\bb\ii\tilde{\mu})$ is the contribution from the Liouville sector (\ref{one-pt-vb}).
We relate $K$ to $K_0$ via equations \eqref{eKT} and (\ref{etension}), which imply that $2 K \eta_c^{1/2} = K_0$.
The final result agrees with the prediction \eqref{eq: disk_one_KPZ} based on KPZ scaling.

\subsection{Disk two-point function} \label{stwopoint}

The disk two-point function receives contribution from the two Feynman diagrams shown in
Fig.~\ref{figthree}. We shall compute the contribution from these two Feynman diagrams 
separately.
\subsubsection{Bulk contribution}
\label{sec-bulkdisk}
We shall first compute the contribution from Fig.~\ref{figthree}(b), which is the contribution from the region away from the boundary of moduli space.
The moduli space of disk with two bulk punctures is one-dimensional. 
We can fix one puncture to be at $z=\ii$ and the other at $z=\ii y$ and then perform the moduli space integral from $y=0$ to $y=1$ to compute the amplitude. 
As described in section \ref{sCCvertex}, the diagram in Fig.~\ref{figthree}(b) covers the region
\be
\varepsilon\le y\le 1, \qquad \text{where } \varepsilon := \lambda^{-2}\, .
\ee
Also, as described in section \ref{sCCvertex}, in this case we choose our PCO locations to be
$\mathcal{X} \widetilde{\mathcal{X}}$ at $\ii y$. 
The action of two PCOs on the cosmological constant vertex operator is given by
\begin{align}
    \mathcal{X} \widetilde{\mathcal{X}}\, c\widetilde{c} \, e^{-\phi}e^{-\widetilde{\phi}} \,  e^{\bb \varphi}
    = \frac{1}{4} \,  \eta \widetilde{\eta} \, e^{\phi} e^{\widetilde{\phi}} \, e^{\bb \varphi} &- \frac{1}{2}\,  c \,  \widetilde{\eta} \,e^{\widetilde{\phi}}\, (-\ii \bb \psi) \,e^{\bb\varphi} \nonumber \\
    &- \frac{1}{2}\, \eta \, e^{\phi}\, \widetilde{c} \,(-\ii \bb \widetilde{\psi}) \, e^{\bb \varphi} + c \widetilde{c} (-\ii \bb \psi)(-\ii \bb \widetilde{\psi}) e^{\bb\varphi}.
    \label{PCO_action}
\end{align}
We also need $b$-ghost insertion in the form of $\mathcal{B}_y$ given in \eqref{eapp0}. 
Since the PCOs are placed on top of the closed-string vertex operator at $\ii y$, in the local coordinate of that closed string they are placed at the origin and hence their positions are independent of $y$. 
Hence in the expression \eqref{eapp0} for $\mathcal{B}_y$, only the part   involving the $b$-ghost survives. 
We now see from \eqref{PCO_action} that  even though it has four terms, the first three contributions vanish as they involve only $\eta$ insertions without any $\partial \xi$ insertions. 
Thus, the last term is the only one contributing non-trivially.

We can now express the contribution from Fig.~\ref{figthree}(b) as\footnote{
The overall sign of this expression can be fixed as follows. 
We can work on the disk $|w|<1$, related to the UHP coordinate $z$ via $w = (1+\ii z)/(1-\ii z)$, and begin with the one-point function of a single closed string on the disk, which according to \eqref{eapp1} and \eqref{ea4new}, carries an extra factor of $\eta_c c_0^-$ compared to other amplitudes.
We now insert another closed-string vertex operator at $w=r e^{\ii \theta}$, and using the fact that $dr\wedge d\theta$ represents positive integration measure, insert the measure factor $drd\theta\BBB_r\BBB_\theta$ into the correlation function. 
It is easy to see that the $\BBB_\theta$ factor produces an operator $\ii(b_0-\tilde b_0)$ acting on the original state. 
This has the effect of removing the $c_0^-$ factor and the $\ii$ combines
with the extra $\eta_c$ to produce a factor of $-1/(2\pi)$. 
The integration over $\theta$ produces a factor of $2\pi$ and at the end we are left with $-\int dr \BBB_r$. 
This can be taken to run along the real $w$ axis which is equivalent to the imaginary $z$ axis, with the variables $r$ and $y$ being related by $r = (1-y)/(1+y)$. 
Thus the range $y\in [\varepsilon,1]$ maps to $r\in [0,1-2\varepsilon]$ to first order in $\varepsilon$.
Now under a change of variable from $r$ to $y$, $\BBB_r dr$ gets mapped to $\BBB_y dy$ and hence $-\int_{0}^{1-2\varepsilon} \BBB_r dr$ is mapped to $-\int_{1}^{\varepsilon} \BBB_y dy =\int_\varepsilon^1 \BBB_y dy$.
} % end footnote
\begin{align}
    A^{(1)}_{\text{disk}}(VV) &=\mathcal{N}_{\text{CC}} \int_\varepsilon^1 
    dy \left\langle \mathcal{B}_y c \widetilde{c} e^{-\phi} e^{-\widetilde{\phi}}  e^{\bb\varphi}(\ii)  c  \widetilde{c}(-\ii \bb \psi)(-\ii \bb \widetilde{\psi}) e^{\bb\varphi}(z, \bar{z}) \right \rangle_{\UHP} \\
    &=-\ii \, \mathcal{N}_{\text{CC}} \int_\varepsilon^1 dy \left\langle c \widetilde{c} e^{-\phi} e^{-\widetilde{\phi}}  e^{\bb\varphi}(\ii) \big( c(z) + \widetilde{c}(\bar{z}) \big)(-\ii \bb \psi)(-\ii \bb \widetilde{\psi}) e^{\bb\varphi}(z, \bar{z}) \right \rangle_{\UHP},
    \label{eq: disk_two_defn}
\end{align}
where it is understood that the second insertion is at $ z = \ii y $ and we used
$\BBB_y=\ii \ointclockwise_{\ii y} b(z) d z - \ii\ointclockwise_{-\ii y} \tilde b(\bar z) d\bar z$
where $\ointclockwise_{\ii y}$ denotes a clockwise contour around $\ii y$.\footnote{
In terms of the general discussion of section \ref{sec-sft-amplitudes}, we take a small circle $C_s$ oriented clockwise around the puncture at $\ii y$ to be the boundary between two coordinate patches.
The coordinate $\sigma_s$ is thus $z$ and the coordinate $\tau_s$ is the local coordinate using which the puncture is inserted. 
We can take $\sigma_s = \tau_s/M + \ii y$ where $M$ is a large positive real number.
Thus, $\partial F/\partial y = \ii$.
} 
Also, we have from \eqref{eapp1}
\be \label{eNCC}
\NNN_{\rm CC} = g_s\, \eta_c^{1/2}\, .
\ee

We shall now  use the equation of motion \eqref{eq:EOM_1} and holomorphicity of $c(z)$ to write the disk two-point function \eqref{eq: disk_two_defn} in a simple form. 
Following the bosonic case \cite{Eniceicu:2022xvk}, we would like to use the equation of motion of Liouville theory to convert the term $\psi\widetilde{\psi} e^{\bb \varphi}$ to $\partial \bar{\partial} \varphi$, a total derivative.
There is an apparent obstruction to his since \eqref{eq:EOM_1} includes a contribution from the auxiliary field $F$.  
Thankfully, the term involving $F$ can be neglected as it corresponds to a contact interaction \cite{Fukuda:2002bv}.
More specifically, the correlation functions involving $F$ have support only when the argument of $F$ coincides with that of another vertex operator or the worldsheet boundary. 
In the present case, the integration range over $y$ in
\eqref{eq: disk_two_defn} already excludes the boundary at $y=0$. It also
implicitly excludes the point $y=1$, where the second vertex operator is
inserted, since that region must be analyzed using closed string field
theory Feynman diagrams. This exclusion was not made explicit in
\eqref{eq: disk_two_defn}, because the integral has no divergence as
$y\to 1$. We may therefore use \eqref{eq:EOM_1}, dropping the term
proportional to $F$, to express the integrand as a total derivative:
\begin{equation}
    A_{\text{disk}}^{(1)}(VV) = -\frac{\ii}{\bb\wt\mu} \, \mathcal{N}_{\text{CC}} \int_\varepsilon^1 dy \left\langle c\widetilde{c} e^{-\phi} e^{-\widetilde{\phi}}  e^{\bb\varphi}(\ii) \big[ \bar{\partial} \big(c\partial\phi(z, \bar{z})\big) + \partial \big(\widetilde{c}\bar{\partial}\phi(z, \bar{z})\big) \big] \right \rangle_{\UHP}\,.
    \label{3.16}
\end{equation}
Using $\p=(\p_x-\ii\p_y)/2$ and $\bar\p=(\p_x+\ii\p_y)/2$, we can
convert the integrand into a linear combination of $y$- and
$x$-derivative terms. The $y$-derivative terms give boundary
contributions. We now discuss how to analyze the terms containing
$x$ derivatives.

If $f\circ W$ denotes the conformal transform of a vertex operator $W$ under
the conformal map $f$, then we have from \eqref{Coord_trs}, that
\begin{equation}
\begin{aligned}
f\circ c\, \p \varphi(z,\bar z) &= c\p\varphi (f(z), \bar f(\bar z)) + {Q\over 2} {f''(z)\over f'(z)^2}
c(f(z))\, , \\
f\circ \tilde c \, \bar\p \varphi(z,\bar z) &= \tilde c\bar \p\varphi (f(z), \bar f(\bar z)) + {Q\over 2} {\bar f''(\bar z)\over \bar f'(\bar z)^2}
\tilde c(\bar f(\bar z))\, .
\end{aligned}
\label{esl2ridentity}
\end{equation}
Let us take $f(z)$ to be the $\text{SL}(2,\mathbb{R})$ transformation $f(z)=(z+a)/(1-az)$.
This is a symmetry of the correlation function on the UHP and leaves the vertex operator
$c\widetilde{c} e^{-\phi} e^{-\widetilde{\phi}}  e^{\bb\varphi}(\ii)$ unchanged. 
It follows that, inside the correlator, we may apply this transformation
to either $c\p\varphi$ or $\tilde c\,\bar\p\varphi$ without changing the result.
For small $a$, we have
\be
f(z) = z + a\, (1+z^2) + \mathcal{O}(a^2), \qquad f'(z)=1 + \mathcal{O}(a),
\qquad f''(z) = 2a + \mathcal{O}(a^2)\, .
\ee
For $z=\ii y$, this gives $\delta x = a(1-y^2)$ and $\delta y=0$.
Hence, using \eqref{esl2ridentity}, we conclude that, inside the correlation function \eqref{3.16},
\begin{equation}
\begin{aligned}
(1-y^2) \p_x (c\p \varphi) + Q\, c &= 0, \\
(1-y^2) \p_x (\tilde c\bar \p \varphi) + Q\, \tilde c &= 0\, .
\end{aligned}
\label{exderiv}
\end{equation}
We now substitute $\p = (\p_x-\ii\p_y)/2$ and $\bar \p = (\p_x+\ii \p_y)/2$, and use (\ref{exderiv}) to express \eqref{3.16} as
{\small
\begin{align}
    A^{(1)}_\text{disk}(VV) &= \frac{1}{2\bb\wt\mu} \, \mathcal{N}_{\text{CC}} \int_{\varepsilon}^1 d y \, 
    \Bigg\langle 
    c \widetilde{c} e^{-\phi} e^{-\widetilde{\phi}}  e^{\bb\varphi}(\ii) \, \bigg[\partial_y \bigg( 
    c\partial{\varphi}(z) 
    - \widetilde{c} \bar{\partial}\varphi(\bar{z})  \bigg) 
    +\ii \, {Q\over 1-y^2} (c(z)+\tilde c(\bar z))\bigg]\Bigg\rangle_{\UHP}\, .
    \label{eq: disk_two_simplified}
\end{align}}
\par\noindent
The term involving the $y$ derivative can be expressed as boundary contributions from $y = \varepsilon$ and $y = 1$.
Using the OPE \eqref{V OPE} between $\partial \varphi$ and $e^{\bb \varphi}$, we can calculate the contribution from the $y=1$ boundary as follows:
\begin{align}
    &\frac{1}{2\bb\wt\mu} \,  \mathcal{N}_{\text{CC}} \, \lim_{z \to \ii} \Big\langle c\widetilde{c}e^{-\phi} e^{-\widetilde{\phi}}(\ii) \Big(c\partial{\varphi}(z)e^{\bb \varphi}(\ii) - \widetilde{c}\bar{\partial}\varphi(\bar{z})e^{\bb \varphi}(\ii)\Big) \Big\rangle_\text{UHP}   \nonumber \\
    &= -\frac{1}{\bb\wt\mu}\,  \mathcal{N}_{\text{CC}}\, \lim_{z \to \ii} \Big\langle c\widetilde{c}e^{-\phi} e^{-\widetilde{\phi}}(\ii) \Big(c(z)\left(\frac{\bb}{z-\ii}\right)e^{\bb \varphi}(\ii) \Big) \Big\rangle_\text{UHP}  
    \nonumber \\
    &=\mathcal{N}_{\text{CC}}\, K\, {\wt\mu}^{-2} \, \frac{Q}{\bb} \, .
    \label{3.31}
\end{align}
In the last line, we also used the one-point function $\langle e^{\bb\varphi}(\ii) \rangle_\text{UHP} = \frac{Q}{2b \, \ii \widetilde{\mu}}$ given in (\ref{one-pt-vb}) and the ghost correlators given in section \ref{Basic conventions for worldsheet CFT}.
The contribution from the $y=\varepsilon$ boundary can be calculated using the ghost correlators, the bulk-boundary OPE given in \eqref{e2.27} and \eqref{e2.28}, and the one-point function of $e^{\bb \varphi}(\ii)$ given in (\ref{one-pt-vb}), with the result\footnote{The first factor of two in the following equation comes from the fact that the terms involving $c \partial \varphi$ and $\tilde{c}\bar{\partial}\varphi$ make equal contributions. The overall minus sign arises because $y=\varepsilon$ is the lower
boundary of the integral.}
\begin{equation}
    - 2 \cdot 
    \frac{\mathcal{N}_{\text{CC}}}{2 \bb\wt\mu} \cdot 
    2\ii K (\varepsilon^2-1) \cdot 
    \frac{1}{2\ii} \cdot 
    \frac{Q}{2 \bb \ii \widetilde{\mu}} \cdot 
    \frac{-Q}{2\ii\varepsilon}
    = \mathcal{N}_{\text{CC}}\, K\, {\wt\mu}^{-2} \, \frac{Q^2}{4b^2} \, 
    \frac{1}{\varepsilon}+ O(\varepsilon)\, .
    \label{3.36}
\end{equation}
The last two terms in \eqref{eq: disk_two_simplified} can be evaluated by using the explicit expressions for the ghost correlators and the one-point function in (\ref{one-pt-vb}).\footnote{The integrand is independent of $y$, so the $y$ integral simply gives $1$.} The result is
\begin{equation}
    - \mathcal{N}_{\text{CC}}\, K\, {\wt\mu}^{-2} \, \frac{Q^2}{2b^2}\, .
    \label{3.33}
\end{equation}
Adding the contributions (\ref{3.31}), (\ref{3.36}) and (\ref{3.33}), we get
\begin{align}
A_\text{disk}^{(1)}(VV) = 
\mathcal{N}_{\text{CC}}\,  K \, \widetilde{\mu}^{-2}
\left( 
\frac{1}{\varepsilon} \frac{Q^2}{4\bb ^2}
+ \frac{Q}{\bb} 
- \frac{Q^2}{2\bb ^2}  
\right) \, .
    \label{ea1vv}
\end{align}

\subsubsection{Open-string exchange contribution} \label{sCOCO}
Next, we have to calculate the contribution from Fig.~\ref{figthree}(a)
representing the degeneration region. 
In the limit of large $\tilde\lambda$ and $\alpha$, the possible contributions come from the exchange of the tachyon and the exchange of the OSG field. 

Since the tachyon vertex operator is GSO odd while the external closed-string state is GSO even, the CO vertex with an internal tachyon vanishes. 
We therefore need only compute the contribution from exchange of the OSG mode.

Since in the definition of the CO interaction vertex we have placed the PCO combination $(\mathcal{X}+\tilde{\mathcal{X}})/2$ on the closed-string operator $V$, we first need to compute $(\mathcal{X}+\tilde{\mathcal{X}})V$.
We find
\begin{equation}
\begin{aligned}
\mathcal{X} V &= \left(-\frac{1}{2} \,\eta \,  \widetilde{c}\, e^{\phi} e^{-\widetilde{\phi}}\,  e^{\bb \varphi} + c\widetilde{c}\,  e^{-\widetilde{\phi}}\,  (-\ii \bb \psi) \,  e^{\bb \varphi}\right) \,, \\
\tilde{\mathcal{X}} V &= \left(-\frac{1}{2} \, \widetilde\eta \,  {c}\, e^{\wt\phi} e^{-{\phi}}\,  e^{\bb \varphi} - c\widetilde{c}\,  e^{-{\phi}}\,  (-\ii \bb \widetilde\psi) \,  e^{\bb \varphi}\right)\, .
\end{aligned}
\label{e6.19a}
\end{equation}
The CO vertex with insertions of $\mathcal{X} V$ and the OSG mode is thus
\begin{equation*}
    \left\langle \left(-\frac{1}{2} \,\eta \,  \widetilde{c}\, e^{\phi} e^{-\widetilde{\phi}}\,  e^{\bb \varphi} + c\widetilde{c}\,  e^{-\widetilde{\phi}}\,  (-\ii \bb \psi) \,  e^{\bb \varphi}\right)\!\! (\ii) \; \partial \xi \, c \partial{c} \, e^{-2\phi}(0) \right\rangle_{\UHP} \, .
\end{equation*}
Here, the second term will not contribute as the correlation function of four $c$-ghosts vanishes on the disk. 
The contribution from the first term gives us
\begin{equation}
    \left\langle 
    \mathcal{X}V(\ii) \, 
    \partial \xi \, c \partial{c} \, e^{-2\phi}(0) 
    \right\rangle_{\UHP} = 
    \left\langle 
    \bigg(-\frac{1}{2} \,\eta \,  \widetilde{c}\,  e^{\phi} e^{-\widetilde{\phi}}\,   e^{\bb \varphi} \bigg)\!(\ii) \; \partial \xi \, c \partial{c} \, e^{-2\phi}(0) 
    \right\rangle_{\UHP} 
    = K\,\frac{Q}{2 \bb\wt\mu}\, .
    \label{CO Out-of-Siegel}
\end{equation}
Similar reasoning gives
\begin{equation}
\left\langle \tilde{\mathcal{X}}V(\ii)
\partial \xi \, c \partial{c} \, e^{-2\phi}(0) 
\right \rangle_\text{UHP}
=
    \left\langle 
    \bigg(-\frac{1}{2}\,c\,\tilde{\eta}\, e^{-\phi}e^{\tilde{\phi}}e^{\bb\varphi}\bigg)(\ii)\, \partial \xi \, c\partial c\,  e^{-2\phi}(0) \right \rangle_\text{UHP}
    =-K\, \frac{Q}{2\bb\wt\mu}\, .\label{CO_OSG_anti_holo}
\end{equation}
Together, the last two equations show that the contribution from
Fig.~\ref{figthree}(a) with OSG exchange vanishes, since the relevant
vertices vanish. Since the tachyon exchange also vanishes, the full
open-string exchange diagram gives no contribution.

\subsubsection{Vertical integration contribution} \label{s6.3}

In computing the contribution from Fig.~\ref{figthree}(b) in section \ref{sec-bulkdisk}, we put the PCOs $\XXX$ and $\wt\XXX$ on the closed-string vertex operator at $z=\ii y$.
On the other hand, in computing the contribution from figure ~\ref{figthree}(a), we put $(\XXX +\wt\XXX)/2$ on each of the two closed-string vertex operators.
Thus the result obtained so far is incomplete: we must add the vertical-integration contribution associated with moving $\XXX$ or $\wt\XXX$ from $\ii$ to $\ii\varepsilon$ at the boundary $y=\varepsilon$ separating the two regions of moduli space.

\begin{figure}[t]
    \centering
    
\tikzset{every picture/.style={line width=0.75pt}} %set default line width to 0.75pt        

\begin{tikzpicture}[x=0.75pt,y=0.75pt,yscale=-0.741,xscale=0.741]
%uncomment if require: \path (0,376); %set diagram left start at 0, and has height of 376

%Shape: Rectangle [id:dp9190227754473517] 
\draw   (259,26) -- (414.91,26) -- (414.91,177.65) -- (259,177.65) -- cycle ;
%Shape: Rectangle [id:dp2937657643509537] 
\draw   (259,177.65) -- (414.91,177.65) -- (414.91,329.29) -- (259,329.29) -- cycle ;
%Straight Lines [id:da7296612113622719] 
\draw    (294.27,171.89) -- (294.17,184.66) ;
%Straight Lines [id:da7079026474779577] 
\draw [color={rgb, 255:red, 255; green, 0; blue, 24 }  ,draw opacity=1 ][line width=1.5]    (259,177.65) -- (414.91,26) ;
%Straight Lines [id:da2550569387547649] 
\draw [color={rgb, 255:red, 255; green, 0; blue, 24 }  ,draw opacity=1 ][line width=1.5]    (294,215.65) -- (414.91,329.29) ;
%Straight Lines [id:da9352696358710064] 
\draw [color={rgb, 255:red, 255; green, 0; blue, 24 }  ,draw opacity=1 ][line width=1.5]    (259,26) -- (294.27,25.89) ;
%Straight Lines [id:da7716701593954303] 
\draw [color={rgb, 255:red, 0; green, 0; blue, 0 }  ,draw opacity=0.3 ] [dash pattern={on 4.5pt off 4.5pt}]  (294.27,25.89) -- (294.24,329.96) ;
%Shape: Rectangle [id:dp14726951562931911] 
\draw   (30.08,26.14) -- (185.99,26.14) -- (185.99,177.78) -- (30.08,177.78) -- cycle ;
%Shape: Rectangle [id:dp1614093856247415] 
\draw   (30.08,177.78) -- (185.99,177.78) -- (185.99,329.43) -- (30.08,329.43) -- cycle ;
%Straight Lines [id:da43806689441840674] 
\draw    (65.35,172.02) -- (65.24,184.79) ;
%Straight Lines [id:da8569705991230978] 
\draw [color={rgb, 255:red, 255; green, 0; blue, 24 }  ,draw opacity=1 ][line width=1.5]    (65.08,144.78) -- (185.99,26.14) ;
%Straight Lines [id:da7858873502430673] 
\draw [color={rgb, 255:red, 255; green, 0; blue, 24 }  ,draw opacity=1 ][line width=1.5]    (30.08,177.78) -- (185.99,329.43) ;
%Straight Lines [id:da39858513665195394] 
\draw [color={rgb, 255:red, 255; green, 0; blue, 24 }  ,draw opacity=1 ][line width=1.5]    (30.08,26.14) -- (65.35,26.02) ;
%Straight Lines [id:da0871995141363896] 
\draw [color={rgb, 255:red, 0; green, 0; blue, 0 }  ,draw opacity=0.3 ] [dash pattern={on 4.5pt off 4.5pt}]  (65.35,26.02) -- (65.32,330.09) ;
%Shape: Rectangle [id:dp058254830196879936] 
\draw   (486.08,26.14) -- (641.99,26.14) -- (641.99,177.78) -- (486.08,177.78) -- cycle ;
%Shape: Rectangle [id:dp6075129212293501] 
\draw   (486.08,177.78) -- (641.99,177.78) -- (641.99,329.43) -- (486.08,329.43) -- cycle ;
%Straight Lines [id:da6951384711279855] 
\draw    (521.35,172.02) -- (521.24,184.79) ;
%Straight Lines [id:da3858465347006991] 
\draw [color={rgb, 255:red, 255; green, 0; blue, 24 }  ,draw opacity=1 ][line width=1.5]    (486.08,177.78) -- (641.99,26.14) ;
%Straight Lines [id:da7299160929293077] 
\draw [color={rgb, 255:red, 255; green, 0; blue, 24 }  ,draw opacity=1 ][line width=1.5]    (521.08,215.78) -- (641.99,329.43) ;
%Straight Lines [id:da5000015981140841] 
\draw [color={rgb, 255:red, 255; green, 0; blue, 24 }  ,draw opacity=1 ][line width=1.5]    (486.08,329.43) -- (521.35,329.31) ;
%Straight Lines [id:da19182200175151887] 
\draw [color={rgb, 255:red, 0; green, 0; blue, 0 }  ,draw opacity=0.3 ] [dash pattern={on 4.5pt off 4.5pt}]  (521.35,26.02) -- (521.32,330.09) ;
%Shape: Rectangle [id:dp46285871583957905] 
\draw   (709.08,26.14) -- (864.99,26.14) -- (864.99,177.78) -- (709.08,177.78) -- cycle ;
%Shape: Rectangle [id:dp7079534460304684] 
\draw   (709.08,177.78) -- (864.99,177.78) -- (864.99,329.43) -- (709.08,329.43) -- cycle ;
%Straight Lines [id:da2880128719881466] 
\draw    (744.35,172.02) -- (744.24,184.79) ;
%Straight Lines [id:da18384534744430792] 
\draw [color={rgb, 255:red, 255; green, 0; blue, 24 }  ,draw opacity=1 ][line width=1.5]    (744.08,144.78) -- (864.99,26.14) ;
%Straight Lines [id:da47261101573466846] 
\draw [color={rgb, 255:red, 255; green, 0; blue, 24 }  ,draw opacity=1 ][line width=1.5]    (709.08,177.78) -- (864.99,329.43) ;
%Straight Lines [id:da9324252778738074] 
\draw [color={rgb, 255:red, 255; green, 0; blue, 24 }  ,draw opacity=1 ][line width=1.5]    (709.05,329.2) -- (744.32,329.09) ;
%Straight Lines [id:da8805488742829696] 
\draw [color={rgb, 255:red, 0; green, 0; blue, 0 }  ,draw opacity=0.3 ] [dash pattern={on 4.5pt off 4.5pt}]  (744.35,26.02) -- (744.32,330.09) ;
%Straight Lines [id:da11398053486914939] 
\draw [color={rgb, 255:red, 255; green, 0; blue, 24 }  ,draw opacity=1 ][line width=1.5]  [dash pattern={on 5.63pt off 4.5pt}]  (65.35,26.02) -- (65.08,144.78) ;
%Straight Lines [id:da8887617438985098] 
\draw [color={rgb, 255:red, 255; green, 0; blue, 24 }  ,draw opacity=1 ][line width=1.5]  [dash pattern={on 5.63pt off 4.5pt}]  (294.27,25.89) -- (294,215.65) ;
%Straight Lines [id:da5809325294800288] 
\draw [color={rgb, 255:red, 255; green, 0; blue, 24 }  ,draw opacity=1 ][line width=1.5]  [dash pattern={on 5.63pt off 4.5pt}]  (521.08,215.78) -- (521.35,329.31) ;
%Straight Lines [id:da41398084688399706] 
\draw [color={rgb, 255:red, 255; green, 0; blue, 24 }  ,draw opacity=1 ][line width=1.5]  [dash pattern={on 5.63pt off 4.5pt}]  (744.08,144.78) -- (744.32,330.09) ;

% Text Node
\draw (296.26,181.32) node [anchor=north west][inner sep=0.75pt]  [font=\small]  {$\varepsilon $};
% Text Node
\draw (370.26,181.32) node [anchor=north west][inner sep=0.75pt]  [font=\small]  {$y\rightarrow $};
% Text Node
\draw (243.59,179.99) node [anchor=north west][inner sep=0.75pt]  [font=\small]  {$0$};
% Text Node
\draw (418.59,179.99) node [anchor=north west][inner sep=0.75pt]  [font=\small]  {$1$};
% Text Node
\draw (243.92,19.32) node [anchor=north west][inner sep=0.75pt]  [font=\small]  {$1$};
% Text Node
\draw (234.26,319.4) node [anchor=north west][inner sep=0.75pt]  [font=\small]  {$-1$};
% Text Node
\draw (67.33,181.46) node [anchor=north west][inner sep=0.75pt]  [font=\small]  {$\varepsilon $};
% Text Node
\draw (141.33,181.46) node [anchor=north west][inner sep=0.75pt]  [font=\small]  {$y\rightarrow $};
% Text Node
\draw (14.67,180.12) node [anchor=north west][inner sep=0.75pt]  [font=\small]  {$0$};
% Text Node
\draw (189.67,180.12) node [anchor=north west][inner sep=0.75pt]  [font=\small]  {$1$};
% Text Node
\draw (15,19.46) node [anchor=north west][inner sep=0.75pt]  [font=\small]  {$1$};
% Text Node
\draw (5,319.54) node [anchor=north west][inner sep=0.75pt]  [font=\small]  {$-1$};
% Text Node
\draw (523.33,181.46) node [anchor=north west][inner sep=0.75pt]  [font=\small]  {$\varepsilon $};
% Text Node
\draw (597.33,181.46) node [anchor=north west][inner sep=0.75pt]  [font=\small]  {$y\rightarrow $};
% Text Node
\draw (470.67,180.12) node [anchor=north west][inner sep=0.75pt]  [font=\small]  {$0$};
% Text Node
\draw (645.67,180.12) node [anchor=north west][inner sep=0.75pt]  [font=\small]  {$1$};
% Text Node
\draw (471,19.46) node [anchor=north west][inner sep=0.75pt]  [font=\small]  {$1$};
% Text Node
\draw (460,319.54) node [anchor=north west][inner sep=0.75pt]  [font=\small]  {$-1$};
% Text Node
\draw (746.33,181.46) node [anchor=north west][inner sep=0.75pt]  [font=\small]  {$\varepsilon $};
% Text Node
\draw (820.33,181.46) node [anchor=north west][inner sep=0.75pt]  [font=\small]  {$y\rightarrow $};
% Text Node
\draw (693.67,180.12) node [anchor=north west][inner sep=0.75pt]  [font=\small]  {$0$};
% Text Node
\draw (868.67,180.12) node [anchor=north west][inner sep=0.75pt]  [font=\small]  {$1$};
% Text Node
\draw (694,19.46) node [anchor=north west][inner sep=0.75pt]  [font=\small]  {$1$};
% Text Node
\draw (684,319.54) node [anchor=north west][inner sep=0.75pt]  [font=\small]  {$-1$};
% Text Node
\draw (240.57,149.12) node [anchor=north west][inner sep=0.75pt]  [font=\scriptsize,rotate=-270.33] [align=left] {PCO Location};
% Text Node
\draw (14.01,150.12) node [anchor=north west][inner sep=0.75pt]  [font=\scriptsize,rotate=-270.33] [align=left] {PCO Location};
% Text Node
\draw (469.01,150.12) node [anchor=north west][inner sep=0.75pt]  [font=\scriptsize,rotate=-270.33] [align=left] {PCO Location};
% Text Node
\draw (691.01,150.12) node [anchor=north west][inner sep=0.75pt]  [font=\scriptsize,rotate=-270.33] [align=left] {PCO Location};
% Text Node
\draw (99.92,349.32) node [anchor=north west][inner sep=0.75pt]  [font=\small]  {$1.$};
% Text Node
\draw (333.92,349.39) node [anchor=north west][inner sep=0.75pt]  [font=\small]  {$2.$};
% Text Node
\draw (558.92,349.39) node [anchor=north west][inner sep=0.75pt]  [font=\small]  {$3.$};
% Text Node
\draw (787.92,349.39) node [anchor=north west][inner sep=0.75pt]  [font=\small]  {$4.$};

\end{tikzpicture}

    \caption{Our choice of PCO locations in the different regions of moduli space for the disk two-point function. At $y=\varepsilon$, the section jumps discontinuously, requiring vertical-integration contributions. We average over the four contributions shown in the panels, corresponding respectively to \eqref{eq: Disk_2pt_VI_1}, \eqref{eq: Disk_2pt_VI_2}, \eqref{eq: Disk_2pt_VI_3}, and \eqref{eq: Disk_2pt_VI_4}.}
    \label{fig: disk_2_pt_VI}
\end{figure}
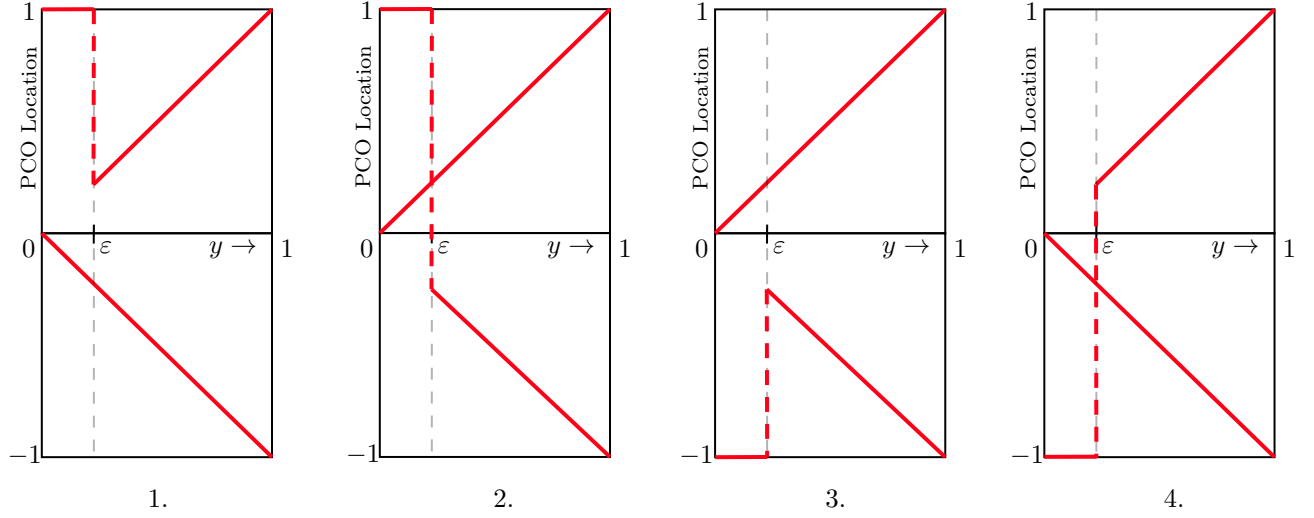

Recalling that $\wt\XXX$ at $z$ can be replaced by $\XXX$ at $\bar z$, we see that we need to consider the following four contributions to vertical integration
\begin{enumerate}
    \item Moving a $\mathcal{X}$ from $\ii$ to $\ii\varepsilon$ with another $\XXX$ fixed at $-\ii\varepsilon$.
    \item Moving a $\mathcal{X}$ from $\ii$ to $-\ii\varepsilon$ with another $\XXX$ fixed at $\ii\varepsilon$.
    \item Moving a $\mathcal{X}$ from $-\ii$ to $-\ii\varepsilon$ with another $\XXX$ fixed at $\ii\varepsilon$.
    \item Moving a $\mathcal{X}$ from $-\ii$ to $\ii\varepsilon$  with another $\XXX$ fixed at $-\ii\varepsilon$.
\end{enumerate} 
See Fig.~(\ref{fig: disk_2_pt_VI}) for a pictorial depiction.
Finally, we sum the four contributions and divide by four. This factor
arises because, in the degeneration region, the two CO vertices together
contribute two factors of $(\mathcal{X}+\widetilde{\mathcal{X}})/2$.

Let us use the procedure described in \cite{deLacroix:2017lif} to compute the first of these contributions.
The PCO $\mathcal{X}$ is moved from $\ii$ to $\ii \varepsilon$ as we enter the bulk region from the degeneration region.
So we get the vertical integration contribution
\begin{align}
    &\mathcal{N}_{\text{CC}} \int_{\ii}^{\ii\varepsilon}d u\left\langle\left(- 2\partial \xi(u)\right) \, \tilde{\mathcal{X}} V(\ii\varepsilon) V(\ii) \right\rangle_{\UHP} \nonumber \\
    &=\mathcal{N}_{\text{CC}} \int_{\ii}^{\ii\varepsilon}d u
    \left\langle \left(-2\partial\xi(u)\right) \, 
    \left( -\frac{1}{2} \tilde{\eta}  c e^{\tilde{\phi}} e^{-\phi} e^{\bb \varphi} + \ii \bb \, c\widetilde{c}  e^{-\phi}\, \tilde{\psi} \,  e^{\bb \varphi} \right) \!\! (\ii \varepsilon) \, c \widetilde{c} e^{-\phi} e^{-\widetilde{\phi}} e^{\bb \varphi} (\ii) \right\rangle_{\UHP} \, \nonumber\\
    &= \mathcal{N}_{\text{CC}} \cdot \frac{\ii(1-\varepsilon)}{2(1 + \varepsilon)}
    \cdot 2\ii K (\varepsilon^2-1) \cdot \left\langle e^{\bb \varphi}(\ii \varepsilon)e^{\bb \varphi}(\ii) \right \rangle
    \label{eq: CC_VI_intermediate}\, .
\end{align}
We used (\ref{e6.19a}) to write the expression for $\tilde{\mathcal{X}} V$.
Note that the contribution from the second term in the middle line vanishes.
We used (\ref{exietaphi}) to compute the correlator involving the $\xi$, $\eta$ and $\phi$ ghosts, which gives the second factor in (\ref{eq: CC_VI_intermediate}).
The $c$-ghost part of the correlator was computed using (\ref{eq:ev_ccc}) and gives the third factor in (\ref{eq: CC_VI_intermediate}).

We now need to evaluate $\left\langle e^{\bb \varphi} (\ii \varepsilon) e^{\bb \varphi} (\ii) \right\rangle$.
For the $(1,1)$ ZZ instanton, the identity operator is the only boundary primary \cite{Zamolodchikov:2001ah, Ahn:2002ev, Fukuda:2002bv}. 
Thus, in the limit of small $\varepsilon$, we can compute this correlator by first using the bulk-boundary OPE given in (\ref{e2.19OPE}) for the operator $e^{\bb \varphi}(\ii \varepsilon)$, which reduces the computation to the one-point function of $e^{\bb \varphi}(\ii)$. 
This one-point function is given in (\ref{one-pt-vb}).
Thus,
\begin{equation}
    \left\langle e^{\bb \varphi} (\ii \varepsilon) e^{\bb \varphi} (\ii) \right\rangle_\text{UHP} = \frac{U(\bb)}{2 \varepsilon} \cdot \frac{U(\bb)}{2} 
    + O(\varepsilon)
    = - \frac{1}{\varepsilon} \frac{Q^2}{4 \bb^2\wt\mu^2} + O(\varepsilon) 
    \, .
\end{equation}
Putting it back in \eqref{eq: CC_VI_intermediate}, we get
\begin{align}
    \label{eq: Disk_2pt_VI_1}
    A_{\text{disk}}^{\text{VI,1}}(VV) = 
    -K\, 
    \mathcal{N}_{\text{CC}}\left(\frac{1}{2\varepsilon}-1\right)\frac{Q^2}{2 \bb^2\wt\mu^2 }
    +\mathcal{O}(\varepsilon)\, .
\end{align}

Analogously, we can find the three remaining vertical integration contributions:
\begin{align}
    A_{\text{disk}}^{\text{VI,2}}(VV) &=-K\, \mathcal{N}_{\text{CC}}\left(\frac{1}{2\varepsilon}+1\right)\frac{Q^2}{2 \bb^2\wt\mu^2 }+\mathcal{O}(\varepsilon)\,,
    \label{eq: Disk_2pt_VI_2} \\
    A_{\text{disk}}^{\text{VI,3}}(VV) &=-K\, 
    \mathcal{N}_{\text{CC}}\left(\frac{1}{2\varepsilon}-1\right)\frac{Q^2}{2 \bb^2\wt\mu^2 }
    +\mathcal{O}(\varepsilon) \,,
    \label{eq: Disk_2pt_VI_3} \\
    A_{\text{disk}}^{\text{VI,4}}(VV) &=-K\,
    \mathcal{N}_{\text{CC}}\left(\frac{1}{2\varepsilon}+1\right)
    \frac{Q^2}{2 \bb^2\wt\mu^2 }+\mathcal{O}(\varepsilon)
    \label{eq: Disk_2pt_VI_4}
\end{align}
Now we need to add all these four contributions and divide by a factor of four.
So we get
\begin{align}\label{eq: Disk_2pt_VI}
    A_{\text{disk}}^{\text{VI}}(VV)=
    -\mathcal{N}_{\text{CC}}\, K \,  \widetilde{\mu}^{-2}\, 
    \frac{Q^2}{4 \bb^2} \frac{1}{\varepsilon}
    +\mathcal{O}(\varepsilon)\, .
\end{align}
Thus, we see that the vertical integration contribution cancels the $1/\varepsilon$ part of the bulk contribution (\ref{ea1vv}) and we are left with a finite result.

\subsubsection{Final result for the disk two-point function}
Adding the contributions \eqref{ea1vv} and \eqref{eq: Disk_2pt_VI}, we get the full disk two-point amplitude
\begin{align}
\label{Adisk_final}
    A_\text{disk}(VV) = \mathcal{N}_{\text{CC}}\, K \,  \widetilde{\mu}^{-2}
\left( \frac{Q}{\bb} 
- \frac{Q^2}{2\bb ^2} 
\right) \, .
\end{align}
Dividing this by the square of the disk one-point function (\ref{3.11}), we get,
\begin{align}
   g_s\,  f = 
   \mathcal{N}_{\text{CC}}\, 
    {2K\over K_0^2}\, \left(\frac{2\bb}{Q} - 1\right)\, .
\end{align}
Finally, using \eqref{eNCC}, \eqref{eKT} and \eqref{etension} to substitute the values $\mathcal{N}_\text{CC} = g_s \eta_c^{1/2}$ and $\frac{K}{K_0} = \frac{1}{2}\eta_c^{-1/2}$, we get
\be
f 
= \frac{1}{K_0}\, \left(\frac{2\bb}{Q}-1\right)\, .
\ee
This agrees perfectly with the prediction \eqref{eq: f and g_KPZ_expectation} from KPZ scaling!

\section{Annulus one-point function} \label{sannulus}
In this section, we compute the annulus one-point function of the cosmological constant operator with the $(1,1)$ ZZ-instanton boundary conditions on both the boundaries. 
As in the case of the disk two-point function, a naive worldsheet expression for this amplitude develops divergences near the boundaries of moduli space, and therefore the computation must be organized using open-closed string field theory. 

Concretely, the final answer receives contributions from several sources. First, there is direct worldsheet contribution from the region of the annulus moduli space associated with the Feynman diagram shown in Fig.~\ref{fig:feynman-diagrams_2}(d). Second, there are 
contributions from the Feynman diagrams shown in Fig.~\ref{fig:feynman-diagrams_2}(a),
(b) and (c) involving open-string propagators. Third, since different regions are described using different induced PCO locations, we must include vertical integration terms at the interfaces between these regions. Finally, for the annulus one-point function, there is an additional contribution arising from the gauge-parameter redefinition, which plays an essential role in obtaining the correct finite answer.

We shall evaluate these contributions one by one and then add them together at the end.
As in \cite{Sen:2020eck,Eniceicu:2022xvk}, during this calculation we shall drop any term that contains a negative power of either $\alpha$ or $\wt\lambda$, even if it contains a positive power of the other quantity, since all such terms are expected to cancel at the end.
In principle, one could retain all such terms, at the cost of a more laborious analysis, and show explicitly that they cancel.

\subsection{Worldsheet expression for the amplitude}
We begin by writing the naive worldsheet expression for the annulus one-point function before decomposing it into the different string-field-theory contributions.

As a first step, we outline the two-dimensional moduli space associated with the annulus. 
We describe the annulus with a complex coordinate $w = 2\pi(x + \ii y),$ introduced in \eqref{annulus_coord}.
The $y$-coordinate is periodic with period $t$ and we define $v := e^{-2\pi t}$.
Since the annulus has translational invariance along the $y$-direction, we use this symmetry to fix the closed-string insertion point to $y=0$.
Thus, the moduli space is parameterized by the variables $ v $ and the $ x $-coordinate of the closed-string vertex operator $V$. 
These variables take values in the ranges\footnote{
Note that the region $ x > \frac{1}{4} $ corresponds to the same geometry since the annulus is invariant under the diffeomorphism $ w \to \pi - w $.}
\begin{equation}
  0 \leq v \leq 1, \quad 0 \leq x \leq \frac{1}{4}\, .
\end{equation}
We will also use the coordinate $z$, related to $w$ via $z = e^{\ii w}$.
The coordinate $z$ lives in the UHP and is subject to the identification $z \equiv v z$.

Our starting point will be the expression for the annulus one-point function
given in \eqref{e2.47}:
\ben \label{e2.47new}
  && -2\pi \mathcal{N}_{\text{C}}\int_0^\infty d t \, \int_0^{1/4} d x \, \Bigg\langle
    \left(\int_0^{\pi}b(w)d w+\int_0^{\pi} \tilde{b}(\bar{w})d\bar{w}\right)
     \left(\oint_{2\pi x}b(w')d w'+\oint_{2\pi x}\tilde{b}(\bar{w}')d \bar{w}'\right)\nonumber \\
     && \hskip 1in \mathcal{X}(w_1)
     \mathcal{X}(w_2) \, V(w, \bar{w}) \Bigg\rangle_{\rm A} \, , \\
  && \mathcal{N}_{\text{C}} = g_s\eta_c\, , \label{encexpression}
\een
where $\ointop_{2\pi x}$ denotes an anti-clockwise contour around the closed-string insertion, and $V=c\widetilde{c} \, e^{-\phi}e^{-\widetilde{\phi}} \,  e^{\bb \varphi}$  is the cosmological constant operator. 
As we shall see, the integral diverges from the $v\to 0$ and the $x\to 0$ regions.
The four Feynman diagrams with internal open-string propagators contributing to this amplitude have been shown in Fig.~\ref{fig:feynman-diagrams_2}, and the regions in the $(v,x)$ plane covered by these four diagrams have been shown in Fig.~\ref{fig:moduli-feynman}.

\subsection{Contribution from region \texorpdfstring{(d)}{}}

The region (d) in Fig.~\ref{fig:moduli-feynman} represents the bulk of moduli space and corresponds to the Feynman diagram in Fig.~\ref{fig:feynman-diagrams_2}(d). 
This requires knowledge of the one-point vertex of a closed-string puncture on the annulus. 
We take the PCOs to be $\XXX\wt\XXX$ on top of the closed-string vertex operator, the resulting operator $\XXX\wt\XXX V$ is given in \eqref{PCO_action}.

On the annulus, we need the total $\phi$-momentum to vanish to get a non-vanishing correlator, and so only the $c \widetilde{c} (-\ii \bb \psi)(-\ii \bb \widetilde{\psi}) e^{\bb \varphi}$ term in \eqref{PCO_action} contributes. 
Therefore we can replace $\mathcal{X}\tilde{\mathcal{X}}V$ by
 \be
 c\tilde c\, \mathcal{V}, \qquad  \text{where} \quad  
 \mathcal{V} := -\bb^2\psi\tilde{\psi}e^{\bb\varphi}\, .
 \ee
To simplify the integrals in (\ref{e2.47new}), we first write the standard mode expansions of $b$ and $c$ in the $z$-coordinate and convert these to the $w$-coordinate. 
The mode expansions in the $w$-coordinate are
\begin{align}\label{eq: bc_mode_exp}
    b(w)=-\sum_nb_ne^{-\ii nw}, \quad\tilde{b}(\tilde{w})=-\sum_nb_ne^{\ii n\bar{w}}\, , \nonumber\\
    c(w)=-\ii\sum_{n}c_ne^{-\ii nw}, \quad \tilde{c}(\bar{w})=\ii\sum_{n}
    c_ne^{\ii n\bar{w}}\, .
\end{align}
Equation (\ref{e2.47new}) now reduces to
\begin{align} \label{e8.11}
    8\pi^2\ii \, \mathcal{N}_{\text{C}}\int_{{(d)}} 
    dt \, dx
    \left\langle b_0c_0\mathcal{V}
  \right\rangle 
  = 4\, \pi \, \ii \, \NNN_{\rm C} \int_{{(d)}} 
    d v \, dx \, \frac{1}{v}
  \left\langle b_0c_0\mathcal{V}
  \right\rangle \, .
\end{align}

Recall that the product $ g_s g $ is defined as the ratio between the annulus and disk one-point amplitudes. Dividing \eqref{e8.11} by the disk one-point function computed in \eqref{3.11}, we get the contribution $g_s g^{(\text{d})}$ to $g_sg$ from region (d):
\begin{align}
g_sg^{(\text{d})} &= \int_{{(\text{d})}} dv\, dx\, F(v,x)\, , \quad \text{where }
\label{egsgdef} \\
F(v,x) &:= g_s \, \wt\mu\, C \, \mathrm{Tr}\left[ \mathcal{V}(w, \bar{w})\, b_0 c_0\, v^{L_0-1} \right]\, , \quad \text{and }
\label{efvxexp} \\
    C
    &:= - 8 \pi \ii \, \frac{\eta_c}{K_0} \, \frac{\bb}{Q}\, .
    \label{e8.12}
\end{align}
Using the equation of motion \eqref{eq:EOM_1} and dropping the terms involving the auxiliary field following the discussion below \eqref{epartial1}, we can express $\mathcal{V}$ as 
\begin{align} \label{eVexpression}
    \mathcal{V} 
      = \frac{1}{\bb\wt\mu}\,\partial_w \partial_{\bar{w}} \varphi
      = \frac{1}{16 \pi^2 \bb\wt\mu} \left( \partial_x^2 + \partial_y^2 \right)\varphi .
\end{align}

The term proportional to $\partial_y^2 \varphi$ in \eqref{eVexpression} does not contribute. 
This is because shifting the vertex operator along the $y$-direction corresponds to a simple translation, for which the anomaly in the transformation of $\partial \varphi$ vanishes (see \eqref{Coord_trs}).
Now rewrite the integrand $ F(v,x) $ as a total $x$-derivative
\begin{equation}\label{eq: G(v,x)_def}
F(v,x) = \partial_x G(v,x), 
\qquad 
G(v,x) := \frac{g_s C}{16 \pi^2 \bb} \, \text{Tr} \!\left[ \partial_x \varphi(w, \bar{w})\, b_0\, c_0\, v^{L_0 - 1} \right].
\end{equation}
Thus, the contribution from region (d) can ultimately be evaluated from boundary terms.

We first determine the behaviour of $F(v,x)$ and $G(v,x)$ in the two potentially singular limits: small $x$ at fixed $v$, and small $v$ at fixed $x$.
Let's start with small $x$ and finite $v$.
The cylinder coordinate $w$ is related to the upper half-plane coordinate by $z=e^{\ii w}$. 
Using the transformation of $ \partial \varphi $ and $ \bar{\partial} \varphi $ is given in \eqref{Coord_trs}, we get
\begin{equation}
\partial_w \varphi(w, \bar{w}) = \ii \frac{Q}{2} + \ii z \partial_z \varphi(z, \bar{z}) ,
\end{equation}
\begin{equation}
\partial_{\bar{w}} \varphi(w, \bar{w}) = -\ii \frac{Q}{2} - \ii \bar{z} \bar{\partial}_{\bar{z}} \varphi(z, \bar{z}) .
\end{equation}
For small $ x $, the insertion approaches the boundary, and we can use the bulk-boundary OPEs \eqref{e2.27} and \eqref{e2.28}. 
So we have
\begin{equation} \label{e7.17xy}
\partial_x \varphi(w, \bar{w}) = 2\pi \left( \partial_w \varphi(w, \bar{w}) + \partial_{\bar{w}} \varphi(w, \bar{w}) \right)
= 2\pi \ii \left( z \partial_z \varphi(z, \bar{z}) - \bar{z} \bar{\partial}_{\bar{z}} \varphi(z, \bar{z}) \right) \approx -\frac{Q}{x}.
\end{equation}
Substituting this into the definition \eqref{eq: G(v,x)_def} of $ G(v,x) $ gives
\begin{align} \label{ssmallx}
G(v,x) &\approx -g_s\, C \frac{Q}{\bb} \frac{1}{16\pi^2 x} \, \text{Tr} \left[ b_0 c_0 \, v^{L_0 - 1} \right] = -g_s\, C \frac{Q}{\bb} \frac{1}{16\pi^2 x} \frac{Z(v)}{v}\, ,
\nonumber\\
F(v,x) &\approx g_s\, C \frac{Q}{\bb} \frac{1}{16\pi^2 x^2} \frac{Z(v)}{v}\, ,
\qquad 
\text{for small }x\, ,
\end{align}
where $ Z(v) $ is the annulus partition function given in \eqref{eq:Zv_def}. 
The small $v$ behavior of $Z(v)$ was given in (\ref{zv-leading}), which we reproduce here for convenience:
\begin{equation}
    Z(v) \approx v^{-\frac{1}{2}}-2 +\mathcal{O}(v^{\frac{1}{2}})\, .
    \label{Annulus_partition}
\end{equation}
As discussed in section \ref{sec-zz-inst}, the term $v^{-1/2}$ is due to the open-string tachyon and the term $-2$ term is due to the pair of $L_0=0$ states with vertex operators $\p\xi c e^{-2\phi}$ and $\eta c$. 

Next, we consider the small-$ v $, finite-$ x $ region. 
We can evaluate this correlation function by separating out the matter and ghost contributions.  
In the matter sector the leading and subleading contributions for small $ v $ come from just the vacuum state $ |0\rangle $, since the next contribution comes from the matter stress tensor of 
dimension 2. In the ghost sector, the leading and subleading contributions to the trace are given in \eqref{Annulus_partition}.
Therefore, for small $v$, we have
\begin{equation} 
\text{Tr} \left[ \partial_x \varphi(w, \bar{w}) \, b_0 c_0 \, v^{L_0 - 1} \right] \approx 
v^{-1} \, \left( 
v^{-\frac{1}{2}}  -2 
+ \mathcal{O}(v^\frac{1}{2})
\right)
\, 
\langle 0 | \partial_x \varphi(w, \bar{w}) | 0 \rangle_{\rm Liouville} \, .
\end{equation}
Since we do not have a trace in the Liouville sector, we no longer need  the identification $ z \equiv vz $. 
Using the third expression of \eqref{e7.17xy} and the one-point functions given in \eqref{ephiexpect}, we get
\begin{equation}
\langle 0 | \partial_x \varphi(w, \bar{w}) | 0 \rangle_{\rm Liouville} = -\ii 2\pi Q \frac{z + \bar{z}}{z - \bar{z}} = -2\pi Q \cot(2\pi x)\,,
\end{equation}
and hence
\begin{align} \label{ssmallv}
G(v,x) &\approx -g_s\, C \frac{Q}{8\pi \bb} \left( v^{-\frac{3}{2}} - 2v^{-1} + \mathcal{O}(v^{-\frac{1}{2}}) \right) \cot(2\pi x),\nonumber\\
F(v,x) &\approx g_s\, C \frac{Q}{\bb} \left( v^{-\frac{3}{2}} - 2v^{-1} + \mathcal{O}(v^{-\frac{1}{2}}) \right) \frac{1}{4 \sin^2(2\pi x)}\, ,  \qquad \text{for small } v.
\end{align}
Equations \eqref{ssmallx} and \eqref{ssmallv} show that 
the integral develops divergences in the limits $ v \to 0 $ and $ x \to 0 $. 
However, we see from \eqref{eq:region_d_approx} and Fig.~\ref{fig:moduli-feynman} that the small-$x$ and small-$v$ regions are excluded from region (d), making the integral finite. 

Now it follows from \eqref{egsgdef} and \eqref{eq: G(v,x)_def} that we have an integral of a total
derivative term and hence the contribution to \eqref{egsgdef}  may be expressed as  a boundary term.
As illustrated in Fig.~\ref{fig:moduli-feynman} and \eqref{eq:region_d_approx},
the boundary of the region (d) consists of four components.  
The first boundary corresponds to the line $ x = \tfrac{1}{4} $. 
However, this boundary gives no contribution due to the symmetry $ w \to \pi - w $ under which $\p_x\varphi$ changes sign, and hence $\text{Tr}\!\left[ \partial_x \varphi(w, \bar{w})\, b_0\, c_0\, v^{L_0 - 1} \right]$  vanishes. 
The second boundary is the line $v=1$. It gives no contribution, since
$v$ is fixed on this boundary while the integrand is a pure
$x$ derivative; moreover, the integrand is nonsingular along this line.
The remaining two boundaries must be taken into account: the boundary separating regions (d) and (b), and the boundary separating regions (d) and (c). We denote their respective contributions by $g^{(\text{d})-(\text{b})}$ and $g^{(\text{d})-(\text{c})}.$

The interface between regions (d) and (b) occurs at small values of $x$, 
specified by 
$
x = \frac{1}{2\pi \tilde{\lambda}},
$
while $v$ lies within the interval 
$
\left(\alpha^2 - \tfrac{1}{2}\right)^{-1} \leq v \leq 1
$ \cite{Sen:2020eck, Eniceicu:2022xvk}.
The small-$x$, finite-$v$ behaviour described in \eqref{ssmallx} 
can then be employed to evaluate $g^{(\text{d})-(\text{b})}$. 
Noting from Fig.~\ref{fig:moduli-feynman} 
that this boundary lies at the lower limit of $x$ integration,
we get
\begin{equation}
\label{eq: g(b-d)}
    g_sg^{(\text{d})-(\text{b})}=-\int_{(\alpha^{2}-\frac{1}{2})^{-1}}^1 dv \ G(v,(2\pi \tilde{\lambda})^{-1})=\frac{ g_s C Q\tilde{\lambda}}{8\pi\bb} \int_{(\alpha^{2}-\frac{1}{2})^{-1}}^1 d v \,\, \frac{Z(v)}{v}\, .
\end{equation}
Using \eqref{e8.12} and \eqref{eKT}, this can be written as
\be\label{e8.18}
\boxed{
g^{(\text{d})-(\text{b})} = {\tilde\lambda\over 2\pi} \, \frac{1}{K_0}\,  \int_{(\alpha^{2}-\frac{1}{2})^{-1}}^1 d v \,\, \frac{Z(v)}{v}\, .}
\ee
We could evaluate this using the known form of $Z(v)$, but we shall not do so here, because
this contribution will be exactly cancelled by the contribution \eqref{e7.61a}
coming from the vertical integration between region (b) and (d).

Next, we examine the boundary between regions (d) and (c), 
which is parametrized by $v(\beta, u)$ and $x(\beta, u)$ as given in
\eqref{eq:xv_simplified}, with $u$ taking the following constant along this boundary:
\be\label{euvalue}
u = \alpha^{-2}\left(1 + \frac{1}{4 \tilde{\lambda}^2}\right)^{-2}\, .
\ee
We again see from Fig.~\ref{fig:moduli-feynman} that 
the integration contour lies at the lower limit of $x$ integration for a given $v$.
Since $v$ remains small along this boundary,  \eqref{ssmallv} 
can be applied to evaluate the contribution $g^{(\text{d})-(\text{c})}$. This gives
\begin{equation}\label{eq: gdc_int_I}
  g_s g^{(\text{d})-(\text{c})}= -\int_{(\text{d})-(\text{c})} dv G(v,x) = \frac{g_sCQ}{8\pi\bb} \int_{(\text{d})-(\text{c})} d v\, \left( v^{-\frac{3}{2}} - 2v^{-1}+\mathcal{O}\left(v^{-\frac{1}{2}}\right)  \right) \cot(2\pi x) \, ,
\end{equation}
where $(\text{d})-(\text{c})$ is the curve separating the regions (d) and (c).
To evaluate this integral, we use the second equation in \eqref{eq:xv_simplified} to change the integration variable from $v$ to $\beta$, with $u$ remaining fixed at \eqref{euvalue}.
This gives
\begin{align}
     g^{(\text{d})-(\text{c})}&=  \frac{CQ}{8\pi\bb} \int_{1}^{\frac{1}{2\tilde{\lambda}}} d \beta\, 
     {dv(\beta)\over d\beta} \, \left[ v(\beta)^{-\frac{3}{2}} -2\,
     v(\beta)^{-1} +\mathcal{O}\left( v(\beta)^{-\frac{1}{2}}\right) \right] \cot(2\pi x(\beta))\, .
\end{align}
When we expand the integrand in inverse powers of $\alpha$, we only retain terms without inverse powers of $\alpha$, since the $\beta$ integral does not produce factors of $\alpha$ in the numerator. 
Using \eqref{e8.12} to express $C Q/\bb$ in terms of $K_0$, and using
$\eta_c = \frac{\ii}{2\pi}$, we find that the integral, after dropping all
terms containing inverse powers of $\alpha$ or $\wt\lambda$, is
\begin{equation} 
\label{eq: g_(d)_(c)_res}
    \boxed{g^{(\text{d})-(\text{c})}
    = \frac{1}{K_0}\,  \left(\frac{\tilde{\lambda} \alpha}{\pi}\log  {2 \tilde{\lambda}}-\frac{\tilde{\lambda} \alpha}{\pi}-\frac{2\tilde{\lambda}}{\pi}+1\right)\, .}
\end{equation}

\subsection{Contribution from region \texorpdfstring{(c)}{}}

We now turn to Fig.~\ref{fig:feynman-diagrams_2}(c), which produces contribution from  region (c) in Fig.~\ref{fig:moduli-feynman}. 
On the COO vertex in Fig.~\ref{fig:feynman-diagrams_2}(c), the PCOs
$\XXX\wt\XXX$ are placed on the closed-string vertex operator, as in
Fig.~\ref{fig:feynman-diagrams_2}(d). Hence the region-(c) integrand is
obtained by analytic continuation of the region-(d) integrand.
Contribution from this region can be calculated by integrating over the region (c) using the $(u, \beta)$ parameterization \eqref{eq:xv_simplified},  expressing this in terms of $q_2$ and $\beta$ variables using \eqref{euqrel}, and finally using the replacement rules \eqref{ereprule} and \eqref{ereprule1}.

We begin with
\begin{align}
    g_sg^{(\text{c})}=\int_{(\text{c})} F(v,x)dv\wedge dx\, .
\end{align}
We can use the small $v$ expansion of $F(v,x)$ given in \eqref{ssmallv} and 
the parametrization \eqref{eq:xv_simplified} to get, 
\begin{align}
    g^{(\text{c})}&=\frac{CQ}{4\bb}\int_{\frac{1}{2\tilde{\lambda}}}^{1} d \beta \int\limits_{0}^{\alpha^{-2}(1+(4\tilde{\lambda}^2)^{-1})^{-2}} d u 
    \,  \left(\frac{1+\beta^2}{4\pi \beta^2 \tilde{\lambda}^2 }\right)\left(\frac{1}{\sin^2(2\pi x)} \right)\left( v^{-\frac{3}{2}} - 2v^{-1} + \mathcal{O}(v^{\frac{1}{2}})\right) \,.
 \end{align} 
 We now change the integration variable from $u$ to $q_2$ using \eqref{eq:xv_simplified}, 
 \eqref{euqrel}, use the
 replacement rule \eqref{ereprule}, \eqref{ereprule1} and carry out the $\beta$ integral.
Further using \eqref{e8.12} and $\eta_c = \frac{\ii}{2\pi}$, the result is, 
\be  \label{eq: g^(c)_contri}
\boxed{
 g^{(\text{c})}  
   = - \frac{1}{K_0}\,  \frac{\alpha \tilde{\lambda}}{\pi}\log{2\tilde{\lambda}}\, .}
   \ee

\subsection{Exchange of the out-of-Siegel gauge mode in region \texorpdfstring{(c)}{}}
We also need to consider the contribution due to the OSG mode $\tau$  flowing along the open string propagator in Fig.~\ref{fig:feynman-diagrams_2}(c).
The vertex operator corresponding to this mode is $V_\tau=\partial \xi c\partial c e^{-2\phi}$.

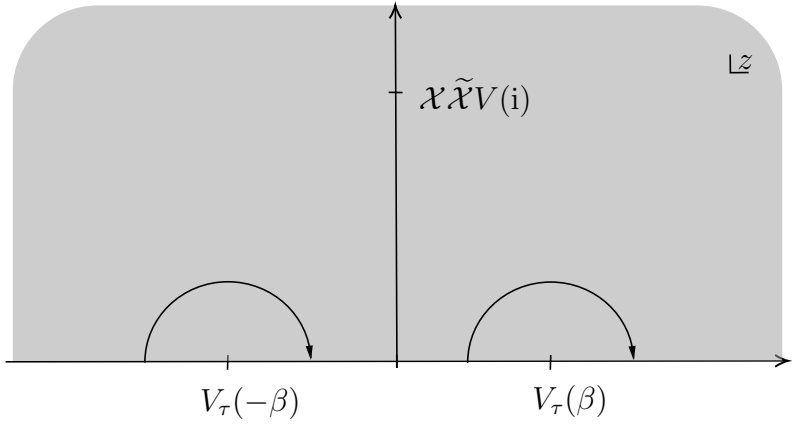
\begin{figure}[t]
    \centering

    \tikzset{every picture/.style={line width=0.6pt}} %set default line width to 0.75pt

    \begin{tikzpicture}[x=0.75pt,y=0.75pt,yscale=-0.6,xscale=0.6]
        %uncomment if require: \path (0,365); %set diagram left start at 0, and has height of 365

        %Straight Lines [id:da7867611659559895]
        \draw (96.59,315.38) -- (749.19,314.6);
        \draw [shift={(751.19,314.6)}, rotate = 179.93] [color={rgb, 255:red, 0; green, 0; blue, 0}][line width=0.75]
            (10.93,-4.9) .. controls (6.95,-2.3) and (3.31,-0.67) .. (0,0)
            .. controls (3.31,0.67) and (6.95,2.3) .. (10.93,4.9);
        \draw [shift={(423.89,314.99)}, rotate = 179.93] [color={rgb, 255:red, 0; green, 0; blue, 0}][line width=0.75]
            (0,5.59) -- (0,-5.59);

        %Straight Lines [id:da641497758944231]
        \draw (282.59,309.76) -- (282.19,322.05);

        %Rounded Same Side Corner Rect [id:dp766733934749634]
        \draw [color={rgb, 255:red, 0; green, 0; blue, 0}, draw opacity=0]
            [fill={rgb, 255:red, 155; green, 155; blue, 155}, fill opacity=0.5]
            (102.56,88.49) .. controls (102.56,49.16) and (134.45,17.27) .. (173.79,17.27)
            -- (674.96,17.27) .. controls (714.3,17.27) and (746.19,49.16) .. (746.19,88.49)
            -- (746.19,314.6) .. controls (746.19,314.6) and (746.19,314.6) .. (746.19,314.6)
            -- (102.56,314.6) .. controls (102.56,314.6) and (102.56,314.6) .. (102.56,314.6)
            -- cycle;

        %Straight Lines [id:da9862913316383304]
        \draw (552.59,309.76) -- (552.19,322.05);

        %Shape: Arc [id:dp015650159553161402]
        \draw [draw opacity=0]
            (213.09,316.07) .. controls (213.09,315.96) and (213.09,315.84) .. (213.09,315.72)
            .. controls (213.09,278.56) and (244.19,248.44) .. (282.55,248.44)
            .. controls (320.28,248.44) and (350.99,277.59) .. (351.98,313.92)
            -- (282.55,315.72) -- cycle;
        \draw
            (213.09,316.07) .. controls (213.09,315.96) and (213.09,315.84) .. (213.09,315.72)
            .. controls (213.09,278.56) and (244.19,248.44) .. (282.55,248.44)
            .. controls (319.72,248.44) and (350.06,276.73) .. (351.91,312.29);
        \draw [shift={(351.98,313.92)}, rotate = 263.26]
            [fill={rgb, 255:red, 0; green, 0; blue, 0}][line width=0.08][draw opacity=0]
            (12,-3) -- (0,0) -- (12,3) -- cycle;

        %Straight Lines [id:da013390080979975627]
        \draw (422.6,19.2) -- (423.89,314.99);
        \draw [shift={(422.59,17.2)}, rotate = 89.75] [color={rgb, 255:red, 0; green, 0; blue, 0}][line width=0.75]
            (10.93,-4.9) .. controls (6.95,-2.3) and (3.31,-0.67) .. (0,0)
            .. controls (3.31,0.67) and (6.95,2.3) .. (10.93,4.9);

        %Straight Lines [id:da15192957358935544]
        \draw (417,90) -- (428.59,90.2);

        %Shape: Right Angle [id:dp3864255893098427]
        \draw (718.09,73.44) -- (702.81,73.59) -- (702.67,58.34);

        %Shape: Arc [id:dp4159706654935944]
        \draw [draw opacity=0]
            (483.09,316.32) .. controls (483.09,316.21) and (483.09,316.09) .. (483.09,315.97)
            .. controls (483.09,278.81) and (514.19,248.69) .. (552.55,248.69)
            .. controls (590.28,248.69) and (620.99,277.84) .. (621.98,314.17)
            -- (552.55,315.97) -- cycle;
        \draw
            (483.09,316.32) .. controls (483.09,316.21) and (483.09,316.09) .. (483.09,315.97)
            .. controls (483.09,278.81) and (514.19,248.69) .. (552.55,248.69)
            .. controls (589.72,248.69) and (620.06,276.98) .. (621.91,312.54);
        \draw [shift={(621.98,314.17)}, rotate = 263.26]
            [fill={rgb, 255:red, 0; green, 0; blue, 0}][line width=0.08][draw opacity=0]
            (12,-3) -- (0,0) -- (12,3) -- cycle;

        % Text Node
        \draw (706.43,56.6) node [anchor=north west][inner sep=0.75pt] [font=\large] {$z$};

        % Text Node
        \draw (535,332.4) node [anchor=north west][inner sep=0.75pt] [font=\large] {$V_{\tau }( \beta )$};

        % Text Node
        \draw (440,75.4) node [anchor=north west][inner sep=0.75pt] [font=\large] {$\mathcal{X}\tilde{\mathcal{X}} V( \ii)$};

        % Text Node
        \draw (257,333.84) node [anchor=north west][inner sep=0.75pt] [font=\large] {$V_{\tau }( -\beta )$};
    \end{tikzpicture}

    \caption{The quantity $\mathcal{B}_{\beta}$ is the sum of two contour integrals, one encircling each of the open-string insertions at $-\beta$ and $\beta$.}
    \label{fig: B_beta_contour}
\end{figure}

The exchange contribution is given by the COO vertex with two $V_\tau$ insertions joined by the $\tau$ propagator:
\begin{align} \label{etaupcont}
    \mathcal{N}_{\text{COO}}\int_{\frac{1}{2\tilde{\lambda}}}^1d\beta\left\langle \mathcal{X}\tilde{\mathcal{X}}V(\ii)
    %F_1\circ\partial \xi c\partial ce^{-2\phi}(0)
    F_1 \circ V_\tau(0)  \, 
    (-\mathcal{B}_{\beta}) \, 
    %F_2\circ\partial \xi c\partial ce^{-2\phi}(0) 
    F_2 \circ V_\tau(0) 
    \right\rangle 
    \cdot \tau_{\text{prop}}\,,
\end{align}
where (\ref{eapp1}) and the discussion about overall signs in section \ref{sec-sft-amplitudes} tells us that
\be \label{eNCOO}
\NNN_{\text{COO}} = - g_s \, \eta_c\, .
\ee
The signs in the previous two expressions are determined according to the discussion below \eqref{ea4new}, with the following additional detail. 
That sign prescription was formulated for the case in which only one open-string insertion depends on the modulus under consideration. 
We now show that the present case, where both the open-string insertions move as $\beta$ varies, is related to that simpler case by a positive Jacobian.
Consider the $\mathrm{SL}(2,\mathbb{R})$ transformation
$z\rightarrow (z+\beta)/(1-\beta z)$, which sends $-\beta$ to $0$ while
keeping $\ii$ fixed. Under this transformation, the point $\beta$ is
mapped to $\beta'=2\beta/(1-\beta^2)$. 
This is a monotonically increasing function for $\beta\in[0,1]$, so as the modulus increases, the open-string insertion moves to the right along the boundary, with the upper half-plane kept on its left. 
Since $d\beta'/d\beta=2(1+\beta^2)/(1-\beta^2)^2$ is positive, $d\beta$ and $d\beta'$ define the same orientation.

The transition functions  in the range $\beta \in [(2\tilde{\lambda})^{-1},1]$ can be obtained by inverting
\eqref{eq: local_coord_COO} and take the form \cite[Eq.~(B.9)]{Sen:2020eck}  
\begin{align}\label{eq: beta_trans-func_2}
F_1(w_1, \beta) & =f_1(\beta) + g_1(\beta) w_1+ \mathcal{O}(w_1^2)
= -\beta + \frac{4 \tilde{\lambda}(1 + \beta^2)}{\alpha(4 \tilde{\lambda}^2 + 1)} w_1+ \mathcal{O}(w_1^2)\, ,\nonumber \\
F_2(w_2, \beta) &= f_2(\beta) + g_2(\beta) w_2+ \mathcal{O}(w_2^2)
= \beta + \frac{4 \tilde{\lambda}(1 + \beta^2)}{\alpha(4 \tilde{\lambda}^2 + 1)} w_2 + \mathcal{O}(w_2^2)\, .
\end{align}

The propagator for $\tau$ can be calculated as follows. 
The BRST current (\ref{eq: J_b_expression}) has a term $-\gamma^2 b/4$, which can be rewritten as $\partial \eta\, \eta \, e^{2\phi} \, b/4$. 
Only this term in $j_B$ will be relevant for us. 
The action of the BRST operator on $V_\tau$ is
\be \label{eq: QB_on_OSG}
  Q_B V_\tau(z_2)=  
  \oint_{z_2}
  \frac{1}{4} \, \partial \eta \, \eta \, e^{2\phi}b(z)\,
  \partial \xi c \partial c e^{-2\phi}(z_2)=
  \frac{1}{2} \, \eta c(z_2)\, .
\ee
Using this we can compute the quadratic term in the action containing $\tau$:
\be
 \frac{1}{2}\tau^2 \langle V_\tau |Q_B|V_\tau\rangle=   \frac{\tau^2}{2}\langle \partial \xi c \partial c e^{-2\phi}(z_1)\eta c(z_2)\rangle =-K\frac{\tau^2}{4}\, .
\ee
This implies that the $\tau$-propagator is
\be \label{eq: tau_prop_result}
\tau_{\text{prop}} =\frac{2}{K} \, .
\ee

Next we compute the correlator appearing in \eqref{etaupcont}
\begin{align} \label{e7.41xx}
    &\left\langle 
    \mathcal{X}\tilde{\mathcal{X}}V(\ii)
    F_1\circ V_\tau(0)\, 
    (-\mathcal{B}_{\beta}) \,  
    F_2\circ V_\tau(0) 
    \right\rangle    
    =
    \left\langle 
    \mathcal{B}_{\beta}  \, 
    \mathcal{X}\tilde{\mathcal{X}}V(\ii) \, 
    F_1\circ V_\tau (0) \, 
    F_2\circ V_\tau (0) 
    \right\rangle\, .
\end{align}
Since $V_\tau$ is a primary of vanishing conformal weight, we simply have
\begin{align}
    F_1 \circ \, V_\tau(0) = V_\tau (-\beta),\qquad 
    F_2 \circ \, V_\tau(0) = V_\tau (\beta)\, .
\end{align}
Further, the only term from $\mathcal{X}\tilde{\mathcal{X}}V$ in (\ref{PCO_action}) that will contribute is $\frac{1}{4}\eta\tilde{\eta} e^{\phi}e^{\tilde{\phi}}e^{\bb \varphi}$.
Using all this information and the definition of
$\BBB_\beta$ given in \eqref{eapp0}, the correlator \eqref{e7.41xx} can be reduced to: 
\begin{align}\label{e7.43}
    \frac{1}{4}\bigg\langle \partial \xi(-\beta)\partial \xi(\beta)\eta \tilde{\eta}(\ii)  e^{\bb \varphi}(\ii) \bigg[-f_1'\partial c(-\beta)  \ c\partial c(\beta) +\frac{g_1'}{g_1}\ c(-\beta)  \ c\partial c(\beta) \nonumber\\
    -f_2'  c\partial c(-\beta)  \ \partial c(\beta) +\frac{g_2'}{g_2}  c\partial c(-\beta)  \ c(\beta) \bigg] e^{\phi}e^{\tilde{\phi}}(\ii)e^{-2\phi}(-\beta)e^{-2\phi}(\beta)\bigg\rangle \,,
\end{align}
where we have used that $\BBB_\beta$ contains clockwise contour integrals around the points $\pm\beta$.
The functions $f_1(\beta)$, $f_2(\beta)$, $g_1(\beta)$, and $g_2(\beta)$ 
are defined via \eqref{eq: beta_trans-func_2}, and so
\begin{align}
    f_1'(\beta) =-1, \quad f_2'(\beta)=1, \quad \frac{g_1'(\beta)}{g_1(\beta)}=\frac{g_2'(\beta)}{g_2(\beta)}
    =\frac{2\beta}{1+\beta^2}\, .
\end{align}
Evaluating the correlators in \eqref{e7.43} and using \eqref{one-pt-vb}, we get
\begin{align}
  -K\, \frac{\ii Q}{4\bb\wt\mu}\frac{(1-\beta^2)^2}{\beta^2(1+\beta^2)}\,.
  \label{eq: g_tau_correl}
\end{align}

Using  \eqref{etaupcont}, \eqref{eNCOO},
\eqref{eq: tau_prop_result}, \eqref{eq: g_tau_correl} and further dividing by disk one-point function (\ref{3.11}), we get the $\tau$ exchange contribution to $g_s g$:
\begin{align}
   g_s g_{\tau}&=-\frac{2\bb}{K_0Q}\cdot {2\over K} \cdot (- g_s \eta_c) \cdot (-K) \frac{\ii Q}{4\bb}\int_{\frac{1}{2\tilde{\lambda}}}^1d\beta \frac{(1-\beta^2)^2}{\beta^2(1+\beta^2)}
   =-g_s  K_0^{-1} 
   \left(\frac{1}{2} -\frac{\tilde{\lambda}}{\pi}\right)+\mathcal{O}(\tilde{\lambda}^{-1}) \, .\label{eq: tau_loop_final}
\end{align}
Dropping terms with $\widetilde{\lambda}$ in the denominator, we record this as
\begin{align}
    \boxed{
    g_\tau = \frac{1}{K_0} \left( \frac{\tilde{\lambda}}{\pi} - \frac{1}{2} \right)\, .
    }
    \label{eq: tau_loop_final_2}
\end{align}

\subsection{Contribution from regions \texorpdfstring{(a)}{} and 
\texorpdfstring{(b)}{}}

The contribution from the Feynman diagrams in  Fig.~\ref{fig:feynman-diagrams_2}(a) and Fig.~\ref{fig:feynman-diagrams_2}(b) contains a CO interaction vertex. 
The relevant open string states that contribute in the limit of large SFT parameters are the tachyon and the OSG state. 
Just as in section \ref{sCOCO}, the tachyon contribution vanishes since it is GSO odd, and the OSG contribution vanishes for our choice of PCO locations on the CO vertex.
Hence the contributions from both of these diagrams vanish.

\subsection{Vertical integration at the boundary 
between regions 
\texorpdfstring{(b)}{} and
\texorpdfstring{(d)}{}
}

We now turn to vertical-integration contributions. 
As noted in section \ref{sec-o-annulus}, no vertical integration is required at the boundary between regions (a) and (b); similarly, as noted in section \ref{sCannulus}, none is required at the boundary between regions (c) and (d). 
Thus there are two remaining possible contributions: we analyze the interface between regions (b) and (d) in this subsection, and the interface between regions (a) and (c) in the next.

In region (d), our choice of PCO insertions was to put the product $\XXX\wt\XXX$ on the closed-string vertex operator $V$, which is located at the point $w_c=2\pi x$ in the $w$-coordinate. 
%introduced in \eqref{annulus_coord}. 

In region (b), the PCO insertions are inherited from the subvertices that are glued together, namely the CO vertex on the UHP and the O vertex on the annulus. 
In the CO vertex on the UHP, we inserted $(\XXX+\wt\XXX)/2$ at the location of the closed-string vertex operator; on the annulus, this location is denoted as $w_c$. 
The second PCO is at the PCO location of the O vertex on the annulus, which was denoted by $\hat{z}_p$ or $w_p$ in section \ref{sec-o-annulus}.

Via the doubling trick, for real $w_c$, $\wt\XXX$ inserted at $w_c$ is equivalent to $\XXX$ inserted at $-w_c$.\footnote{
This is clearest if we transform the annulus geometry to the upper half plane via the map $w \to e^{\ii w}$.} 
So we conclude that  as we move from region (b) to region (d), we have to move a PCO from $w_p$ to $w_c$ or $-w_c$, keeping the other PCO fixed at $-w_c$ or $w_c$, respectively.
From \eqref{eq:x_annulus_b_approx}, we see that on the boundary 
between (b) and (d), the value of  $w_c$ is given by 
%the value of $2\pi x$ , is given by
\be\label{ewcvalue}
w_c = \tilde{\lambda}^{-1} \left( 1 - \alpha^{-2} \right)\, .
\ee
We do not need an explicit expression for $w_p$ since the result will turn out to be independent of it.

The boundary between the regions (b) and (d) occurs at a constant value of $x$. 
Hence by the rule of vertical integration, if we want to move a PCO from $w_p$ to $\pm w_c$, we need to replace the $\int dx \, \BBB_x$ insertion into the correlation function by $-2(\xi(\pm w_c)-\xi(w_p))$
and drop the $\XXX(\pm w_c)$ from the $\XXX(w_c)\XXX(-w_c)V(w_c,\bar w_c)$ factor in the correlation function.
It follows from the definition \eqref{eapp0} that
\be\label{ebfactor}
\BBB_x dx =-2\pi dx \left(\oint_{2\pi x}b(w')d w'+\oint_{2\pi x}\tilde{b}(\bar{w}')d \bar{w}'\right)
+ \cdots\, ,
\ee
where the $-2\pi$ factor can be traced to the fact that the local coordinate at the closed-string puncture is $w-2\pi x$, and the $\cdots$ denote the terms involving $\xi,\wt\xi$. 
Hence the result of vertical integration can be obtained by replacing the factor \eqref{ebfactor} in \eqref{e2.47new} by $-2(\xi(\pm w_c) -\xi(w_p))$, and dropping a PCO and the integration over $x$:
\be \label{e2.47newvertical}
\mathcal{N}_{\text{C}}\int d t \,  \Bigg\langle
    \left(\int_0^{\pi}b(w)d w+\int_0^{\pi} \tilde{b}(\bar{w})d\bar{w}\right)
     \left[ - 2(\xi(\pm w_c) -\xi(w_p))\right]
     \mathcal{X}(\mp w_c) \, V(w_c, \bar{w_c}) \Bigg\rangle_{\rm A} \, .
  \ee
  
Let us first consider the bottom sign. 
We use the mode expansions \eqref{eq: bc_mode_exp} to replace the contour integrals over $b$, $\widetilde{b}$ by $-2\pi b_0$. 
We also divide by the disk one-point function (\ref{3.11}) to arrive at the contribution to  $g_s g$:
\begin{align} \label{e7.51}
    g_sg^{(\text{b})-(\text{d})}_{\text{VI}, \mathcal{X}}&=  
    -\frac{2\bb\wt\mu}{K_0 Q}\cdot
    \mathcal{N}_{\text{C}}\int_0^{\frac{1}{2\pi}\log(\alpha^2-\frac{1}{2})} \left\langle (-2\pi b_0) \left(-2(\xi(-w_c)-\xi(w_p))\right)\mathcal{X}V(w_c)\right\rangle_{\rm A} d t \\
    &=- \frac{4 \pi \bb \widetilde{\mu}  \mathcal{N}_{\text{C}}}{K_0 Q}\int_0^{\frac{1}{2\pi}\log(\alpha^2-\frac{1}{2})} \left\langle b_0 \left(\xi(w_p)-\xi(-w_c)\right)\eta \tilde{c}e^{\phi}e^{-\tilde{\phi}}e^{\bb\varphi}(w_c)\right\rangle_{\rm A} d t  \\
    &=- \ii \frac{4\pi \bb \widetilde{\mu}  \mathcal{N}_{\text{C}}
    }{K_0 Q}\int_0^{\frac{1}{2\pi}\log(\alpha^2-\frac{1}{2})} \left\langle b_0 c_0 \left(\xi(w_p)-\xi(-w_c)\right)\eta e^{\phi}e^{-\tilde{\phi}}e^{\bb\varphi}(w_c)\right\rangle_{\rm A} d t
    \, . \label{eannend}
\end{align}
In the second line, we used the fact that  $-\frac{1}{2}\eta \tilde{c}e^{\phi}e^{-\tilde{\phi}}e^{b\varphi}$  is the only term from  $\mathcal{X}V$ in \eqref{e6.19a} that contributes, as the rest of the terms violate $\phi$-momentum conservation. 
In the third line, we used the mode expansion for $\widetilde{c}$ given in \eqref{eq: bc_mode_exp}, and the fact that we need a $c_0$ in the correlator to get a non-zero result.

Next we need to find the reflection rules on the annulus that will relate the anti-chiral fields to the chiral fields on a torus. 
Defining $z=e^{\ii w}$, the region $0<\text{Re}(w)<\pi$ gets mapped to the $\text{Im}(z)>0$ region, and we can use the known reflection rules on the UHP to find the reflection rules on the annulus. 
This 
gives\footnote{In these equations the arguments of the operators do not reflect the  chirality of the
operator. They simply describe where the operator is inserted.}
\begin{align}
     e^{\wt\phi}(w,\bar w) &= \left(\frac{\partial \bar z}{\partial \bar w}\right)^{-\frac{3}{2}} 
     e^{\wt\phi}(z,\bar z)
    =\left(\frac{\partial \bar z}{\partial \bar w}\right)^{-\frac{3}{2}} e^{\phi}(\bar z, z)
    = \left(\frac{\partial \bar z}{\partial \bar w}\right)^{-\frac{3}{2}} 
    \left(\frac{\partial z}{\partial w}\bigg|_{z\to \bar z}\right)^{\frac{3}{2}}  e^{\phi}(-\bar w,-w) \nonumber \\
    & = \ii^3 \, 
     e^{\phi}(-\bar w,-w)
    \nonumber \\
       e^{-\tilde{\phi}}(w,\bar w)&=\left(\frac{\partial \bar{z}}{\partial \bar{w}}\right)^{
       \frac{1}{2}}e^{-\tilde{\phi}}(z,\bar z)=
       \left(\frac{\partial \bar{z}}{\partial \bar{w}}\right)^{
       \frac{1}{2}}e^{-{\phi}}(\bar z, z) = \left(\frac{\partial \bar{z}}{\partial \bar{w}}\right)^{
       \frac{1}{2}}
        \left(\frac{\partial {z}}{\partial {w}}\bigg|_{z\to \bar z}\right)^{-
       \frac{1}{2}}
       e^{-{\phi}}(-\bar w, -w) \nonumber \\ &= -\ii\,  e^{-{\phi}}(-\bar w,-w) \nonumber \\
       \wt\eta(w,\bar w) &= \left(\frac{\partial \bar{z}}{\partial \bar{w}}\right) \wt\eta(z,\bar z)=
       \left(\frac{\partial \bar{z}}{\partial \bar{w}}\right)\eta (\bar z, z) = 
       \left(\frac{\partial \bar{z}}{\partial \bar{w}}\right) 
       \left(\frac{\partial {z}}{\partial {w}}\bigg|_{z\to \bar z}\right)^{-1}
       \eta(-\bar w, -w) \nonumber \\ &= -    \eta(-\bar w,-w) \, .\label{e7.53}
    \end{align}
For $e^{\bb\varphi}(w, \bar{w})$ we follow a different strategy. We write
    \be
    e^{\bb\varphi}(w, \bar{w})= \left(\frac{\partial z}{\partial w}\right)^{\frac{1}{2}}\left(\frac{\partial \bar{z}}{\partial \bar{w}}\right)^{\frac{1}{2}}e^{\bb\varphi}(z, \bar{z})=e^{\ii (w-\bar w)/2}\, e^{\bb\varphi}(z, \bar{z})
    \, .
    \ee
We need to evaluate it at $w=w_c$, which from \eqref{ewcvalue}, can be seen to be small for large
$\wt\lambda$. Hence its image $z_c=e^{\ii w_c}$ is close to the real axis. Hence we can use the bulk
boundary OPE given in \eqref{e2.19OPE}, \eqref{eq: U_alpha_expression} 
to replace $e^{\bb\varphi}(w_c, \bar{w_c})$ by
\be \label{e754xx}
{Q\over \ii\bb\wt\mu} \, \frac{1}{|e^{\ii w_c}-e^{-\ii w_c}|} \left(1 +\mathcal{O}(w_c^2)\right)
= {Q\over 2\ii\bb\wt\mu} \, \wt\lambda \, \left(1 - \alpha^{-2}
\right)^{-1}  + \mathcal{O}\left(\wt\lambda^{-1}\right)\, .
\ee
Using these results, and the fact that $w_c$ is real,
we can express \eqref{eannend} as the integral of a torus correlation function
{\small
\ben \label{etorusbeg}
g_sg^{(\text{b})-(\text{d})}_{\text{VI}, \mathcal{X}} =
\frac{2\ii \pi \wt\lambda  \mathcal{N}_{\text{C}}}{K_0 } 
(1-\alpha^{-2})^{-1}
\int_0^{\frac{1}{2\pi}\log(\alpha^2-\frac{1}{2})} dt
\left\langle b_0 c_0 \left(\xi(w_p)-\xi(-w_c)\right)
\eta(w_c) e^{\phi}(w_c) e^{-{\phi}}(-w_c) \right\rangle_{T^2} 
\een
}
\par \noindent
where $\langle ~\rangle_{T^2}$ denotes a torus correlation function, normalized the same way as the annulus
\be\label{e756aa}
\langle b_0 c_0 \rangle_{T^2} = Z(v)\, .
\ee
In \eqref{etorusbeg}, we have dropped terms from \eqref{e754xx} that carry inverse powers of $\wt\lambda$ since the integral over $t$ cannot produce terms carrying positive powers of $\wt\lambda$.

Now let us analyze the contribution associated with the top sign in \eqref{e2.47newvertical}.  
The analog of \eqref{e7.51} for this contribution is
\be\label{e7.51new}
    g_sg^{(\text{b})-(\text{d})}_{\text{VI}, \wt{\mathcal{X}}}    
    =- \frac{4\bb\pi  \mathcal{N}_{\text{C}}\wt\mu}{K_0 Q}
    \int_0^{\frac{1}{2\pi}\log(\alpha^2-\frac{1}{2})} dt \left\langle b_0 \left(\xi(w_p)-\xi(w_c)\right)\, \wt\eta \,
     {c}\, e^{\tilde\phi}\, e^{-{\phi}}\, e^{\bb\varphi}(w_c)\right\rangle_{\rm A}\, .
\ee
Using \eqref{e7.53} and manipulations similar to the one performed earlier, we can bring this to the form
{\small
\ben \label{etorusend}
g_sg^{(\text{b})-(\text{d})}_{\text{VI}, \wt\XXX} 
=
\frac{2\ii \pi \wt\lambda\mathcal{N}_{\text{C}}}{K_0 } 
(1-\alpha^{-2})^{-1}
\int_0^{\frac{1}{2\pi}\log(\alpha^2-\frac{1}{2})} dt 
\left\langle b_0 c_0 \left(\xi(w_p)-\xi(w_c)\right)
\eta(-w_c) e^{\phi}(-w_c) e^{-{\phi}}(w_c) \right\rangle_{T^2} \, .
\een
}

Now since $w_c\sim \wt\lambda^{-1}$ and hence small, we can examine the correlators in \eqref{etorusbeg} and \eqref{etorusend} for small $w_c$. 
Using the $\xi$-$\eta$ OPE (\ref{xi-eta-ope}), we can see that the leading term in the correlator is $\langle b_0 c_0 \rangle_{T^2} = Z(v)$ for small $w_c$.
Furthermore, since \eqref{etorusbeg} and \eqref{etorusend} are related by $w_c\to -w_c$, the average of the two expressions is an even function of $w_c$, and hence the first subleading term in the expansion of the correlator in powers of $w_c$ is of order $w_c^2\sim \wt\lambda^{-2}$.
Even after being multiplied by the overall factor of $\wt\lambda$ outside the integral, this is of order $\wt\lambda^{-1}$ and hence vanishes in the large $\wt\lambda$ limit. 
Therefore we can express the average of \eqref{etorusbeg} and \eqref{etorusend} as
\be\label{e7.60a}
g_sg^{(\text{b})-(\text{d})}_{\text{VI}}=
\frac{1}{2} \left( 
g_sg^{(\text{b})-(\text{d})}_{\text{VI},\mathcal{X}}+g_sg^{(\text{b})-(\text{d})}_{\text{VI}, \tilde{\mathcal{X}}}
\right)
=\frac{2\ii \pi \wt\lambda \mathcal{N}_{\text{C}}}{K_0 } 
(1-\alpha^{-2})^{-1}
\int_0^{\frac{1}{2\pi}\log(\alpha^2-\frac{1}{2})} Z(v) dt\, .
\ee
Changing the integration variable to $v=e^{-2\pi t}$,  ignoring terms that have powers of $\alpha$ or $\wt\lambda$ in the denominator,  and using $\mathcal{N}_\text{C} = g_s \eta_c$ with $\eta_c = \frac{\ii}{2\pi}$, we get
\begin{align}
    \boxed{
        g^{(\text{b})-(\text{d})}_{\text{VI}} = 
        -{\tilde\lambda\over 2\pi}\, \frac{1}{K_0}\, 
        \int_{\left(\alpha^2-\frac{1}{2}\right)^{-1}}^1 \frac{d v}{v} Z(v)\, .
    }
    \label{e7.61a}
\end{align}
Note that this exactly cancels the contribution \eqref{e8.18}.

\subsection{Vertical integration at the boundary 
between regions 
\texorpdfstring{(a)}{} and
\texorpdfstring{(c)}{}
}

Next, we consider the vertical integration contribution $g^{(\text{a})-(\text{c})}_{\text{VI}}$ from the interface of regions (a) and region (c)
in Fig.~\ref{fig:moduli-feynman}.  
In region (c), we have placed the PCO combination $\mathcal{X}\widetilde{\mathcal{X}}$ on C, while in region (a), there is a
$(\XXX+\wt\XXX)/2$ insertion on C and the location of the second PCO is
induced from the OOO interaction vertex.
Going from region (a) to (c), we need to take an average of two vertical integrations
\begin{enumerate}
    \item When we have $\mathcal{X}V$ inserted in the CO vertex, we move the $\mathcal{X}$ from the OOO factor to a $\tilde{\mathcal{X}}$ on top of the C insertion in the CO factor. 
    In the $w$ coordinate on the annulus, this means that 
    we move the $\mathcal{X}$ from $w_p$ to $-w_c$, 
    \item Similarly, when we have $\tilde{\mathcal{X}}V$ inserted in the CO vertex, we will move the $\mathcal{X}$ from $w_p$ to $w_c$.
\end{enumerate}

We considered the exact same movement of PCOs at the interface  of region (b) and region (d). 
Hence, we could directly use the result \eqref{e7.60a} and then carry out the $v$ integral using the SFT replacement rules \eqref{ereprule} and \eqref{ereprule1}. 
There is however one additional point that needs to be addressed. 

Let us examine the relation between the parameters $x,v$ and the Schwinger parameters $q_1,q_2$ associated with the two propagators in  Fig.~\ref{fig:feynman-diagrams_2}. 
This was given in \eqref{eq: reg(a)_xv}.
The boundary between the regions (a) and (c) is at $q_1=1$. 
Using (\ref{eq: reg(a)_xv}),  this translates to
\be\label{eboundary}
2\pi x = \wt\lambda^{-1} (1-v + \mathcal{O}(\alpha^{-4}))\, .
\ee
By contrast, in the analysis leading to \eqref{e7.60a}, the boundary was at a constant value of $x$.
Due to this difference the result of vertical integration will not be the same as \eqref{e7.60a}.
However, we note that $v$ is of order $\alpha^{-2}$ in the region of interest, and since $\alpha$ is large, the correction is small. 
For this reason, we shall first ignore this difference and later estimate the correction.
So we start with
\begin{align}
    g_sg^{(\text{a})-(\text{c})}_{\text{VI}}=\ii \tilde{\lambda} \mathcal{N}_{\text{C}}K_0^{-1}\int_0^{\left(\alpha^2-\frac{1}{2}\right)^{-1}} \frac{d v}{v} Z(v)+\mathcal{O}(\tilde{\lambda}^{-1})\, .
\end{align}
Using $Z(v)\approx v^{-\frac{3}{2}}-2 v^{-1}+\mathcal{O}(v^{-\frac{1}{2}})$ and \eqref{eq: reg(a)_xv}, we get
\begin{align}\label{eq: g^ac_VI_fin}
    g_sg^{(\text{a})-(\text{c})}_{\text{VI}}&=\ii \tilde{\lambda} \mathcal{N}_{\text{C}}K_0^{-1}
    \int_0^{1} d q_2 \left(\alpha q_2^{-\frac{3}{2}}-2 q_2^{-1}+\mathcal{O}(\alpha^{-1}q_2^{-\frac{1}{2}})\right)+\mathcal{O}(\tilde{\lambda}^{-1}) \nonumber\\
    &= g_s\, K_0^{-1} {\alpha\tilde\lambda\over \pi} +\mathcal{O}(\tilde{\lambda}\alpha^{-1})\, ,
\end{align}
where in the last step we have used the replacement rules \eqref{ereprule} and \eqref{ereprule1}.

Let us now address the effect of the correction terms proportional to $v$ in \eqref{eboundary}. 
Since $v$ is of order $\alpha^{-2}$ in the region of interest, this will generate fractional corrections of order $\alpha^{-2}$. 
Since the leading term in \eqref{eq: g^ac_VI_fin} is of order $\wt\lambda\alpha$, this means that the corrections will be of order $\wt\lambda/\alpha$. 
Since in our analysis we are systematically dropping all terms with $\alpha$ or $\wt\lambda$ in the denominator, we can ignore the correction terms and take \eqref{eq: g^ac_VI_fin} as the answer. Thus,
\begin{align}
    \boxed{
        g^{(\text{a})-(\text{c})}_{\text{VI}} = 
        \frac{1}{K_0}\, {\alpha\tilde\lambda\over \pi} 
        \, .
    }
    \label{eq: g^ac_VI_fin_2}
\end{align}

\subsection{Contribution due to gauge parameter redefinition}\label{subsec: gauge_param_redef}

In this section, we determine the relation between the SFT gauge transformation parameter $\theta$ and the rigid $U(1)$ transformation parameter $\tilde{\theta}$. 
This rigid symmetry is defined as the $U(1)$ symmetry that 
multiplies by $e^{i\wt\theta}$ the open string fields with one endpoint attached to the ZZ-instanton.
The gauge parameter $\theta$ arises from the Siegel-gauge ghost zero mode associated with the state $\beta_{-\frac{1}{2}}\,c_1\,|-1\rangle$, equivalently represented by the vertex operator $\partial \xi \, c \, e^{-2\phi}$.
Once the relation between $\theta$ and $\wt\theta$ is found, we can compute $\int d\theta$ using the fact that the rigid $U(1)$ symmetry with parameter $\tilde{\theta}$ has period $2\pi$. 
The SFT path integral is supposed to be divided by $\int d\theta$ in order to take into account the fact that we do not
fix the gauge symmetry generated by $\theta$ \cite{Sen:2021qdk}.

Following \cite{Sen:2020eck, Sen:2021qdk}, in order to find the relation between $\theta$ and $\wt\theta$, we shall introduce a spectator instanton so that open strings stretching between the spectator instanton and the original instanton are charged under the rigid U(1) symmetry under consideration.
Let $\Psi$ be the Faddeev-Popov `c-ghost' field corresponding to the gauge parameter $\theta$. 
It is a Grassmann odd field multiplying the same ghost number zero vertex operator $\partial \xi \, c\, e^{-2\phi}$ as the gauge transformation parameter. 
In the presence of the spectator instanton, $\Psi$ carries the Chan-Paton factor $\begin{pmatrix}1&0\\ 0 &0\end{pmatrix}$ since it corresponds to
an open string state with both ends on the original instanton.
We shall analyze the $U(1)$ transformation law of the field multiplying the same vertex operator $\partial \xi \,c\, e^{-2\phi}$, but multiplied by a Chan-Paton factor $\begin{pmatrix}0 &1\\ 0&0 \end{pmatrix}$. 
We denote this field by $\Xi$. This is just a convenient choice; one could instead analyze the transformation law of any other field and obtain the same result.
Since the vertex operator is Grassmann even, $\Xi$ must be a
Grassmann-odd field. 
Let $\Xi^*$ be the Grassmann-even antifield of $\Xi$, which multiplies the vertex operator $\eta c\partial c$ and carries Chan-Paton factor $\begin{pmatrix}0 &0\\ 1&0 \end{pmatrix}$. 
Let us define
 \begin{align}
    V_\Psi :=\partial \xi c e^{-2\phi}, \quad 
    V_\Xi :=\partial \xi c e^{-2\phi} , \quad 
    V_{\Xi^*} :=\eta c\partial c \, .
    \label{vops_cooo}
\end{align}
In a disk amplitude, these vertex operators must be inserted in the
cyclic order $V_\Xi V_{\Xi^*}V_\Psi$; otherwise, the amplitude vanishes
because of the trace over Chan-Paton factors.

The SFT action contains the following cubic coupling proportional to $\Xi\Xi^*\Psi$:
\begin{align}
S &\supset \NNN_{\rm OOO}
\left\langle
\Xi\, f_1\circ \partial \xi c e^{-2\phi}(0)\;
\Xi^*\, f_2\circ \eta c\partial c(0)\;
\Psi\, f_3\circ \partial \xi c e^{-2\phi}(0)\;
\XXX(z_p)
\right\rangle_\text{UHP} \nonumber  \\
&= - \NNN_{\rm OOO}\, \Xi \Xi^* \Psi \, 
\left\langle
\partial \xi c e^{-2\phi}(0)\;
\eta c\partial c(1)\;
\partial \xi c e^{-2\phi}(\infty)\;
\XXX(z_p)
\right\rangle_\text{UHP} \,,\label{exipsicoupling2}
\end{align}
where $f_1,f_2,f_3$ are obtained by inverting
\eqref{O-O-O coord}, and the PCO is placed at
$z_p=e^{i\pi/3}$, as in (\ref{def-zp}). 
The normalization is
\be \label{eNOOO}
\NNN_{\rm OOO} = g_s^{1/2} \eta_c^{3/4}\,,
\ee
as follows from \eqref{eapp1}.
In the second line of (\ref{exipsicoupling2}), we have moved the fields $\Xi$, $\Xi^*$ and $\Psi$ outside the worldsheet correlation function, keeping track of the signs produced in commuting the Grassmann-odd fields $\Xi$ and $\Psi$ past the Grassmann-odd worldsheet operators.
Note also that there are no extra factors due to the conformal transformations $f_a$ as all the vertex operators are dimension zero primaries.
Evaluation of the correlation function in (\ref{exipsicoupling2}) gives
\be\label{eleadingOOO}
S \supset \frac{1}{2}\, K\, \NNN_{\rm OOO}\, \Xi \Xi^* \Psi \, .
\ee

Let us denote by $\Phi_C$ the closed-string field that multiplies the
vertex operator
$V=c\tilde c e^{-\phi} e^{-\wt\phi} e^{\bb\varphi}$, whose annulus
one-point function we are studying.
Suppose that after integrating out the massive open-string modes and also the OSG mode, the SFT effective action has a term of the form
\be\label{eclosedOOO}
S \supset \mathcal{A}_{\rm COOO} \, \Phi_C \, \, \Xi \Xi^* \Psi \, .
\ee
Then the sum of \eqref{eleadingOOO} and \eqref{eclosedOOO} gives
\be
\frac{1}{2}\, K\, \NNN_{\rm OOO}\, \Xi \Xi^* \Psi \left(1 + {2\, \mathcal{A}_{\rm COOO}\over K\, \NNN_{\rm OOO}} \Phi_C\right)\, .
\ee
It follows from the rules of SFT that the gauge transformation law of
$\Xi$ to this order takes the form
\be\label{ecomp1}
\delta \Xi = \theta \cdot  \frac{1}{2}\, K\, \NNN_{\rm OOO}\, \left(1 + {2\, \mathcal{A}_{\rm COOO}\over K\, \NNN_{\rm OOO}} \Phi_C\right) \, \Xi\, .
\ee
On the other hand, under the infinitesimal rigid $U(1)$ gauge transformation generated by $\wt\theta$, the gauge transformation
law is supposed to take the form
\be\label{ecomp2}
\delta\Xi= \ii\, \wt\theta\, \Xi\, .
\ee
Comparing \eqref{ecomp1} and \eqref{ecomp2} we see that to this order
\be
\ii\, \wt \theta =  \frac{1}{2}\, K\, \NNN_{\rm OOO}\, \left(1 + {2\, \mathcal{A}_{\rm COOO}\over K\, \NNN_{\rm OOO}} \, \Phi_C\right) \theta\, .
\ee
Since the range of $\wt\theta$ is $2\pi$
and we are supposed to divide the path integral by $\int d\theta$, we effectively get a factor of $\left(1 + {2\, \mathcal{A}_{\rm COOO}
\over K\, \NNN_{\rm OOO}} \Phi_C\right)$
in the numerator of the path integral. 
Writing this as $\exp\left( {2\, \mathcal{A}_{\rm COOO} \over K\, \NNN_{\rm OOO}} \Phi_C\right)$, we see that we effectively have an extra term in the action given by
\be
{2\, \mathcal{A}_{\rm COOO}\over K\, \NNN_{\rm OOO}} \Phi_C\, .
\ee
This gives an additional contribution to the annulus one-point function of $\Phi_C$. 
Dividing the disk one-point function $-K_0Q/2\bb\wt\mu$ (see (\ref{3.11})), we get an additional contribution to $g_s g$: 
\be \label{e7.77}
g_s \, g_{\rm gauge} = -{2\bb\wt\mu\over K_0 Q} \cdot {2\, \mathcal{A}_{\rm COOO}\over K\, \NNN_{\rm OOO}}\, .
\ee
Thus our goal is to compute $\mathcal{A}_{\rm COOO}$, the amplitude with one external closed string and three external open strings. 
It receives contributions from the four Feynman diagrams shown in Fig.~\ref{figcooo}, as well as possible vertical-integration terms.

\subsubsection{Contribution from Fig.~\ref{figcooo}(d)} \label{esbulkcooo}

We first compute the contribution  from Fig.~\ref{figcooo}(d), which forms the bulk of the moduli space.

The COOO interaction vertex has two moduli that we shall call $\beta_1$ and $\beta_2$, as in \cite{Sen:2020eck,Eniceicu:2022xvk}.
We insert the closed-string vertex operator $V$ at $z=\ii$ on the UHP, and the open-string vertex operators $V_\Xi$, $V_{\Xi^*}$, and $V_\Psi$ at $f_1(\vec\beta)$, $f_2(\vec\beta)$, and $f_3(\vec\beta)$, respectively, with the cyclic ordering held fixed.
%The $z_a$'s are functions of $\beta_1,\beta_2$.
We take the  local coordinates around the punctures  to be $w_1, w_2$ and $w_3$ with transition functions given in \eqref{elocalCOOO}, which we repeat here for convenience
\begin{equation} \label{elocalCOOOrep}
    z=F_a(w_a, \vec{\beta})=f_a(\vec{\beta})+g_a(\vec{\beta}) \, w_a+ \frac{1}{2}h_a(\vec{\beta})\, w_a^2 +\mathcal{O}(w_a^3), \qquad a=1,2,3\, .
\end{equation}
According to (\ref{eq: PCO_number}), we also need to insert three PCOs.
We choose to put one $\mathcal{X}$ on $V_\Psi$, another on $V_\Xi$, and the combination $(\XXX+\wt\XXX)/2$ on $V$.\footnote{
One might worry that the interaction vertex in not symmetric under the permutations of the external open strings. 
However we note that since $\Xi$ and $\Xi^*$ have been introduced as spectator fields and carry different Chan-Paton factors compared to the dynamical fields, we do not need to symmetrize the interaction term with respect to all the open string fields. 
In particular the PCO locations (and possibly also choice of local coordinates at the punctures) could depend on the Chan-Paton factors carried by the external states. 
Such an action will still have the full gauge invariance generated by the SFT parameter $\theta$ and can be used to determine the relation between $\theta$ and $\wt\theta$ to all orders.
\label{foot1}
}
A short calculation shows that
\begin{align}
    \mathcal{X}V_\Psi=-\frac{\mathbb{I}}{2}, \quad \mathcal{X}V_\Xi=-\frac{\mathbb{I}}{2}, 
\end{align}
with $\mathbb{I}$ being the identity operator.
The expressions for $\XXX V$ and $\wt\XXX V$ are given in \eqref{e6.19a}. 

Hence, the worldsheet expression for the term in the SFT action proportional to $\Phi_C \, \Xi\Xi^*\Psi$ can be written as 
\begin{align} 
S &\supset - \mathcal{N}_{\text{\rm COOO}} \, \Phi_C \, \Xi\Xi^*\Psi\,
\int d \beta_1 \wedge d \beta_2 \,
\Bigg\langle 
 \, 
 \frac{-\mathbb{I}}{2}(f_1) \, 
 (-\mathcal{B}_{\beta_1}) \,
(\eta c \partial c)(f_2) \, 
(-\mathcal{B}_{\beta_2}) \,
\frac{\mathbb{-I}}{2}(f_3) \,
\, \nonumber \\
& \frac{1}{2} \,
\left(-\frac{1}{2} \,\eta \,  \widetilde{c}\, e^{\phi} e^{-\widetilde{\phi}}\,  e^{\bb \varphi} + c\widetilde{c}\,  e^{-\widetilde{\phi}}\,  (-\ii \bb \psi) \,  e^{\bb \varphi}
-\frac{1}{2} \,  \widetilde\eta \,  {c}\, e^{\wt\phi} e^{-{\phi}}\,  e^{\bb \varphi} - c\widetilde{c}\,  e^{-{\phi}}\,  (-\ii \bb \widetilde\psi) \,  e^{\bb \varphi}
\right)(\ii)
\Bigg\rangle\, ,
\label{ghost contribution bulk}
\end{align}
where the minus sign is due to the Grassmann-odd nature of $\Xi$ and $\Psi$, just as in (\ref{exipsicoupling2}).
From \eqref{eapp1} and the sign prescription below \eqref{ea4new}, we have the normalization
\be \label{eNCOOO}
\NNN_{\rm COOO} = - g_s^{3/2} \eta_c^{7/4}\, .
\ee
See section \ref{s796} for more discussion about the overall sign of this amplitude.
Since the PCOs are inserted at the locations of the vertex operators, their locations
in the local coordinate system are independent of $\beta_1$, $\beta_2$, and hence terms
proportional to $\p\xi$ are absent in ${\mathcal B}_{\beta_1}$, ${\mathcal B}_{\beta_2}$.

The correlator in \eqref{ghost contribution bulk}, and hence the contribution from Fig.~\ref{figcooo}(d), vanishes. 
This is because in the small Hilbert space a nonzero correlator requires equal numbers of $\eta$ and $\xi$ insertions; see \eqref{exietaphi} and the discussion below it. None of the terms in \eqref{ghost contribution bulk} satisfies this condition.

\subsubsection{Contributions from Fig.~\ref{figcooo}(a) and \ref{figcooo}(c)}

Both these diagrams involve a CO interaction vertex, with O being either a tachyon or the OSG field. 
We argued in section \ref{sCOCO} that, with our choice of PCO locations, these CO interaction vertices vanish.
Hence there are no contributions from Fig.~\ref{figcooo}(a) and \ref{figcooo}(c).

\subsubsection{Contribution from Fig.~\ref{figcooo}(b)}
\label{sec-4b}

Fig.~\ref{figcooo}(b) consists of a COO interaction vertex connected to an OOO interaction vertex by an open-string propagator.
Since the closed-string insertion $V$ and the three external open-string
states in \eqref{vops_cooo} are GSO even, while the tachyon is GSO odd,
both the CO and OOO interaction vertices vanish if the open-string
propagator carries a tachyon state. Thus only the OSG state can propagate.
The Siegel gauge states like $\p\xi c e^{-2\phi}$ cannot propagate since they are associated with improper gauge fixing and have been removed from the open string spectrum (see discussion above (\ref{ereprule1})). 

The OSG state has ghost number one. For the OOO vertex to be
nonvanishing, the total ghost number must be three; hence the two
external open-string states on the OOO vertex must have ghost numbers
summing to two. This is impossible for the external states in
\eqref{vops_cooo}, whose ghost numbers are $0$, $0$, and $3$.

Thus the contribution from Fig.~\ref{figcooo}(b) also vanishes.

\subsubsection{Vertical integration between Fig.~\ref{figcooo}(d) and \ref{figcooo}(c)}

In Fig.~\ref{figcooo}(d), the PCOs are chosen to be $\XXX$ on $V_\Psi$, $\XXX$ on $V_\Xi$, and $(\XXX+\wt\XXX)/2$ on $V$.
In Fig.~\ref{figcooo}(c), the insertion $(\XXX+\wt\XXX)/2$ on $V$ is inherited from the CO interaction vertex, while the other two PCOs are inherited from the OOOO interaction vertex and are placed away from
the open-string punctures, see (\ref{eperminv}). 
Thus the PCO locations mismatch at the boundary between the two regions, obtained by setting the Schwinger parameter $q_2$ of Fig.~\ref{figcooo}(c) to $1$, and vertical integration is required. 
Note that the closed-string insertion carries the PCO factor
$(\XXX+\wt\XXX)/2$ throughout this process.

We approach the $q_2=1$ boundary from the side of Fig.~\ref{figcooo}(d). 
For large $\lambda$, the configuration at $q_2=1$ is near a degeneration. 
We can therefore insert a complete set of open-string states with $L_0\leq 0$ along the long strip, which becomes the open-string propagator in the region covered by Fig.~\ref{figcooo}(c).

Since $(\XXX+\wt\XXX)V/2$ has ghost number $2$ and picture number $-1$, only open-string states with ghost number $1$ and picture number $-1$ can propagate along this strip.
There is no restriction on this state to be in the Siegel gauge since during the vertical integratiion at $q_2=1$ the $b_0$ factor in the propagator is absent.\footnote{Due to the $b_0$ factor in the propagator, the states between which the propagator is sandwiched must carry $c_0$ factors. Hence their conjugate states that are inserted into the interaction vertex have no $c_0$ and satisfy the Siegal gauge condition.}
The only such states with $L_0\leq 0$ are the tachyon and the OSG state, both of which give vanishing contributions to the disk CO two-point function, as shown in section \ref{sCOCO}.\footnote{
It is instructive to try to apply this reasoning to the vertical-integration contribution obtained in section \ref{s6.3} for the disk two-point amplitude.
In this case, approaching the boundary between the vertex region and the Feynman diagram region from the side of the vertex region, it can be seen that we need states of ghost number $1$, but with picture number $0$ and $2$ to flow on the long strip.
A conjugate pair of such states is $c$ and $c \, e^{-2\phi}$, which have $L_0 = -1$, and give the terms proportional to $\varepsilon^{-1}$ in (\ref{eq: Disk_2pt_VI_1})-(\ref{eq: Disk_2pt_VI_4}).
} 

This argument shows that this vertical-integration contribution is
suppressed by inverse powers of $\lambda$, which is equivalent to positive powers of $\alpha/\widetilde{\lambda}$. 
Since the OOOO vertex depends only on $\alpha$, the powers of $\widetilde{\lambda}$ in the denominator cannot be canceled. 
In our approximation, where all terms with negative powers of either $\alpha$ or $\widetilde{\lambda}$ are dropped, this contribution can therefore be set to zero.

\subsubsection{Vertical integration between Fig.~\ref{figcooo}(c) and \ref{figcooo}(a)}

The moduli-space regions associated with Fig.~\ref{figcooo}(a) and
Fig.~\ref{figcooo}(c) share the boundary $q_1=1$, across which the PCO
locations jump. This jump, however, lies in the four-open-string
subsector, and the corresponding vertical-integration contribution is
multiplied by the CO interaction vertex. Since the CO interaction vertex
vanishes, as shown in section \ref{sCOCO}, the vertical-integration
contribution between Fig.~\ref{figcooo}(a) and Fig.~\ref{figcooo}(c)
vanishes.

\subsubsection{Vertical integration between Fig.~\ref{figcooo}(a) and \ref{figcooo}(b)} \label{s797}

The moduli-space regions associated with Fig.~\ref{figcooo}(a) and
Fig.~\ref{figcooo}(b) share the boundary $q_2=1$. Across this boundary,
the PCO locations jump, so a vertical-integration contribution must be
included.

To analyze this contribution, note that the subdiagram containing the
OOO interaction vertex with two external O insertions is common to both
regions, and therefore appears as a common factor. 
Since all three external open-string states have picture number $-1$, and there is one PCO in the OOO vertex, the open-string propagator with Schwinger parameter $q_1$ must carry a state with picture number $-1$.
The only such states with $L_0\leq 0$ are the tachyon and the OSG state. The tachyon is excluded by GSO parity conservation, and the OSG state by ghost-number conservation, just as in section \ref{sec-4b}.
This common factor is therefore suppressed by an inverse power of
$\alpha$.

The remaining factor, associated with the vertical integration itself,
describes the transition between the COO vertex and the degeneration in
which a CO vertex is connected to an OOO vertex by an open-string
propagator. This factor depends only on $\widetilde\lambda$. Since we
drop all terms suppressed by inverse powers of $\alpha$, the
vertical-integration contribution at this boundary can be discarded.

\subsubsection{Vertical integration between Fig.~\ref{figcooo}(b) and \ref{figcooo}(d)} \label{s796}

In Fig.~\ref {figcooo}(d) the PCOs are arranged as $(\XXX+\wt\XXX)/2$ on the closed-string vertex operator $V$ placed at $\ii$, $\XXX$ on $V_\Psi$, and $\XXX$ on $V_\Xi$. 
On the other hand, in Fig.~\ref{figcooo}(b), the PCOs are arranged as
$\XXX\wt\XXX$ on $V$ and an additional insertion $\XXX(a)$, where $a$
denotes the image of the OOO PCO location $z_p$ on the UHP for the
COOO amplitude.
Thus a vertical integration is required across the boundary between the moduli-space regions associated with the two diagrams.

The boundary is obtained by setting the Schwinger parameter $q_1$ associated with open-string propagator in Fig.~\ref{figcooo}(b) to 1. 
Hence, near the boundary, we can use $\beta$---the modulus associated with the COO interaction vertex---and $q_1$ as the moduli of the COOO amplitude. 
So we choose $\beta_1=\beta$ and $\beta_2=q_1$ in \eqref{ghost contribution bulk}. 
Keeping the PCO locations to be at generic points $y_\alpha$, we can express the bulk integral as
\begin{align} 
&
\mathcal{N}_{\text{\rm COOO}} \, \Phi_C \, \Xi\Xi^*\Psi\,
\int d \beta \wedge d q_1 
\Bigg\langle \BBB_\beta \, \BBB_{q_1} \, 
\left(
\prod_{\alpha=1}^3
\XXX(y_\alpha) \right) 
V (\ii)\, %\nonumber\\
%& \hskip 1in
 \p\xi c e^{-2\phi}(f_1)\, 
\eta c \partial c(f_2)  \, 
\p\xi c\, e^{-2\phi}(f_3) 
\Bigg\rangle\, .
\label{ghost contribution bulkqbeta}
\end{align}
Note that, compared to \eqref{ghost contribution bulk}, we have brought the $\mathcal{B}$ insertions to the leftmost position in the correlator. 

The general rule for vertical integration is that, if the PCO location $y_\alpha$ jumps from $y_\alpha^{(1)}$ to $y_\alpha^{(2)}$ across the boundary $q_1=1$, we omit the integration over $q_1$, replace $\BBB_{q_1}$ by $-2\bigl(\xi(y_\alpha^{(2)})-\xi(y_\alpha^{(1)})\bigr)$, and remove the corresponding factor $\XXX(y_\alpha)$ from the product. 
To apply this rule, we must determine the orientation of the $(\beta,q)$ integration measure, which fixes whether the degeneration region or the bulk region corresponds to $y_\alpha^{(1)}$ or $y_\alpha^{(2)}$.
If $d\beta\wedge dq_1$ is the positive measure, then $y_\alpha^{(1)}$ and $y_\alpha^{(2)}$ denote the PCO locations in the $q_1<1$ degeneration region and the $q_1>1$ bulk region, respectively.
If instead $dq_1\wedge d\beta$ is the positive measure, these two assignments are reversed.

To determine the orientation of the $(\beta,q)$ integration measure, we
use the following observation. 
After an appropriate $\mathrm{SL}(2,\mathbb{R})$ transformation, we fix the third open-string puncture and place the first two at points $z_1'$ and $z_2'$ on the real axis. 
Then, by the sign prescription given below \eqref{ea4new}, $dz_1'\wedge dz_2'$ is the positive integration measure.
It was shown in \cite[Eq.~(D.27)]{Sen:2020eck} that, when the internal open-string propagator in Fig.~\ref{figcooo}(b) is attached to the puncture at $\beta$ on the COO interaction vertex,\footnote{We also need to consider the case when the propagator attaches to the puncture at $-\beta$. This is treated below.} $d\beta\wedge dq_1$ is a positive multiple of $dz'_1\wedge dz'_2$. 
Thus $d\beta\wedge dq_1$ is the positive integration measure, and $y_\alpha^{(1)}$ and $y_\alpha^{(2)}$ denote the PCO locations for $q_1<1$ and $q_1>1$, respectively.

With this orientation fixed, we describe the PCO motion from the degeneration region to the bulk region in two steps. 
We use the UHP coordinate associated with the COOO interaction vertex throughout.
\begin{enumerate}
    \item \textbf{Step 1.}
    We keep $\XXX\wt\XXX$ fixed at the location $\ii$ of
    the closed-string vertex operator $V$, and move a PCO from $a$ to
    the location of $V_\Psi$.
    \item \textbf{Step 2.}
    We average over two vertical integrations that move
    a PCO to the location of $V_\Xi$. In the first, $\XXX$ is moved
    from $\ii$, while $\wt\XXX$ remains fixed at $\ii$; in the second,
    $\XXX$ is moved from $-\ii$, while $\XXX$ remains fixed at $\ii$.
    In both cases, another $\XXX$ remains fixed at the location of
    $V_\Psi$.
\end{enumerate}
We discuss Step 2 first since it is slightly easier.

\paragraph{Step 2:}
The locations of the open-string vertex operators in the UHP associated with the COOO amplitude are at $f_1$, $f_2$ and $f_3$, see \eqref{elocalCOOOrep}. 
Let us begin with the configuration where $V_\Xi$ is inserted at $f_1$, $V_{\Xi^*}$ is inserted at $f_2$
and $V_\Psi$ is inserted at $f_3$. 
Later we will sum over cyclic permutations. 

First we move a PCO from $\ii$ to $f_1$, keeping fixed the other two PCOs at $-\ii$ and $f_3$.
The contribution to the effective action from this vertical integration
is given by
\begin{align} 
    \text{VI}_{\ii \to f_1}^{(2)}
    &=
    \mathcal{N}_{\text{\rm COOO}} \, \Phi_C\Xi\Xi^*\Psi\,
    \int d\beta \Bigg\langle \left(\sum_{a=1}^3\ointclockwise_{z_a}\frac{\partial F_a}{\partial \beta 
    }b(z)dz\right)
    \left(-2(\xi(f_1)-\xi(\ii))\right)\, 
    \tilde{\mathcal{X}}V(\ii) \nonumber \\
    &\hspace{2in}
    \partial \xi ce^{-2\phi}(f_1) 
    \eta c \partial c(f_2) \frac{-\mathbb{I}}{2}(f_3)\Bigg\rangle\\
    &= \mathcal{N}_{\text{COOO}}  
    \,\Phi_C\Xi\Xi^*\Psi\, 
    \frac{Q}{4\ii \bb\wt\mu}
    \int d\beta 
    \bigg(f_1'\left\langle\left(\xi(f_1) - \xi(\ii)\right)\tilde{\eta}c e^{\tilde{\phi}}e^{-\phi}(\ii)\partial \xi e^{-2\phi}(f_1)\eta c \partial c(f_2)\right \rangle \nonumber \\ 
    &  \hspace{150 pt}+f_2'\left\langle\left(\xi(f_1) - \xi(\ii)\right)\tilde{\eta}c e^{\tilde{\phi}}e^{-\phi}(\ii)\partial \xi ce^{-2\phi}(f_1)\eta \partial c(f_2)\right \rangle\nonumber \\ 
    &\hspace{150 pt}-\frac{g_2'}{g_2}\left\langle\left(\xi(f_1) - \xi(\ii)\right)\tilde{\eta}c e^{\tilde{\phi}}e^{-\phi}(\ii)\partial \xi c e^{-2\phi}(f_1)\eta c (f_2)\right \rangle\bigg)\, .
    \label{edeg1}
\end{align}
Here $\ointclockwise_{z_a}$ denotes a \emph{clockwise} contour around $z_a$, the prime denotes derivative with respect to $\beta$, and the functions $f_a,g_a$ are as in \eqref{elocalCOOOrep}.
We have used the fact that $-\frac{1}{2}\tilde{\eta}c e^{\tilde{\phi}}e^{-\phi}e^{\bb\varphi}$ is the only term 
from $\wt\XXX V$ that contributes, see (\ref{e6.19a}).
We also used \eqref{one-pt-vb} for the disk one-point function of $e^{\bb \varphi}$. 
We have evaluated the correlation function appearing in (\ref{edeg1}) in the \texttt{Mathematica} file accompanying the arXiv submission, yielding
{\small
\begin{align}\label{eq: Vert_2_CO_OOOO_2}
     \text{VI}_{\ii \to f_1}^{(2)}
     =&-K \, \mathcal{N}_{\text{COOO}}  
    \Phi_C\Xi\Xi^*\Psi\, 
   \frac{Q}{4\ii \bb\wt\mu} 
    \nonumber \\ &
     \int d\beta \left(\frac{\left(f_2+\ii\right) \left[ \left(f_2-\ii\right){}^2 f_1' g_2+\left(f_1-\ii\right) \left(\left(f_1-2 f_2+\ii\right) f_2' g_2
      +\left(f_2-\ii\right) \left(f_2-f_1\right) g_2'\right) \right]}{\left(f_1-f_2\right){}^2 \left(f_2-\ii\right) g_2}\right)\, .
\end{align}
}

The second part of Step 2 is to move a PCO from $-\ii$ to $f_1$, keeping the other two PCOs at $\ii$ and $f_3$.
This contribution is similarly evaluated as
{\small
\begin{align} \label{edeg2}
    \text{VI}_{-\ii \to f_1}^{(2)}
    &= -K \, \mathcal{N}_{\text{COOO}}  
    \Phi_C\Xi\Xi^*\Psi\, 
   \frac{Q}{4\ii \bb\wt\mu} 
    \nonumber \\ &
     \int d \beta
     \left(
     \frac{\left(f_2-\ii\right) \left[\left(f_2+\ii\right){}^2 g_2 f_1'+\left(f_1+\ii\right) \left(\left(f_1-2 f_2-\ii\right) g_2 f_2'+\left(f_2+\ii\right) \left(f_2-f_1\right) g_2'\right)\right]}{\left(f_1-f_2\right){}^2 \left(f_2+\ii\right) g_2}\right)\, .
\end{align}
}

To proceed further we need to determine the $f_a$'s and the $g_a$'s. 
If the internal open-string propagator in Fig.~\ref{figcooo}(b) is
attached to the puncture at $\beta$ on the COO interaction vertex, the
relation between the COO coordinate $z_{\rm coo}$ and the OOO coordinate
$z_{\rm ooo}$ is obtained by a plumbing fixture with parameter $q_1=1$.
This fixture glues the local coordinate $w_1$ in \eqref{O-O-O coord} to
the local coordinate $w_1$ in \eqref{eq: local_coord_COO}.
Identifying $z_{\rm coo}$ with the coordinate $z_{\rm cooo}$ on the
resulting UHP for the COOO amplitude, we find
\be \label{e787b}
\alpha \, \frac{2 z_{\rm ooo}}{2 - z_{\rm ooo}} \cdot 
\alpha \wt\lambda
\frac{4\wt\lambda^2+1}{4\wt\lambda^2}
\frac{z_{\rm cooo}-\beta}
{(1+\beta z_{\rm cooo})+\wt\lambda f(\beta)(z_{\rm cooo}-\beta)} =-1\, .
\ee
Then the function $F_1$ may be identified as the function that relates $z_{\rm cooo}$ to $w_2$ in \eqref{eq: local_coord_COO} and $F_2$ and $F_3$ may be obtained by relating $z_{\rm cooo}$ to the coordinates $w_2$ and $w_3$ in \eqref{O-O-O coord} via \eqref{e787b}. 
The resulting functions $F_1,F_2,F_3$ can be found in \cite[Eq.~(D.23)]{Sen:2020eck} and are given below:
\begin{equation}
\label{e7.87a}
\begin{aligned}
F_1(w_1,\beta) &= -\beta
+ {4\wt\lambda^2\over 4\wt\lambda^2+1}\,
{1+\beta^2\over \alpha\wt\lambda} w_1
+\mathcal{O}(w_1^2)\, , \\
F_2(w_2,\beta)
&= {\beta+ u_0(\beta\wt\lambda f-1)\over
1 + u_0(\beta+\wt\lambda f)}
+ {2\, u_0\over \alpha}
{(1+\beta^2) \over (1+ u_0(\beta+\wt\lambda f))^2}\, w_2
+\mathcal{O}(w_2^2)\, , \\
F_3(w_3,\beta)
&= {\beta- u_0(\beta\wt\lambda f-1)\over
1 - u_0(\beta+\wt\lambda f)}
+ {2\, u_0\over \alpha}
{(1+\beta^2) \over (1- u_0(\beta+\wt\lambda f))^2}\, w_3
+\mathcal{O}(w_3^2)\, , \\
u_0 &:= \frac{1}{2\alpha^2\wt\lambda}
{4\wt\lambda^2\over 4\wt\lambda^2+1}\, .
\end{aligned}
\end{equation}
Comparison of \eqref{elocalCOOOrep} with \eqref{e7.87a} determines
$f_a$ and $g_a$ for $a=1,2,3$, which can then be used to evaluate (\ref{eq: Vert_2_CO_OOOO_2}) and \eqref{edeg2} and take their average.

Next, we sum over the three cyclic permutations of
$(f_1,g_1)$, $(f_2,g_2)$, and $(f_3,g_3)$, corresponding to cyclic
permutations of the external open-string states in
Fig.~\ref{figcooo}(b) and Fig.~\ref{figcooo}(d).

We must also include the contribution in which the internal open string
is attached to the puncture at $-\beta$ on the UHP of the COO interaction
vertex. This is accounted for by extending the range of the $\beta$
integration from $[1/(2\wt\lambda),1]$ to
$[1/(2\wt\lambda),2\wt\lambda]$, and extending $f(\beta)$ beyond
$\beta=1$ using
$f(1/\beta)=f(-\beta)=-f(\beta)$ \cite{Sen:2020eck}.

Adding these contributions, we obtain the result for \textbf{Step 2}:
\begin{align}\label{eq: COO_OOO_VI_2}
    \text{VI}^{(2)} &=-K\, \mathcal{N}_{\text{COOO}}  \, 
    \Phi_C\Xi\Xi^*\Psi\,
    \frac{Q}{4\ii \bb\wt\mu} \, \tilde{\lambda}\int_{\frac{1}{2\tilde{\lambda}}}^{2\tilde{\lambda}} d \beta \, f'(\beta)+\mathcal{O}(\alpha^{-4})\nonumber\\
    &= K\, \mathcal{N}_{\text{COOO}}  \, 
    \Phi_C\Xi\Xi^*\Psi\,
    \frac{Q}{4\ii \bb\wt\mu} \, \tilde{\lambda}\, ,
\end{align}
where we used the values of $f(\frac{1}{2\widetilde{\lambda}})$ and 
$f(2\widetilde{\lambda})$ from (\ref{f-values}) and dropped all terms with inverse powers of $\widetilde{\lambda}$ and $\alpha$.

\paragraph{Step 1:}
In this step, we move the $\mathcal{X}$ from $a$ to $f_3$. 
Recall that $a$ denotes the image of the OOO PCO location $z_p$ on the
upper half-plane describing the COOO amplitude.
The integrand can be obtained similarly to \eqref{edeg1},
{\small
\begin{align}
    \text{VI}^{(1)}&= \mathcal{N}_{\text{COOO}} \, \Phi_C\Xi\Xi^*\Psi\, 
    \int d\beta\, 
    \Bigg\langle 
    \left(\sum_{a=1}^3\ointclockwise_{z_a}\frac{\partial F_a}{\partial \beta}b(z)dz\right) \nonumber \\
    &\hspace{2in}\left(-2(\xi(f_3)-\xi(a))\right)\,
    \mathcal{X}\tilde{\mathcal{X}}V(\ii)\, 
    \partial \xi ce^{-2\phi}(f_1)\, 
    \eta c \partial c(f_2)\,
    \partial \xi ce^{-2\phi}(f_3)\Bigg \rangle\nonumber\\
    &=\mathcal{N}_{\text{COOO}} \, \Phi_C\Xi\Xi^*\Psi\, \frac{Q}{4\ii \bb\wt\mu} \int d\beta\,  
    \Bigg[f_1'\left\langle (\xi(f_3) - \xi(a))\eta \widetilde{\eta} e^{\phi} e^{\widetilde{\phi}}(\ii)\partial \xi e^{-2\phi}(f_1)\,\eta c \partial c(f_2) \, \partial \xi c e^{-2\phi}(f_3)\right \rangle\nonumber\\
    &\hspace{135 pt} +f_2'\left\langle (\xi(f_3) - \xi(a)) \eta \widetilde{\eta} e^{\phi} e^{\widetilde{\phi}}(\ii)\partial \xi ce^{-2\phi}(f_1)\,\eta \partial c(f_2) \, \partial \xi c e^{-2\phi}(f_3)\right \rangle\nonumber\\
    &\hspace{135 pt} -\frac{g_2'}{g_2}\left\langle (\xi(f_3) - \xi(a)) \eta \widetilde{\eta} e^{\phi} e^{\widetilde{\phi}} (\ii)\partial \xi c e^{-2\phi}(f_1)\,\eta c (f_2)\, \partial \xi c e^{-2\phi}(f_3)\right \rangle \nonumber\\
    &\hspace{135 pt}-f_3'\left\langle (\xi(f_3) - \xi(a))\eta \widetilde{\eta} e^{\phi} e^{\widetilde{\phi}}(\ii) \partial \xi c e^{-2\phi}(f_1)\,\eta c \partial c(f_2) \, \partial \xi  e^{-2\phi}(f_3)\right \rangle\Bigg]\, .
\end{align}
}
\par\noindent
Here only the term $\frac{1}{4} \eta \widetilde{\eta} e^{\phi} e^{\widetilde{\phi}} e^{\bb \varphi}\subset\mathcal{X}\tilde{\mathcal{X}}V(\ii)$ contributes (see \eqref{PCO_action}) and we have used \eqref{one-pt-vb} for the disk one-point of $e^{\bb \varphi}$. 
The location $a$ may be found from \eqref{e787b} with $z_{\text{ooo}} = z_p$:
\begin{align}
    a= \beta +\frac{2 \tilde{\lambda}  (z_p-2) (\beta  z_p+1)}{\alpha ^2 \left(4 \tilde{\lambda} ^2+1\right) z_p-2 \tilde{\lambda} ^2 (z_p-2) f(\beta )} \, .
\end{align}

Again we need to sum over cyclic permutations of $F_1,F_2,F_3$ and extend the integration range over $\beta$ to $[1/2\wt\lambda, 2\wt\lambda]$ to take into account the case where the open-string propagator is attached to the puncture at $-\beta$ on the COO vertex. 
The result, ignoring terms carrying powers of $\alpha$ or $\wt\lambda$
in the denominator, is given by
\begin{align}
    \text{VI}^{(1)} &=-K\,  \mathcal{N}_{\text{COOO}}  \,
    \Phi_C\Xi\Xi^*\Psi\,  
    \frac{Q}{4\ii\bb\wt\mu}\, 
    \int_{\frac{1}{2\tilde{\lambda}}}^{2\tilde{\lambda}}
    d\beta\, \frac{1}{2}\left(-1-\frac{1}{\beta^2}-2\tilde{\lambda} f'(\beta)\right)
    \nonumber \\ 
     &=
    K\, \mathcal{N}_{\text{COOO}} \, 
    \Phi_C\Xi\Xi^*\Psi \, 
    \frac{Q}{4\ii\bb\wt\mu}\, \tilde{\lambda}\, 
    \label{eq: COO_OOO_VI_3}\, .
\end{align}

\subsubsection{Total contribution to \texorpdfstring{$g_{\rm gauge}$}{}}

To summarize, with our choice of interaction vertices and PCO locations, the only nonzero contribution to $\mathcal{A}_{\text{COOO}}$ comes from the vertical integration between moduli space regions corresponding to Fig.~\ref{figcooo}(b) and Fig.~\ref{figcooo}(d).  
Adding \eqref{eq: COO_OOO_VI_2} and \eqref{eq: COO_OOO_VI_3}, we get
\begin{align}\label{eq: COOO_contribution_final}
    \mathcal{A}_{\text{COOO}}=K\,  \mathcal{N}_{\text{COOO}} \, 
    \frac{Q}{2\ii\bb\wt\mu}\, 
    \tilde{\lambda} \, .
\end{align}
Plugging this into \eqref{e7.77}, and substituting the values $\mathcal{N}_{\text{OOO}} = g_s^{1/2}\eta_c^{3/4}$ and $\mathcal{N}_{\text{COOO}} = - g_s^{3/2} \eta_c^{7/4}$ with $\eta_c = \frac{\ii}{2\pi}$, we get
\begin{align}
\boxed{
    g_{\text{gauge}}= \frac{1}{K_0}\, {\tilde\lambda\over\pi}\, .}
    \label{eq: g_gauge_fin}
\end{align}

\subsection{Final result for the annulus one-point function}\label{sec: ann_1_pt_final}

The nonzero contributions to $g$ derived in this section are collected in the boxed equations, namely
(\ref{e8.18}), 
\eqref{eq: g_(d)_(c)_res}, 
\eqref{eq: g^(c)_contri}, 
\eqref{eq: tau_loop_final_2}, 
\eqref{e7.61a}, 
\eqref{eq: g^ac_VI_fin_2}, 
\eqref{eq: g_gauge_fin}.
Adding all these contributions together, we get
\begin{align}
    g=\frac{1}{2 \, K_0} \, ,
\end{align}
in perfect agreement with the result \eqref{eq: f and g_KPZ_expectation} obtained from the DDK-KPZ scaling argument.

\section{Extension to the type 0B theory}
\label{sec:0B}

In this section, we extend our results to minimal superstring theory with type 0B GSO projection.
In the ungapped phase, the simplest ZZ branes are the R-charge conjugates $\ket{(1,1)}$ and $\ket{\overline{(1,1)}}$.
The super-Liouville sector of these states have the following expansion as a sum over Ishibashi states \cite{Ahn:2002ev, Fukuda:2002bv, Irie:2007mp}:
\begin{align}
    \ket{(1,1)} &= \int_{0}^\infty d P \, \Big( 
    \Psi^\ns_{1,1}(P) \,\vert \text{NS}, P, + \rrangle
    + 
    \Psi^\text{R}_{1,1}(P) \,\vert \text{R}, P, + \rrangle \Big) \,, 
    \label{11-state-0b}\\
    \ket{\overline{(1,1)}} &= \int_{0}^\infty d P \, \Big( 
    \Psi^\ns_{1,1}(P) \,\vert \text{NS}, P, + \rrangle
    -
    \Psi^\text{R}_{1,1}(P) \,\vert \text{R}, P, + \rrangle \Big) 
    \,, \label{11bar-state-0b}\\
    \Psi^\ns_{1,1}(P) &= \left( \pi \mu \gamma\left( \frac{\bb Q}{2} \right) \right)^{-\ii P \bb^{-1}} \frac{\sqrt{2}\,\pi}{P \,\Gamma \left( -\ii P  \bb^{-1} \right)\,\Gamma \left( -\ii P \bb \right)} \, ,\\
    \Psi^\text{R}_{1,1}(P) &= \left( \pi \mu \gamma\left( \frac{\bb Q}{2} \right) \right)^{-\ii P \bb^{-1}}  \frac{\sqrt{2}\,\pi}{\Gamma \left( \frac{1}{2}-\ii P \bb^{-1} \right)\,\Gamma \left( \frac{1}{2}-\ii P \bb \right)}\, 2^\frac{1}{4} \, .
\end{align}
Let $\mathcal{T}^{0\text{B}}$ denote the tension of the ZZ instanton in the type 0B theory. 
The tension of the anti-instanton is also the same.
It follows from (\ref{11-state-0b}), (\ref{11bar-state-0b}) and (\ref{11-state-0a}) that
\begin{align}
    \mathcal{T}^{0\text{B}} = \frac{\mathcal{T}}{\sqrt{2}}\, .
\end{align}

In the 0B case, the leading nonperturbative contribution comes from an instanton-anti-instanton pair,\footnote{This is true for the ungapped phase of the theory, which is where the $(1,1)$ ZZ instanton is relevant \cite{Seiberg:2003nm, Chakrabhavi:2024szk}.} and thus the exponential prefactor is $\exp(-2\, \mathcal{T}^{0\text{B}}) = \exp(-\sqrt{2}\, \mathcal{T})$.
Furthermore, we now have two sets of ghost zero modes, one on the instanton and one on the
anti-instanton, with each pair giving a contribution of $g_s^{1/2}$ to the partition
function. 
Thus, we can write the analog of (\ref{eq: one_instanton_partition_fn}) as
\begin{align} \label{e711xy}
    Z^{(1,1)}_{\rm 0B}(\widetilde{\mu},g_s) &= 
    Z^{(0)}_{\rm 0B}(\widetilde{\mu},g_s) \, \exp 
    \left( 
        g_s^{-1} \,  A^{\rm 0B}(\widetilde{\mu}) + \log g_s + B^{\rm 0B}(\widetilde{\mu}) + \ldots
    \right) \,,\\ 
    A^{\rm 0B}(\widetilde{\mu}) &= \sqrt 2\, A(\widetilde{\mu}) \, .
\end{align}
We have not included the term proportional to $C$ since it is not relevant for our analysis.
The analog of (\ref{etension}) now takes the form
\be \label{etension0B}
2\, \TTT^{\rm 0B} = K_0^{\rm 0B}\, g_s^{-1}, \qquad \text{with } \quad K_0^{\rm 0B} := - A^{\rm 0B}
(\tilde\mu)  =\sqrt 2\, K_0 \, .
\ee
The same KPZ scaling arguments that led to (\ref{eq: disk_one_KPZ}) now gives:
\begin{align}
    A_\text{disk}^{0\text{B}}(V) = - \frac{Q}{2\bb\widetilde{\mu}}\, 
    K_0^{0\text{B}}\,,\quad 
    f^{0\text{B}} = \frac{1}{K_0^{0\text{B}}} \left( \frac{2\bb}{Q} - 1\right)\,,\quad 
    g^{0\text{B}} = 2 \cdot \frac{1}{2K_0^{0\text{B}}}\, .
    \label{0B-results-2}
\end{align}
The extra factor of 2 in the expression for $g^{0\text{B}}$ compared to what was in 
(\ref{eq: f and g_KPZ_expectation}) can be traced to the replacement
of $(\log g_s)/2$ in the exponent of (\ref{eq: one_instanton_partition_fn}) by $\log g_s$ in the 
exponent of (\ref{e711xy}).

We shall now briefly discuss how these relations are reproduced by direct string theory computations.
Since the boundary state of the instanton-anti-instanton pair in 0B theory is $\sqrt 2$ times that of the boundary state of the instanton in the type 0A theory,\footnote{This is true since we are only considering amplitudes involving closed strings, and the instanton and anti-instanton under consideration don't have any moduli. Therefore, we can just add the two sources \eqref{11-state-0b} and \eqref{11bar-state-0b} of closed-string fields.} all closed-string disk amplitudes in the 0B theory will be $\sqrt 2$ times the corresponding amplitudes in the 0A theory. 
Furthermore, all closed-string annulus amplitudes in the 0B theory will
be twice the corresponding amplitudes in the 0A theory. This gives the results:
\begin{align} \label{e0A0Brel}
    A_\text{disk}^{0\text{B}}(V) &= \sqrt{2}\, \frac{\partial A}{\partial \widetilde{\mu}}\, ,\quad 
    f^{0\text{B}} = \frac{1}{\sqrt{2}} \, f \,,\quad 
    g^{0\text{B}} = \sqrt{2}\, g\, ,
\end{align}
where we used the fact that $A_\text{disk}^{0\text{B}}(V)$ represents a disk one-point function,
$f^{0\text{B}}$ represents the ratio of a disk two-point function and the square of a disk one-point function, 
and $g^{0\text{B}}$ represents the ratio of an annulus one-point function and  a disk one-point function. 
The quantity $g^{0\text{B}}$ also receives contribution from the Jacobian factor associated with the gauge parameter redefinition.
This also yields a factor of two, since the gauge group is now $U(1) \times U(1)$.
Indeed, the factor $1 + 2 \mathcal{A_\text{COOO}}/(K \, \mathcal{N}_{\text{OOO}}) \, \Phi_C$ will be squared, which after the Taylor expansion to linear order, yields a factor of two enhancement in the contribution to the one-point function of $\Phi_C$.

Using (\ref{etension0B}), (\ref{eq: disk one-point definition}), (\ref{eq: disk_one_KPZ})  and (\ref{eq: f and g_KPZ_expectation}) in (\ref{e0A0Brel}), we recover (\ref{0B-results-2}).

\paragraph{Acknowledgments.}
RM thanks the Leinweber Institute for Theoretical Physics at Stanford
for its hospitality during the completion of part of this work.
CM is supported in part by the DOE through DE-SC0013528 and the QuantISED grant DE-SC0020360.
AS is supported by the ICTS Infosys Madhava Chair Professorship.
We acknowledge support of the Department of Atomic Energy, Government of India, under project no. RTI4019.
This research was supported in part by the International Centre for Theoretical Sciences (ICTS) for participating in the program - ``Quantum Information, Quantum Field Theory and Gravity" (code: ICTS/qftg2024/08).

\appendix

\section{Relation between \texorpdfstring{$\TTT$}{} and \texorpdfstring{$K$}{}} 
\label{sa}

In this appendix, we derive the relation between the brane tension $\TTT$ and the constant $K$ that appears in the normalization 
\be
\langle c(z_1) c(z_2) c(z_3) e^{-2\phi}(0)\rangle = - K\, (z_1-z_2) (z_1-z_3) (z_2-z_3)\, .
\ee
Note that this correlator is written in the small Hilbert space, see the discussion around
\eqref{exietaphi} for more details.

Ref. \cite{Sen:2024npu} derived the analogous relation in bosonic string theory:
\be \label{ebosonic}
K= - \frac{1}{2}\, \eta_c^{-1/2}\,  g_s \TTT, \qquad \text{where } \quad  \eta_c := \frac{\ii}{2\pi}\, \qquad 
\text{(bosonic string)} \, .
\ee
In the bosonic string, $K$ is the constant that appears in the normalization of the disk amplitude of three $c$-ghosts:
\be
\langle c(z_1) c(z_2) c(z_3)\rangle = - K\, (z_1-z_2) (z_1-z_3) (z_2-z_3)\, \qquad 
\text{(bosonic string)} \, .
\ee

To set up the normalization convention, we consider a critical superstring theory where the matter SCFT consists of ten scalars $X^\mu$ and ten pairs of Majorana-Weyl fermions
$\psi^\mu$, $\tilde\psi^\mu$. 
We normalize their OPEs as
\be
\p X^\mu (z) \, \p X^\nu(w) \sim -{\eta^{\mu\nu}\over 2(z-w)^2}, \qquad 
\psi^\mu (z) \, \psi^\nu(w) \sim -{\eta^{\mu\nu}\over 2(z-w)},
\label{ope-norm-app}
\ee
with similar OPEs for the anti-holomorphic fields.
In the bosonic string, the OPE of $\partial X$ with itself is normalized the same way as in \eqref{ope-norm-app}. 

We begin by reviewing the derivation of \eqref{ebosonic} and then discuss what changes for superstrings.
In the bosonic string, a background closed string field of the form
\be
\frac{1}{g_s} \, h_{\mu\nu} \, c\tilde c \, \p X^\mu \bar\p X^\nu (0)|0\rangle
\label{unintegrated-operator}
\ee
corresponds to deforming the worldsheet CFT action by an operator \cite{Sen:2021tpp, Sen:2024nfd}
\be
- \int {d^2 z\over \pi} \, h_{\mu\nu} \, \p X^\mu \bar\p X^\nu\, .
\label{integrated-action}
\ee
This in turn corresponds to a deformation of the target space background metric by $h_{\mu\nu}$.
Using the effective action of massless closed string fields in the presence of a D$p$-brane of
tension $\TTT$,
\be
-\TTT \int d^{p+1} x \, e^{-\Phi} \sqrt{\det g}\, ,
\ee
where $\Phi$ is the dilaton field and $g_{\mu\nu}$ is the string-frame metric,
we can compute the one-point function of the field $h_{\mu\nu}$ in the presence of the brane.
However, this one-point function can also be computed from the one-point function of the operator $g_s^{-1} c\bar c \p X^\mu \bar\p X^\nu $ on the disk. 
Since the former is proportional to $\TTT$ while the latter is proportional to $K/g_s$, we get the relation \eqref{ebosonic} between $K/g_s$ and $\TTT$.

Let us now review what changes in the case of superstring theory. 
Here we start with the string field in the $(-1,-1)$ picture
\be\label{eB7}
\frac{1}{g_s} \, h_{\mu\nu} \, c\tilde c \, e^{-\phi}  \psi^\mu e^{-\tilde\phi}\tilde \psi^\nu (0)|0\rangle\, .
\ee
This field has picture number $(-1,-1)$, so we need to multiply it by $\mathcal{X}\tilde{\mathcal{X}}$ to obtain a picture number $0$ state, which can be added to the worldsheet CFT action.
After acting with the PCOs as normalized in \eqref{pco}, and with the supercurrent $T_F$ normalized as in \eqref{T_FT_F OPE}, the transformed state is
\be
\label{eB8}
    \frac{1}{g_s} \, h_{\mu\nu} \, c\tilde c \, \p X^\mu \bar\p X^\nu (0)|0\rangle\, .
\ee
From here on the analysis proceeds as in the case of bosonic string and the one-point function of the state, expressed in terms of the brane tension $\TTT$, will have the same form in both theories.

The difference between the bosonic and superstring theories appears 
when we compute the one-point function in terms of $K$.
To see this, note that in bosonic string theory the one-point function of 
$c\tilde c \p X^\mu \bar\p X^\nu$ is given by
\be\label{eboson2}
\frac{1}{2} \langle (\p c - \bar\p \tilde c) c \tilde c \, \p X^\mu \bar\p  X^\nu\rangle
= \frac{1}{2} \cdot 8 K \cdot \frac{1}{8} \, \eta^{\mu\nu} = 
\frac{K}{2}\, \eta^{\mu\nu} \,,
\ee
where we have chosen $\mu,\nu$ to be directions tangential to the brane.
We contrast this with the one-point function of $c\wt{c} \, e^{-\phi}  \psi^\mu 
e^{-\tilde\phi}\tilde \psi^\nu$
in the superstring theory, which is given by
\be\label{esuper2}
\frac{1}{2} \langle (\p c - \bar\p \tilde c) c \tilde c e^{-\phi}  \psi^\mu e^{-\tilde\phi}\tilde \psi^\nu
\rangle
= \frac{1}{2} \cdot 8 K \cdot (-1) \cdot \frac{1}{2\ii} \cdot \frac{-1}{4\ii} \eta^{\mu\nu}
= -\frac{K}{2} \, \eta^{\mu\nu} \, .
\ee
The difference in sign between \eqref{eboson2} and \eqref{esuper2} leads to an extra sign in the
relation between $K$ and $\TTT$ in superstring theory compared to \eqref{ebosonic}, and we get
\be \label{esuper}
K= \frac{1}{2}\, \eta_c^{-1/2}\,  g_s \TTT, \qquad \text{where } \quad  \eta_c := \frac{\ii}{2\pi}\, .
\ee
The above derivation makes it clear that in superstring theory, the relation between $K$ and $\TTT$ depends on the precise normalization of the PCO. 
For example, if we had included an extra factor of $\ii$ in the definition of $\mathcal{X}$ and $\tilde{\mathcal{X}}$, then \eqref{eB7} will have an 
extra minus sign in order to reproduce \eqref{eB8} after picture changing. 
This will give an extra minus sign in the computation of the one-point function given in \eqref{esuper2},  and the relation between $K$ and $\TTT$ will take the same form as in the bosonic case.

Let us spell out one implicit assumption in the above derivation. 
The argument in this appendix assumes the existence of at least one flat target-space direction, an assumption not satisfied in minimal bosonic strings or
minimal superstrings. 
Nevertheless, since the derivation leads to a universal relation between $K$ and $\mathcal{T}$, it seems natural to assume that it holds even in backgrounds without any flat directions. 
It would be interesting to understand this point better in future work.

\section{Alternative choice of PCO locations}
\label{sec: diff_PCO_choice}

In the main text, the CO and the COOO interaction vertices were defined with the ``symmetric" PCO insertion $(\XXX+\wt\XXX)/2$ at the closed-string puncture. 
This choice is convenient, but it is not unique, {\it e.g.} one could instead define these vertices by placing only the holomorphic PCO $\XXX$ at the closed-string insertion. 
We make this alternative choice in this appendix for the CO and the COOO interaction vertices. The PCO locations for all other vertices, summarized in table~\ref{ta2}, remain unchanged.
Once the appropriate vertical integration contributions are included, the final answers must be unchanged, although the cancellation need not occur diagram by diagram.
In this appendix, we verify that the final amplitudes are indeed independent of this choice of PCO prescription.

With the symmetric PCO prescription, any diagram containing a CO vertex with the out-of-Siegel-gauge state $\tau$ vanishes because the two contributions in
\eqref{CO Out-of-Siegel} and \eqref{CO_OSG_anti_holo} cancel against each
other.  
With the alternative prescription, this cancellation no longer occurs,
so these diagrams must be included explicitly.  We will see that the new
nonzero contributions are precisely cancelled by the corresponding changes in
the vertical integration terms.

The notable differences in the two calculations are as follows:
\begin{enumerate}
    \item In the computation of the disk two-point function, 
    the contribution from Fig.~\ref{figthree}(a)
    with a $\tau$ propagator joining two CO vertices, becomes nonzero.
    The vertical-integration contribution changes to compensate for this, leaving the final disk two-point function unchanged.

    \item In the computation of the annulus one-point function, the moduli
    space regions (a) and (b) in Fig.~\ref{fig:moduli-feynman} now give
    non-vanishing contributions.  These are cancelled by the corresponding
    changes in vertical integration between regions (a)$-$(c) and
    (b)$-$(d).

    \item There are also extra intermediate contributions in the computation of
    the gauge-parameter redefinition.  The vertical integration terms change
    accordingly so that the final answer remains the same.
\end{enumerate}

\subsection{Disk two-point function}

The contribution from Fig.~\ref{figthree}(b), the bulk of the moduli space, is unchanged, since it still uses the original insertion $\XXX\wt\XXX$ on the closed-string vertex operator at $\ii y$.

There is, however, a new contribution from Fig.~\ref{figthree}(a), in which
the out-of-Siegel-gauge open-string state $\tau$ propagates between two CO
interaction vertices:
\begin{align}
    A_{\text{disk}}^{\text{OSG}}(VV)=\mathcal{N}_{\text{CO}}^2\tau_{\text{prop}}\left\langle \mathcal{X}V(\ii)\partial \xi c\partial ce^{-2\phi}(0) \right \rangle_{\text{UHP}}\left\langle \mathcal{X}V(\ii)\partial \xi c\partial ce^{-2\phi}(0) \right \rangle_{\text{UHP}}\, ,
\end{align}
where, from \eqref{eapp1} and the sign rules described below \eqref{ea4new},
\begin{align}\label{eq: NCO} 
    \mathcal{N}_{\text{CO}}=-g_s^{\frac{1}{2}}\eta_c^{\frac{1}{4}}\, .
\end{align}
Using \eqref{CO Out-of-Siegel} and \eqref{eq: tau_prop_result}, we get, 
\begin{align}\label{eq: disk_2pt_OSG}
    A_{\text{disk}}^{\text{OSG}}(VV)=\mathcal{N}_{\text{CO}}^2\tau_{\text{prop}}\frac{Q^2}{4\bb^2\wt\mu^2}=K\,\mathcal{N}_{\text{CO}}^2\frac{Q^2}{2\bb^2\wt\mu^2}\, .
\end{align}
The vertical integration contribution discussed in section \ref{s6.3} is also
modified.  In the present prescription, the relevant vertical integration segment moves
$\XXX$ from $\ii$ to $-\ii\varepsilon$, while keeping the second
$\XXX$ fixed at $\ii\varepsilon$.
This contribution has already been computed in \eqref{eq: Disk_2pt_VI_2}. Adding all the contributions \eqref{ea1vv}, \eqref{eq: Disk_2pt_VI_2} and \eqref{eq: disk_2pt_OSG}, we get the final result
\begin{align} \label{eB4}
    A_{\text{disk}}(VV)
    &= K \, \mathcal{N}_{\text{CC}} \, \widetilde{\mu}^{-2} \left( \frac{Q}{\bb} - \frac{Q^2}{2\bb^2} \right) \, . 
\end{align}
We used the fact that $\mathcal{N}_{\text{CO}}^2=\mathcal{N}_{\text{CC}}$ which can be seen from \eqref{eNCC} and \eqref{eq: NCO}. 
This matches the result \eqref{Adisk_final} obtained in the main text.

\subsection{Annulus one-point function}

We now turn to the annulus one-point function.  
The contributions from Fig.~\ref{fig:feynman-diagrams_2}(c) and Fig.~\ref{fig:feynman-diagrams_2}(d) are unchanged; in both cases, the PCOs $\XXX\wt\XXX$ are located at closed-string insertion, exactly as in the main text.
The only changes occur in the degeneration regions in which a CO vertex appears with a $\tau$ insertion, together with the vertical integration terms at the corresponding boundaries of moduli space.  
Thus, we only need to compute the non-vanishing contributions from regions (a) and (b) as well as the modified vertical-integration terms between regions (a)$-$(c) and (b)$-$(d).

\def\vt{\vartheta}

\subsubsection{Contribution from region (b)}
We now consider the contribution from Fig.~\ref{fig:feynman-diagrams_2}(b), where a CO disk interaction vertex and an O annulus interaction vertex are joined by an open-string propagator.  
This diagram covers the small-$x$ region labeled (b) in Fig.~\ref{fig:moduli-feynman}.  
The tachyon cannot propagate in this channel because it has odd GSO parity, so the only relevant contribution comes from the out-of-Siegel-gauge mode $\tau$.

The O annulus vertex contains a PCO $\mathcal{X}$ inserted at the point
$w=w_p$ in the bulk of the annulus, together with the $\tau$ open-string
insertion on the boundary.  
Since $\phi$-momentum must be conserved on the annulus,
the only terms in $\mathcal{X}$ that contribute are
\be
    -\frac{1}{2}\partial\eta e^{2\phi}b
    -\frac{1}{2}\partial(\eta e^{2\phi}b)\, .
\ee
The CO disk contribution with $\XXX V$ and $\tau$ insertions is given in
\eqref{CO Out-of-Siegel} to be $K\,Q/(2\bb\wt\mu)$.  Dividing by the disk
one-point function, $-K_0 Q/(2\bb\wt\mu)$, we obtain the ratio
$-K/K_0$.
The $\tau$-propagator is given in \eqref{eq: tau_prop_result} to be $\frac{2}{K}$.
Therefore, using \eqref{e253xx} and \eqref{eq: bc_mode_exp}, we obtain the contribution
to $g$ 
\begin{align}\label{eq: g^(b)_Ist_step}
    g_sg^{(\text{b})}
    &=\frac{1}{K_0}\,  \mathcal{N}_{\text{CO}}\mathcal{N}_{\text{O}}\int d t \left\langle 2\pi b_0 \left(\partial\eta e^{2\phi}b+\partial(\eta e^{2\phi}b)\right)(w_p)\partial \xi c\partial ce^{-2\phi}(0)\right\rangle_{T^2} \, .
\end{align}
We have doubled the annulus geometry to a rectangular torus with modular parameter $\ii t$.
The holomorphic and the anti-holomorphic integrals over the $b$-ghost in \eqref{e253xx} combine to yield $2\pi b_0$ on the torus.
For convenience, we will work in the $u=w/2\pi$ plane with the
identification $u\equiv u+1\equiv u+\ii t$. 
Since $b_0$ as well as the vertex operators are invariant under scaling, the form of the correlation function does not change.
After defining
\be
u_p = \frac{w_p}{2\pi}\, ,
\ee
we can express \eqref{eq: g^(b)_Ist_step} in the same form with $w_p$ replaced by $u_p$.
We now rewrite the correlator appearing in the integrand as
\begin{align} \label{ecorr112}
    \p_z \left\langle b_0 \eta(z) e^{2\phi}b(u_p)\p \xi c \p ce^{-2\phi}(0)\right\rangle_{T^2}\bigg|_{z=u_p}
    + \p_z \left\langle b_0 \eta e^{2\phi} b (z) \p \xi c \p ce^{-2\phi}(0) \right\rangle_{T^2}\bigg|_{z=u_p} \, .
\end{align}
To compute this, we need to evaluate the correlator
\be
    \left\langle b_0 \eta(z_1) e^{2\phi} (z_2) b(z_3)\partial \xi c\partial ce^{-2\phi}(w)\right\rangle_{T^2}\, .
\ee
The general $\eta \xi \phi$ correlator on the torus was first obtained in \cite[Eq.~(36)]{Verlinde:1987sd}.
We use the general formula to obtain the following correlator, up to an overall constant which is independent of the positions of the operators:
{\small
\begin{align}
\label{eq: eta_xi_first_steprep}
    &\langle \xi(x_1)\p\xi(x_2) \eta(y_1)  e^{-2\phi}(z_1) e^{2\phi}(z_2)\rangle^{\text{large}, \delta}_{T^2}\nonumber \\
     &= 
    \frac{\p}{\p x_2} \left(
    \frac{\vartheta[\delta](x_1+x_2-2y_1-2z_1+2z_2-2\Delta)}{\vartheta[\delta](x_2-y_1-2z_1+2z_2-2\Delta)\vartheta[\delta](x_1-y_1-2z_1+2z_2-2\Delta)}\,
    \frac{E(x_1,x_2)E(z_1,z_2)^4}{E(x_1,y_1) E(x_2,y_1)}
    \frac{\sigma(z_1)^4}{\sigma(z_2)^4} \right)
    \, .
\end{align}
}
In this expression, we have
\begin{align}
E(x,y) :=\frac{\vartheta_1(x-y)}{\vartheta_1'(0)} \,.   
\end{align}
Furthermore, $\delta$ denotes the spin structure on the torus, and
$\Delta=\frac{1+\ii t}{2}$ (see  \cite[Eq.~(6.37)]{DHoker:1988pdl}). 
For us, $\delta$ will correspond to the NS-NS spin structure for which 
\begin{align}
[\delta] =  \begin{bmatrix} 0\\0 \end{bmatrix} \,,\quad \text{and}
\quad 
\vartheta\begin{bmatrix} 0\\0 \end{bmatrix}(z) = \vartheta_3(z). 
\end{align}
The factor $\sigma(z_1)^4/\sigma(z_2)^4$ can be computed using \cite[Eq.~(7.56)]{DHoker:1988pdl} 
\begin{align}
    \frac{\sigma(z_1)}{\sigma(z_2)}=\frac{\vartheta_3(z_1-q+\Delta)}{\vartheta_3(z_2-q+\Delta)}\frac{\vartheta_1(z_2-q)}{\vartheta_1(z_1-q)}=e^{-\pi \ii (z_1-z_2)} \,,
\end{align}
where the second equality follows from the relation $\vartheta_{3}(z+\Delta)=e^{-\pi \ii(z+\frac{\ii t}{4})}\vartheta_1(z)$.
Further using $\vartheta_3(z-2\Delta)=e^{\pi t +2\pi \ii z}\vartheta_3(z)$, we see that right side of \eqref{eq: eta_xi_first_steprep} reduces to
\begin{align}
    \p_{x_2}
    \frac{\vartheta_3(x_1+x_2-2y_1-2z_1+2z_2) }{\vartheta_3(x_2-y_1-2z_1+2z_2)\vartheta_3(x_1-y_1-2z_1+2z_2)
    }
    \frac{E(x_1,x_2) E(z_1,z_2)^4}{E(x_1,y_1) E(x_2,y_1)}\,,
\end{align}
up to a position-independent factor. 
This factor is not important since the overall normalization of the correlator can be fixed from the leading singular term in the limit $x_2\to y_1$, $z_2\to z_1$. 
We will fix it at the end after combining the contributions from all the sectors.
 
It is known that the $x_1$ dependence drops out eventually, so, for convenience, we can set $x_1 = x_2$ after taking the derivative $\partial_{x_2}$.
Since $E(x_1,x_1) = 0$, the only non-trivial contribution comes from the term where $\p_{x_2}$ acts on the $E(x_1,x_2)$.
Evidently, $\p_{x_2} E(x_1,x_2) |_{x_1=x_2} = -1$, so we get
\be
\label{eq: eta_xi_first_stepreprep}
    \langle \p\xi(x_1) \eta(y_1) e^{-2\phi}(z_1) e^{2\phi}(z_2)\rangle^{\text{small}, \delta}_{T^2}=-
   \frac{\vartheta_3(2x_1-2y_1-2z_1+2z_2)E(z_1,z_2)^4}{\vartheta_3(x_1-y_1-2z_1+2z_2)^2 E(x_1,y_1)^2}\, .
\ee
From now on, we will write all formulas in the small Hilbert space.

Next, we compute the correlation function $\langle b_0 b(x) c \p c(y) \rangle$.
Since we need a $b_0 c_0$ factor to get a non-trivial result, it follows from the mode expansion in \eqref{eq: bc_mode_exp} that the $c_0$ comes from $c(y)$ and 
\be
   \langle b_0 b(x) c \p c(y) \rangle_{T^2} = \ii \langle b_0 c_0 b(x) \p c(y) \rangle_{T^2} \, .
\ee
The correlation function on the right side should be a function of $x-y$, be doubly periodic, and have a double pole at $x=y$.
It follows that
\be
    \langle  b_0 c_0 b(x) \p c(y) \rangle_{T^2} = C_1 \p_y^2 \log \vt_1(x-y) + C_2 \, ,
\ee
for some $C_1$, $C_2$ that only depend on $t$. 
To determine $C_2$, we can integrate both sides
of this equation over the range
$0\le y<1$, i.e. along a non-contractible cycle of the torus.
Since this integration contour does not intersect the integration contour that defines $b_0$, the integral of the left side vanishes because $c(y)$ is continuous along this circle and satisfies the periodicity condition $c(y+1) = c(y)$. On the other hand, the integral of the first term on the right side vanishes while the second term gives $C_2$. Matching these, we get $C_2 = 0$.
We will fix the overall multiplicative constant in the full string theory expression below, so we do not worry about fixing $C_1$ at this stage. Thus we have
\be
    \langle  b_0 b(x) c\p c(y)\rangle_{T^2} = 
    C_1\, \p_y^2 \log \vt_1(x-y)\, .
    \label{eq: bc_first_stepreprep}
\ee

Combining the results in \eqref{eq: eta_xi_first_stepreprep} and \eqref{eq: bc_first_stepreprep}, with the matter part only contributing to the overall normalization, we get the full string theory correlator
\be
    \left\langle b_0 \eta(z_1) e^{2\phi}(z_2) b(z_3)\partial \xi c\partial ce^{-2\phi}(w) \right\rangle_{T^2}
    = \NNN \,  \frac{\vartheta_3(2z_2-2z_1) E(z_2,w)^4}{\vartheta_3(2z_2-z_1-w)^2 E(z_1,w)^2} \, \p_{w}^2\,\log\vartheta_1(z_3-w)\, .
\ee
Here $\NNN$ is a normalization constant, which we determine as follows.
Setting $z_1=z_2=z_3$ and taking the limit $z_1\to w$, we can compare the leading term  i.e. the term proportional to $(z-w)^{0}$ on both sides to get $\ii \langle b_0 c_0 \rangle_{T^2} = \mathcal{N}/\vartheta_3(0)$.
Here we used \eqref{eq: bc_mode_exp} to replace the $c$ insertion with a factor of $\ii c_0$. 
Now using \eqref{e756aa}, we get $\NNN =  \ii \, \vartheta_3(0) \, Z(v)$,
and hence
{\small
\begin{align}
    \left\langle b_0 \eta(z_1) e^{2\phi}(z_2) b(z_3)\partial \xi c\partial ce^{-2\phi}(w) \right\rangle_{T^2}
    &= \ii\, Z(v)  \frac{\vartheta_3(0) \,  \vartheta_3(2z_2-2z_1) E(z_2,w)^4}{\vartheta_3(2z_2-z_1-w)^2 E(z_1,w)^2} \, \p_{w}^2\,\log\vartheta_1(z_3-w)\, .
\end{align}
}

In order to produce the correct PCO terms from \eqref{eq: g^(b)_Ist_step}, we first set $z_3=z_2$. 
For the first PCO term in \eqref{eq: g^(b)_Ist_step}, we take the $z_1$ derivative and then set $z_1=z_2=u_p$. 
For the second PCO term in \eqref{eq: g^(b)_Ist_step}, we first set $z_1=z_2=u_p$ and then take $u_p$ derivative. 
Further using $\vartheta_3'(0)=0$,  \eqref{ecorr112} evaluates to 
\be
    \left\langle b_0 \left(\partial\eta e^{2\phi}b+\partial(\eta e^{2\phi}b)\right)(u_p)\partial \xi c\partial ce^{-2\phi}(0)\right\rangle_{T^2} 
    = \ii \, Z(v) \frac{ \vartheta_3(0)^2 E(0,u_p)^2}{\vartheta_3(u_p)^2} \p_{u_p}^3 \log \vartheta_1(u_p) \, .
\ee
Using the identity\footnote{One can prove this using $\partial_v^3 \log \vartheta_1(v) = -\wp'(v)$ \cite{WikipediaThetaWeierstrass}, differentiating the relation 
$\wp(v) = e_2 + \frac{\vartheta_1'(0)^2}{\vartheta_3(0)^2}
\left(
\frac{\vartheta_3(v)}{\vartheta_1(v)}
\right)^2$,
where \(e_2=\wp((1+\tau)/2)\) is independent of $v$ \cite{DLMF:23.6.6}, and then using the identity (\ref{dx-theta1-theta3}) given below.}
\begin{align}
\label{eq: main_identity}
    \partial_v^3\log{\vartheta_1(v)} = 2\pi \, \vt_1'(0)^2 \,\frac{\vartheta_2(v) \vartheta_3(v)\vartheta_4(v)}{\vartheta_1(v)^3}\, ,
\end{align}
we obtain the final result\footnote{Very similar results were obtained in \cite{Agmon:2023zhi} while studying type IIB string theory in 10d flat space.}
\be
    \left\langle b_0 \left(\partial\eta e^{2\phi}b+\partial(\eta e^{2\phi}b)\right)(u_p)\partial \xi c\partial ce^{-2\phi}(0)\right\rangle_{T^2} = 2 \pi \ii \, Z(v) \frac{\vt_3(0)^2 \vt_2(u_p)\vt_4(u_p)}{\vt_3(u_p) \vt_1(u_p)}\, .
\ee
Substituting this into \eqref{eq: g^(b)_Ist_step}, and using \eqref{eq: NCO} and \eqref{eq: NO}, we get
\be
    \label{eq: gb_alt_PCO_final}
     g_sg^{(\text{b})}=\frac{g_s}{K_0}\int_{\left(\alpha^2-\frac{1}{2}\right)^{-1}}^1 \frac{d v}{v}\frac{\vartheta_3(0)^2\vartheta_2(u_p)\vartheta_4(u_p)}{\vartheta_1(u_p)\vartheta_3(u_p)}Z(v)\  .
\ee

\subsubsection{Vertical integration at the boundary between regions (b) and (d)}

With the choice of PCO location in the CO vertex in this appendix, the PCO section changes when we pass from region (b) to (d).
Therefore, we must include the contribution from the corresponding vertical integration.

Compared to the CO vertex, the C vertex on the annulus has an extra $\wt\XXX$ inserted at $w=w_c$ or, equivalently, $u=u_c$.
Recall that $u=w/ 2\pi$, so the periods are $1$ and $\ii t$.
In going from region (b) to (d), we thus have to move the PCO $\XXX$ from the point $u=u_p$ to $u = -u_c$, since $\tilde{\mathcal{X}}(u_c)=\mathcal{X}(-u_c)$ in cylinder coordinates. 
This is exactly the $g^{(\text{b})-(\text{d})}_{\text{VI}, \XXX}$ in \eqref{etorusbeg}
\begin{align}
    g_sg^{(\text{b})-(\text{d})}_{\text{VI}} &=2 \ii\, \frac{ \wt\lambda\pi  \mathcal{N}_{\text{C}}}{K_0 } 
    (1-\alpha^{-2})^{-1}
    \int_0^{\frac{1}{2\pi}\log(\alpha^2-\frac{1}{2})} dt \nonumber \\
    & \hskip 1in  \left\langle b_0 c_0 \left( \xi(u_p)-\xi(-u_c) \right)
    \eta(u_c) e^{\phi}(u_c) e^{-{\phi}}(-u_c) \right\rangle_{T^2}  \, .
\label{gvi-bd-app}
\end{align}
To evaluate this correlator, convert to the large Hilbert space by inserting a $\xi(u_p)$ at the leftmost position in the correlator.
Then the term involving $\xi(u_p) \xi(u_p)$ vanishes, so we only have to evaluate the term involving $-\xi(u_p) \xi(-u_c)$. 
Evaluating the correlators by the same steps as in the previous subsection, we find
\be
    \left\langle b_0 c_0 \left(\xi(u_p)-\xi(-u_c)\right) \eta(u_c) e^{\phi}(u_c) e^{-{\phi}}(-u_c) \right\rangle_{T^2}
    = Z(v) \frac{ \vartheta_3(u_p-u_c) \vartheta_1(u_p+u_c) }{ \vartheta_3 (u_p+u_c)  \vartheta_1(u_p-u_c) }\,,
    \label{b22-correlator}
\ee
Here, the overall normalization has been fixed by considering the limit $u_c\to 0$.
Since $u_c$ is small, we expand the right side to first order in $u_c$ as\footnote{We use the identity \cite[Eq.(1.9.8)]{lawden2013elliptic}
\be
    \partial_x \left(\frac{\vartheta_1(x)}{\vartheta_3(x)}\right)=\pi\vartheta_3(0)^2\frac{\vartheta_2(x)\vartheta_4(x)}{\vartheta_3(x)^2} \, .
\label{dx-theta1-theta3}
\ee
}
\begin{align}
\nonumber
    Z(v)\left(1+2\pi \vartheta_3(0)^2u_c\frac{\vartheta_2(u_p)\vartheta_4(u_p)}{\vartheta_1(u_p)\vartheta_3(u_p)}+\mathcal{O}(u_c^2)\right)\,,
\end{align}
Substituting this into \eqref{b22-correlator}, we can write \eqref{gvi-bd-app} as
\begin{align}
    g_sg^{(\text{b})-(\text{d})}_{\text{VI}} &=2\ii\, \frac{\wt\lambda\pi  \mathcal{N}_{\text{C}}}{K_0 } 
    \frac{1}{1-\alpha^{-2}}
    \int_0^{\frac{1}{2\pi}\log(\alpha^2-\frac{1}{2})} dt \,
    Z(v)\left(1+2\pi \vartheta_3(0)^2u_c\frac{\vartheta_2(u_p)\vartheta_4(u_p)}{\vartheta_1(u_p)\vartheta_3(u_p)}+\mathcal{O}(u_c^2)\right)\, .
\end{align}
Since, $u_c\simeq x_c\simeq (2\pi\tilde{\lambda})^{-1}$ we can drop the $\mathcal{O}(u_c^2)$ terms in the integrand. 
Moreover, terms with negative powers of $\alpha$ and $\tilde{\lambda}$ can be neglected.
After changing the integration variable to $v=e^{-2\pi t}$, we obtain 
\begin{align}\label{eq: gbdVI_alt_PCO_final}
    g_s g^{(\text{b})-(\text{d})}_{\text{VI}} &=-\frac{g_s \tilde{\lambda}}{2\pi K_0}\int_{(\alpha^2-\frac{1}{2})^{-1}}^1\frac{d v}{v}Z(v)-\frac{g_s}{K_0}\int_{(\alpha^2-\frac{1}{2})^{-1}}^1\frac{d v}{v}\frac{\vartheta_3(0)^2\vartheta_2(u_p)\vartheta_4(u_p)}{\vartheta_1(u_p)\vartheta_3(u_p)}Z(v)\,,
\end{align}
where we have used $\mathcal{N}_{\text{C}}=g_s\eta_c$.
It is evident that the second term cancels against \eqref{eq: gb_alt_PCO_final}, so
\be 
    g_s g^{(\text{b})} + g_s g^{(\text{b})-(\text{d})}_{\text{VI}} = -\frac{g_s \tilde{\lambda}}{2\pi K_0}\int_{(\alpha^2-\frac{1}{2})^{-1}}^1\frac{d v}{v}Z(v)\, .
\ee
This matches the result obtained with the original choice of PCO locations, for which $g^{(\text{b})}$ vanishes and $g^{(\text{b})-(\text{d})}_{\text{VI}}$ is given in \eqref{e7.61a}.

\subsubsection{Contribution from region (a) and vertical integration between regions (a) and (c)}

Having dealt with the change in contributions from the (b) and (d) regions, we now proceed to the (a) and (c) regions.
First, note that only $\tau$ propagates in the small-$x$ channel, i.e. the propagator labeled $q_2$ in Fig.~\ref{figcooo}(a).
Furthermore, there is no change in PCO prescription between regions (a) and (b), so the contribution from region (a) can be obtained by simply changing the integration region to $0\leq v \leq (\alpha^2-\frac{1}{2})^{-1}$ in \eqref{eq: gb_alt_PCO_final}
\begin{align}\label{eq: g_(a)_result}
    g_sg^{(\text{a})}=\frac{g_s}{K_0}\int_0^{\left(\alpha^2-\frac{1}{2}\right)^{-1}}\frac{d v}{v}\frac{\vartheta_3(0)^2 \vartheta_2(u_p) \vartheta_4(u_p)}{\vartheta_1(u_p) \vartheta_3(u_p)} Z(v) \, .
\end{align}
Similarly, we can obtain the vertical integration contribution from the (a)$-$(c) interface by changing the integration domain in \eqref{eq: gbdVI_alt_PCO_final}
\be
\label{eq: g_(a)_(c)_VI_1}
    g_sg^{(\text{a})-(\text{c})}_{\text{VI}}=-\frac{g_s \tilde{\lambda}}{2\pi K_0}\int_0^{(\alpha^2-\frac{1}{2})^{-1}}\frac{d v}{v}Z(v)-\frac{g_s}{K_0}\int_0^{(\alpha^2-\frac{1}{2})^{-1}}\frac{d v}{v}\frac{\vartheta_3(0)^2 \vartheta_2(u_p)\vartheta_4(u_p)}{\vartheta_1(u_p)\vartheta_3(u_p)}Z(v)\, .
\ee
Hence,
\be
\label{eq: g_(a)_(c)_VI_2}
     g_sg^{(\text{a})} + g_s g^{(\text{a})-(\text{c})}_{\text{VI}} =-{g_s\wt\lambda\over 2\pi K_0}\int_0^{\left(\alpha^2-\frac{1}{2}\right)^{-1}} \frac{d v}{v}Z(v)+\mathcal{O}(\tilde{\lambda}^{-1})
     ={g_s\wt\lambda\alpha\over \pi K_0} +\mathcal{O}(\tilde{\lambda}\alpha^{-1})\,,
\ee
where we have performed a small-$v$ expansion in the integrand and then used the replacement rules given in \eqref{ereprule} and \eqref{ereprule1}. 
This matches the result obtained with the original choice of PCO locations, for which $g^{(\text{a})}$ vanishes and $g^{(\text{a})-(\text{c})}_{\text{VI}}$ is given in \eqref{eq: g^ac_VI_fin_2}.

\subsubsection{Contribution due to gauge parameter redefinition}

In this subsection, we repeat the analysis of section \ref{subsec: gauge_param_redef} with the alternative choice of PCO locations. 
We show that although some contributions to $\mathcal{A}_{\text{COOO}}$ change, the final result is unchanged.

\subsubsection*{Contribution from Fig.~\ref{figcooo}(d)}
Let us first discuss the contribution from the COOO interaction vertex, which 
contributes to the Feynman diagram shown in Fig.~\ref{figcooo}(d). 
The PCOs are placed as $\XXX$ on $V_{\Psi}$, $\XXX$ on $V_{\Xi}$, and $\XXX$ on the closed-string vertex operator $V$, rather than the symmetric insertion $(\XXX+\widetilde{\XXX})/2$ used in the main text.
The first two terms in the second line of \eqref{ghost contribution bulk} correspond to $\mathcal{X}V$.
Thus, the correlator in \eqref{ghost contribution bulk} needs to be changed to
\be
    \Bigg\langle \frac{-\mathbb{I}}{2}(f_1) \, (-\mathcal{B}_{\beta_1}) \, (\eta c \partial c)(f_2) \,  (-\mathcal{B}_{\beta_2}) \, \frac{\mathbb{-I}}{2}(f_3) \,
    \left(-\frac{1}{2} \,\eta \,  \widetilde{c}\, e^{\phi} e^{-\widetilde{\phi}}\,  e^{\bb \varphi} + c\widetilde{c}\,  e^{-\widetilde{\phi}}\,  (-\ii \bb \psi) \,  e^{\bb \varphi} \right)(\ii) \Bigg\rangle\, ,
\ee
As earlier, this vanishes because there are no terms with an equal numbers of $\eta$ and $\xi$ insertions.
 
\subsubsection*{Contribution from Fig.~\ref{figcooo}(a) and Fig.~\ref{figcooo}(b)} 
Fig.~\ref{figcooo}(a) represents the Feynman diagram where one CO and a pair of OOO interaction vertices are connected  by a pair of open string propagators, one between CO and OOO, and another between the two OOOs.  Fig.~\ref{figcooo}(b) represents the Feynman diagram where a COO and a OOO interaction vertices are connected by an open-string propagator. 
The leading contribution carrying non-negative powers of $\alpha$ and $\wt\lambda$ arises from either the out-of-Siegel-gauge mode $\tau$ or the tachyon flowing along the open-string propagators. 
In particular, either the OSG mode or the tachyon flows in the $q_1$ channel. 

Recall that the three external open-string states $V_{\Psi}, V_{\Xi}$ and $V_{\Xi^*}$ are GSO even. 
If the tachyon flows in the $q_1$ channel, the OOO factor with two external O insertions will vanish because the tachyon is GSO odd.

If $\tau$ flows in the $q_1$ channel, then the other two external open-string states connected to the OOO vertex must have total ghost number two. But this is not possible since the three external open-string states carry ghost numbers 0, 0 and 3. Hence this contribution vanishes.

This implies that the non-trivial contributions from Fig.~\ref{figcooo}(a) and (b) will have negative powers 
of $\tilde{\lambda}$ or $\alpha$.
Hence, the contributions from these diagrams can be ignored. This is the same result that we obtained with the original choice of PCOs.

\subsubsection*{Contribution from Fig.~\ref{figcooo}(c)}

Now we evaluate the contribution from Fig.~\ref{figcooo}(c), where a CO
vertex is joined to an OOOO vertex by an open-string propagator. 
If the tachyon propagates in the $q_2$ channel, the CO factor vanishes because the C insertion is GSO even while the tachyon is GSO odd.
Thus, we only need to consider $\tau$ propagating in the $q_2$ channel. 
Therefore, OOOO vertex will have insertions of $\Xi\partial\xi c e^{-2\phi}, \Psi\partial \xi c e^{-2\phi}, \Xi^*\eta c \partial c $ and $\tau\partial \xi c\partial c e^{-2\phi}$. 
Note that the $\tau$ can carry either the Chan-Paton factor $\text{diag}(1,0)$ or $\text{diag}(0,1)$.

The OOOO interaction vertex described in section \ref{soooo} has cyclic symmetry, but we have to sum over the six inequivalent cyclic orderings of these vertex operators. 
The trace over the Chan-Paton factors is nonzero only for the three of these orderings which we discuss below. 
In each of these cases, the CO vertex contributes $\frac{KQ}{2b \tilde{\mu}}$, as computed in \eqref{CO Out-of-Siegel}. 
Additionally, we also have a factor of $\tau_{\text{prop}}$ coming from the $\tau$ propagator. 
We will need the details of local coordinates around $0,x,1$ and $\infty$ given in \eqref{eq: OOOO_trans_func_explicit} and the PCO locations in \eqref{eperminv}. 
\begin{itemize}
    \item \textbf{Permutation I $(\Xi \tau \Xi^* \Psi)$:} 
    Here and in the permutations discussed below, an ordering such as $\Xi\,\tau\,\Xi^*\,\Psi$ denotes the left-to-right order of the corresponding vertex operators on the real axis of the UHP for the OOOO vertex.
    For this permutation, we will have the following contribution
\begin{align} 
       &\frac{KQ\tau_{\text{prop}}\mathcal{N}_{\text{CO}}\mathcal{N}_{\text{OOOO}}}{2\widetilde{\mu}\bb} \nonumber \\ & \times \ \int_{\epsilon}^{1-\epsilon} dx \left\langle  \mathcal{X}(\hat{y}_1) \ \mathcal{X}(\hat{y}_2) \ \Xi\partial \xi c e^{-2\phi}(0) \ (-\mathcal{B}_x)\partial \xi c \partial c e^{-2\phi}(x) \ \Xi^*\eta c \partial c(1) \ \Psi \partial \xi c e^{-2\phi}(\infty)\right\rangle\, ,
\end{align}
where, $\hat{y}_{1,2}$ are the PCO locations in the OOOO UHP \eqref{eperminv}. 
A new feature of this calculation is that the PCO locations $\hat{y}_{1,2}$ depend on the modulus $x$ and are not placed on the insertions. 
As a consequence, we need to consider the terms proportional to $\dfrac{\partial \xi}{\mathcal{X}}$ in $\mathcal{B}_x$ given in the last line of \eqref{eapp0}. 
However, in our computation, these terms in $\mathcal{B}_x$ violate
$\eta\xi$ charge conservation on the disk and hence do not contribute.
For this reason, we will ignore these terms in the other permutations as well.

Now, we can, in principle, glue the OOOO to CO at $0,x, 1$ or $\infty$ in the OOOO UHP, i.e., we can put $\tau$ at $0,x,1$ or $\infty$ maintaining the order $\Xi \tau\Xi^* \Psi$ around the OOOO disk. Since the OOOO vertex we have constructed in section \ref{soooo} is symmetric under the cyclic permutations of the four open string insertions, we can just consider any one of the above four possibilities and work with it. 
We choose to glue at the insertion point $x$, so that $\tau$ is placed
at $x$, giving
\begin{align}
    A_{\text{I}, x} = &-\frac{KQ\tau_{\text{prop}}\mathcal{N}_{\text{CO}}\mathcal{N}_{\text{OOOO}}}{2\widetilde{\mu}\bb}\, \Xi\Xi^* \Psi \int_{\epsilon}^{1-\epsilon} dx \nonumber\\
    \bigg( 
    &-\tilde{f}_2'\left\langle  \mathcal{X}(\hat{y}_1) \ \mathcal{X}(\hat{y}_2) \ \partial \xi c e^{-2\phi}(0) \ \partial \xi  \partial c e^{-2\phi}(x) \ \eta c \partial c(1)  \ \partial \xi c e^{-2\phi}(\infty)\right\rangle\nonumber\\
    &+\frac{\tilde{g}_2'}{\tilde{g}_2}\left\langle  \mathcal{X}(\hat{y}_1) \ \mathcal{X}(\hat{y}_2) \ \partial \xi c e^{-2\phi}(0) \ \partial \xi c e^{-2\phi}(x) \ \eta c \partial c(1) \ \partial \xi c e^{-2\phi}(\infty)\right\rangle\nonumber\\
    &-\frac{\tilde{g}_3'}{\tilde{g}_3}\left\langle \mathcal{X}(\hat{y}_1) \ \mathcal{X}(\hat{y}_2) \ \partial \xi c e^{-2\phi}(0) \ \partial \xi c \partial c e^{-2\phi}(x) \ \eta c (1) \ \partial \xi c e^{-2\phi}(\infty)\right\rangle\bigg)\, .\label{eq: integrand_AIx}
\end{align}
From the product $\XXX(\hat{y}_1)\XXX(\hat{y}_2)$, only the following term contributes
\begin{align}
    \frac{1}{4}\left(\partial \eta e^{2\phi}b(\hat{y}_1)+\partial( \eta e^{2\phi}b)(\hat{y}_1)\right)\left(\partial \eta e^{2\phi}b(\hat{y}_2)+\partial( \eta e^{2\phi}b)(\hat{y}_2)\right)\, .\label{eq: XX_prod_term_2}
\end{align} 
The $\widetilde{f}_a$ and $\widetilde{g}_a$ can be read from the small-$w_a$ expansion of \eqref{eq: OOOO_trans_func_explicit}. 
Using the expressions for $\hat{y}_{1,2}$ in \eqref{eperminv}, we find
\begin{align} \label{eB42}
    A_{\text{I}, x}=&-\frac{KQ\tau_{\text{prop}}\mathcal{N}_{\text{CO}}\mathcal{N}_{\text{OOOO}}}{8\widetilde{\mu}\bb}\Xi\Xi^*\Psi \int_{\epsilon}^{1-\epsilon}d x \, \left(-\frac{3}{x}\right) \, .
\end{align}
We have calculated the quantity in the bracket in \eqref{eq: integrand_AIx} in the \texttt{Mathematica} notebook accompanying the arXiv submission.
We can analogously proceed for the other two permutations that lead to non-vanishing results.
 \item \textbf{Permutation II ($\Psi \tau \Xi \Xi^*$):} For this permutation, we get
\begin{align}
   A_{\text{II},x} &=\frac{KQ\tau_{\text{prop}}\mathcal{N}_{\text{CO}}\mathcal{N}_{\text{OOOO}}}{2\widetilde{\mu}\bb}\Xi\Xi^*\Psi \int_{\epsilon}^{1-\epsilon} dx \nonumber\\
   \bigg(&+\tilde{f}_2'\left\langle  \mathcal{X}(\hat{y}_1) \ \mathcal{X}(\hat{y}_2) \ \partial \xi c e^{-2\phi}(0) \ \partial \xi  \partial c e^{-2\phi}(x) \ \partial \xi c e^{-2\phi}(1) \ \eta c \partial c(\infty) \right\rangle\nonumber\\
    &-\frac{\tilde{g}_2'}{\tilde{g}_2}\left\langle  \mathcal{X}(\hat{y}_1) \ \mathcal{X}(\hat{y}_2) \ \partial \xi c e^{-2\phi}(0) \ \partial \xi c e^{-2\phi}(x)  \ \partial \xi c e^{-2\phi}(1) \ \eta c \partial c(\infty)\right\rangle\bigg)\, .\label{eq: integrand_AIIx}
\end{align}
Uing \eqref{eq: OOOO_trans_func_explicit} and \eqref{eperminv}, we have evaluated the integrand using \texttt{Mathematica} with the result
\begin{align} \label{eB44}
    A_{\text{II}, x}=&\frac{KQ\tau_{\text{prop}}\mathcal{N}_{\text{CO}}\mathcal{N}_{\text{OOOO}}}{8\widetilde{\mu}\bb}\Xi\Xi^*\Psi \int_{\epsilon}^{1-\epsilon}d x \, \left(\frac{3}{1-x} - \frac{3}{x}\right)\, .
\end{align}
\item \textbf{Permutation III ($\Xi^* \tau \Psi\Xi$):} Similar to the previous two cases, we get
\begin{align}
   A_{\text{III},x}&=\frac{KQ\tau_{\text{prop}}\mathcal{N}_{\text{CO}}\mathcal{N}_{\text{OOOO}}}{2\widetilde{\mu}\bb}\Xi\Xi^*\Psi \int_{\epsilon}^{1-\epsilon} d x \nonumber\\
   &\hspace{10 pt}\bigg(\frac{\tilde{g}_1'}{\tilde{g}_1}\left\langle  \mathcal{X}(\hat{y}_1) \ \mathcal{X}(\hat{y}_2) \eta c  (0)\,\partial \xi c \partial ce^{-2\phi}(x) \, \partial \xi c e^{-2\phi}(1)  \,\partial \xi c e^{-2\phi}(\infty) \right\rangle\nonumber\\
    &\hspace{30 pt}+\tilde{f}_2'\left\langle  \ \mathcal{X}(\hat{y}_1) \ \mathcal{X}(\hat{y}_2) \eta c \partial c (0)\,\partial \xi \partial ce^{-2\phi}(x) \, \partial \xi c e^{-2\phi}(1)  \,\partial \xi c e^{-2\phi}(\infty)\right\rangle \nonumber \\
    &\hspace{30 pt}-\frac{\tilde{g}_2'}{\tilde{g}_2}\left\langle  \ \mathcal{X}(\hat{y}_1) \ \mathcal{X}(\hat{y}_2) \eta c \partial c (0)\,\partial \xi ce^{-2\phi}(x) \, \partial \xi c e^{-2\phi}(1)  \,\partial \xi c e^{-2\phi}(\infty)\right\rangle\bigg)\, ,\label{eq: integrand_AIIIx}
\end{align}
which evaluates to
\begin{align} \label{eB46}
    A_{\text{III}, x}=&\frac{KQ\tau_{\text{prop}}\mathcal{N}_{\text{CO}}\mathcal{N}_{\text{OOOO}}}{8\widetilde{\mu}\bb}\Xi\Xi^*\Psi \int_{\epsilon}^{1-\epsilon} d x \,\left(-\frac{3}{1-x}\right)\, .
\end{align}
\end{itemize}

Adding \eqref{eB42}, \eqref{eB44} and \eqref{eB46}, we see that the net contribution of Fig.~\ref{figcooo}(c) vanishes.

\subsubsection*{Vertical integration between Fig.~\ref{figcooo}(a) and (c)}

The vertical integration across these regions corresponds to the movement of PCOs on the OOOO factor only. Since only $\tau$ (with vertex operator $\partial \xi c\partial c e^{-2\phi}$) can propagate between the CO and the OOOO interaction vertices, the OOOO vertex will have the open string insertions $\partial\xi c e^{-2\phi}, \partial \xi c e^{-2\phi}, \eta c \partial c $ and $\partial \xi c\partial c e^{-2\phi}$. 
There are two PCO's at $\hat{y}_{1,2}$ on the OOOO UHP and we will move one of them at a time. 
This will require us to remove the corresponding PCO and replace it with $\int\partial \xi$, keeping the other PCO fixed.
Hence, leaving aside the PCO, we have four factors of $\p\xi$ and one factor of $\eta$ in the correlator. 
Since a PCO can supply at most one $\eta$, the correlator vanishes by
$\xi\eta$ charge conservation.

\subsubsection*{Vertical integration between Fig.~\ref{figcooo}(a) and (b)}
The same argument as in section \ref{s797} tells us that the contribution from
this vertical integration vanishes when we drop all terms carrying negative powers of either $\alpha$
or $\wt\lambda$.

\subsubsection*{Vertical integration between Fig.~\ref{figcooo}(b) and (d)}

The analysis of this contribution will be identical to that in section \ref{s796} except
that in step 2, instead of averaging over the contributions $V^1_{\beta,\XXX}$ and
$V^1_{\beta,\wt\XXX}$, we only have to consider the contribution $V^1_{\beta,\XXX}$.
However the net contribution remains unchanged, and we get the same result as
\eqref{eq: COOO_contribution_final}:
\begin{align}\label{eq: COOO_contribution_final_appendix}
    K\,  \mathcal{N}_{\text{COOO}}  \frac{Q}{2\ii\bb\widetilde{\mu}}
    \tilde{\lambda} \, .
\end{align}

\subsubsection*{Vertical integration between Fig.~\ref{figcooo}(c) and (d)}

In Fig.~\ref{figcooo}(d), two PCOs are located on top of two O insertions ($\Xi$ and $\Psi$).
In Fig.~\ref{figcooo}(c), they are induced from the symmetric points 
$\hat{y}_{1,2}$ in the OOOO upper half plane. 
Hence at the boundary between the moduli space regions covered by these two diagrams, we have to move the two $\mathcal{X}$s from $\Xi$ and $\Psi$ to the points $\hat{y}_{1}$ and $\hat{y}_{2}$ on the OOOO UHP one by one.
These vertical integrations happen on a dimension-one slice, parametrized by $x$. For brevity, we will omit writing the explicit integral $\int_{\epsilon}^{1 - \epsilon} dx$ until later.

\paragraph{Step 1.}
In the first step of vertical integration, we move the PCO from $\Xi$ at $f_1$
to the point $\hat{y}_1$ in the bulk of the OOOO factor. 
From the perspective of CO factor, it is a small movement of PCO near the boundary. 
The integrand can be written analogous to \eqref{edeg1}
\begin{align}
    V_{x,\mathcal{X}}^1 &= 
    -\mathcal{N}_{\text{COOO}}\Bigg\langle \left(\sum_{a=1}^3\oint_{z_a}\frac{\partial F_a}{\partial x}b(z)dz\right)  \nonumber \\
    & \hspace{70pt}\left(-2(\xi(f_1)-\xi(y_1))\right)\Phi_C \mathcal{X} V(\ii) \ \Xi \partial \xi ce^{-2\phi}(f_1)\ \Xi^*\eta c \partial c(f_2)\left(-\Psi \frac{\mathbb{I}}{2}\right)(f_3)\Bigg\rangle \,.
\end{align}
Here $y_1$ denotes the image, on the upper half-plane for the CO vertex, of the bulk OOOO point $\hat y_1$ from which the PCO is moved to $f_1$.
We have also checked that the measure $dx\wedge dq_1$ is positive, by an argument analogous to the one below \eqref{ghost contribution bulkqbeta}, so no extra sign arises.
We now use the fact that only the term $-\frac{1}{2}\eta\tilde{c}\,e^{\phi}e^{-\tilde{\phi}}e^{\bb\varphi}$ in $\mathcal{X}V$ contributes; the remaining term in \eqref{e6.19a} violates $\phi$-momentum conservation. 
We now use \eqref{one-pt-vb} to get
\begin{align}
    V_{x,\mathcal{X}}^1 &= \mathcal{N}_{\text{COOO}}\Xi\Xi^*\Psi \frac{Q}{4\ii \bb\widetilde{\mu}}
    \bigg(f_1'\left\langle\xi(y)\eta\tilde{c} e^{\phi}e^{-\tilde{\phi}}\partial \xi e^{-2\phi}(f_1)\eta c \partial c(f_2)\right \rangle \nonumber \\
    &\hspace{90pt} +f_2'\left\langle\xi(y)\eta\tilde{c} e^{\phi}e^{-\tilde{\phi}}\partial \xi ce^{-2\phi}(f_1)\eta \partial c(f_2)\right \rangle\nonumber\\
    &\hspace{90 pt}-\frac{g_2'}{g_2}\left\langle\xi(y)\eta\tilde{c} e^{\phi}e^{-\tilde{\phi}}\partial \xi c e^{-2\phi}(f_1)\eta c (f_2)\right \rangle\bigg)\bigg|_{y=y_1}^{y=f_1}\, .
\end{align}
Evaluating the correlators, we get
\begin{align}\label{eq: Vert_2_CO_OOOO_1}
    V_{x,\mathcal{X}}^1=&-\mathcal{N}_{\text{COOO}}\Xi\Xi^*\Psi \frac{KQ}{4\ii \bb\widetilde{\mu}}
    \Bigg(-2 \ii (f_2-\ii) (f_1-y_1)^2 \bigg(\frac{(f_2' g_2 (f_1-2 f_2-\ii)+(f_2+\ii) g_2' (f_2-f_1))}{(f_1+\ii) g_2 (y_1-\ii) (f_1-f_2)^2 (f_2-y_1)} \nonumber\\
    &\hspace{100 pt}+\frac{f_1' (f_2+\ii)^2}{(f_1+\ii)^2 (y_1-\ii) (f_1-f_2)^2 (f_2-y_1)}\bigg)\Bigg)\, .
\end{align}

\paragraph{Step 2.}
Next, we move the PCO on $\Psi$ (at $f_3$) to the point $\hat{y}_2$ in the OOOO UHP. 
We denote by $y_2$ the image of $\hat y_2$ on the CO, equivalently COOO,
upper half-plane.
We get the following contribution 
\begin{align}
    V_{x,\mathcal{X}}^2 &=
    -\mathcal{N}_{\text{COOO}} 
    \Bigg\langle 
    \left(\sum_{a=1}^3\oint_{z_a}
        \frac{\partial F_a}{\partial x}b(z)dz
    \right) \nonumber \\
    & \left(-2\xi(y)\right)\mathcal{X}(y_1)\mathcal{X}V(\ii)\, \Xi \partial \xi ce^{-2\phi}(f_1)\,\Xi^*\eta c \partial c(f_2)\,\Psi \partial \xi ce^{-2\phi}(f_3)\Bigg\rangle\Bigg|_{y=y_2}^{y=f_3}\, .
\end{align}
From the product $\mathcal{X}(y_1)\mathcal{X}V(\ii)$, only the following terms contribute:
\begin{align}
    \left(-\frac{1}{2}\partial \eta e^{2\phi}b-\frac{1}{2}\partial( \eta e^{2\phi}b)\right)(y_1)\left(-\frac{1}{2}\eta\tilde{c}e^{\phi}e^{-\tilde{\phi}}e^{b\varphi}(\ii)\right) \, , 
\end{align}
as all the other terms violate $\phi$-momentum conservation. 
Again, using \eqref{one-pt-vb}, we get 
{\small
\begin{align}
   V_{x,\mathcal{X}}^2&=\mathcal{N}_{\text{COOO}}\Xi\Xi^*\Psi \frac{Q}{4\ii\bb\widetilde{\mu}}\nonumber\\
   &\hspace{10 pt}\times\Bigg(f_1'\left\langle\xi(y) \left(\partial \eta e^{2\phi}b+\partial( \eta e^{2\phi}b)\right)(y_1)\eta\tilde{c}e^{\phi}e^{-\tilde{\phi}}(\ii) \partial \xi e^{-2\phi}(f_1)\,\eta c \partial c(f_2) \, \partial \xi c e^{-2\phi}(f_3)\right \rangle\nonumber\\
    &\hspace{30 pt} +f_2'\left\langle \xi(y)\left(\partial \eta e^{2\phi}b+\partial( \eta e^{2\phi}b)\right)(y_1)\eta\tilde{c}e^{\phi}e^{-\tilde{\phi}}(\ii)\partial \xi ce^{-2\phi}(f_1)\,\eta \partial c(f_2) \, \partial \xi c e^{-2\phi}(f_3)\right \rangle\nonumber\\
    &\hspace{30 pt} -\frac{g_2'}{g_2}\left\langle\xi(y)\left(\partial \eta e^{2\phi}b+\partial( \eta e^{2\phi}b)\right)(y_1)\eta\tilde{c}e^{\phi}e^{-\tilde{\phi}}(\ii)\partial \xi c e^{-2\phi}(f_1)\,\eta c (f_2)\, \partial \xi c e^{-2\phi}(f_3)\right \rangle \nonumber\\
    &\hspace{30 pt}-f_3'\left\langle\xi(y)\left(\partial \eta e^{2\phi}b+\partial( \eta e^{2\phi}b)\right)(y_1)\eta\tilde{c}e^{\phi}e^{-\tilde{\phi}}(\ii)\partial \xi c e^{-2\phi}(f_1)\,\eta c \partial c(f_2) \, \partial \xi  e^{-2\phi}(f_3)\right \rangle\Bigg)\Bigg|_{y=y_2}^{y=f_3}\, . 
\end{align}
}
\par\noindent
Since the resulting expression for the correlator is lengthy and not particularly
illuminating, we do not display it here.
Now, for each of these steps, we have to sum over three cyclic permutations as before. 
These calculations are carried out in the \texttt{Mathematica} file included with the arXiv submission.
\begin{itemize}
    \item \textbf{Permutation I} $(\Xi \tau \Xi^* \Psi)$: We will define the vertical integration contributions from steps 1 and 2 for the first permutation as follows
    \begin{align}
        V^{\alpha}_{\text{I}}=V^{\alpha}_{x, \mathcal{X}}|_{f_{\{1, 2, 3\}}\rightarrow f^x_{\{0,1,\infty\}}, \, g_{\{1, 2, 3\}}\rightarrow g^x_{\{0,1,\infty\}}},\quad  \alpha=1,2\, ,
    \end{align}
    where $f^x_{\{0,1,\infty\}}$ and $g^x_{\{0,1,\infty\}}$ are the leading and subleading coefficients in the near-insertion expansion of the transition functions given in \eqref{eq: SymmOOOO_vertex_gluing}. Using this definition, we get 
    \begin{align}\label{eq: perm_1_VI}
        V_{\text{I}}^1+V_{\text{I}}^2=-\mathcal{N}_{\text{COOO}}\Xi\Xi^*\Psi \frac{KQ}{4\ii\bb\widetilde{\mu}}\int_{\epsilon}^{1-\epsilon}dx\left(-\frac{2\ii}{x}+\mathcal{O}(\lambda^{-1})\right)\, .
    \end{align}
    \item \textbf{Permutation II} ($\Psi \tau \Xi \Xi^*$): For the second  permutation, the respective contributions from step 1 and step 2 are defined as
    \begin{align}
        V^{\alpha}_{\text{II}}=V^{\alpha}_{x, \mathcal{X}}|_{f_{\{1, 2, 3\}}\rightarrow f^x_{\{1,\infty,0\}}, \, g_{\{1, 2, 3\}}\rightarrow g^x_{\{1,\infty,0\}}},\quad  \alpha=1,2\, ,
    \end{align}
    leading to the following contribution at leading order in $\lambda^{-1}$,
    \begin{align}\label{eq: perm_2_VI}
        V^1_{\text{II}}+V^2_{\text{II}}=-\mathcal{N}_{\text{COOO}}\Xi\Xi^*\Psi \frac{KQ}{4\ii\bb\widetilde{\mu}}\int_{\epsilon}^{1-\epsilon}dx\left(\frac{4 \ii (2 x-1)}{(x-1) x}+\mathcal{O}(\lambda^{-1})
        \right)\, .
    \end{align}
    \item \textbf{Permutation III} ($\Xi^* \tau \Psi\Xi$): For the third permutation, we can analogously define
    \begin{align}
        V^{\alpha}_{\text{III}}=V^{\alpha}_{x, \mathcal{X}}|_{f_{\{1, 2, 3\}}\rightarrow f^x_{\{\infty,0,1\}}, \, g_{\{1, 2, 3\}}\rightarrow g^x_{\{\infty,0,1\}}},\quad  \alpha=1,2\, ,
    \end{align}
    leading to the following contribution at leading order in $\lambda^{-1}$
    \begin{align}\label{eq: perm_3_VI}
        V^1_{\text{III}}+V^2_{\text{III}}=-\mathcal{N}_{\text{COOO}}\Xi\Xi^*\Psi \frac{KQ}{4\ii\bb\widetilde{\mu}}\int_{\epsilon}^{1-\epsilon}dx\left(\frac{2 \ii}{1-x}+\mathcal{O}(\lambda^{-1})\right)\, .
    \end{align}
\end{itemize}
Summing \eqref{eq: perm_1_VI}, \eqref{eq: perm_2_VI} and \eqref{eq: perm_3_VI}, we get 
\begin{align}
    \mathcal{N}_{\text{COOO}}\Xi\Xi^*\Psi \frac{Q}{4\ii\bb}\int_{\epsilon}^{1-\epsilon}d x \left(2 \ii \left(\frac{1}{x}-\frac{1}{1-x}\right)+\mathcal{O}(\lambda^{-1})\right)=\mathcal{O}(\lambda^{-1})\, .
\end{align}
Since $\lambda = \widetilde{\lambda}/\alpha$, inverse powers of $\lambda$ contain inverse powers of $\widetilde{\lambda}$, and we have been consistently setting such terms to zero.
Thus \eqref{eq: COOO_contribution_final_appendix} gives the only non-vanishing contribution to the COOO amplitude, yielding
\begin{align}\label{eq: COOO_contribution_final_appendixtot}
    \mathcal{A}_{\text{COOO}}=K\,  \mathcal{N}_{\text{COOO}}  \frac{Q}{2\ii\bb}
    \tilde{\lambda} \, .
\end{align}
This is the same as what we got with the earlier choice of PCO locations,  showing that $g_{\rm gauge}$ remains unchanged from the earlier value given in \eqref{eq: g_gauge_fin}.

Combining the contributions \eqref{eq: gb_alt_PCO_final}, \eqref{eq: gbdVI_alt_PCO_final}, \eqref{eq: g_(a)_result}, \eqref{eq: g_(a)_(c)_VI_2} and \eqref{eq: COOO_contribution_final_appendixtot}, we see that the annulus one-point function is independent of whether the CO and the COOO vertices are defined with $(\XXX+\wt\XXX)/2$ or with $\XXX$ at the closed-string
punctures.
This is a strong consistency check on our computations.

\bibliographystyle{apsrev4-1long}
\bibliography{main}
\end{document}